\documentclass[12pt]{article}
\usepackage{amsmath}
\usepackage{amssymb}
\usepackage{graphicx}
\usepackage{cite}

\numberwithin{equation}{section}

\begin{document}
\baselineskip=18pt
\begin{titlepage}
\begin{flushright}
{\small KYUSHU-HET-92}\\[-1mm] hep-ph/0604096%
\end{flushright}
\begin{center}
\vspace*{11mm}

{\large\bf%
Neutrino-induced Electroweak Symmetry Breaking\\[2mm]%
in Supersymmetric SO(10) Unification%
}\vspace*{8mm}

Kenzo Inoue, Kentaro Kojima, and Koichi Yoshioka%
\vspace*{1mm}

{\it Department of Physics, Kyushu University, Fukuoka 812-8581, Japan}

\vspace*{3mm}

{\small (April, 2006)}
\end{center}
\vspace*{5mm}

\begin{abstract}\noindent%
The radiative electroweak symmetry breaking, the unification of
third-generation Yukawa couplings, and flavor-changing rare decay are
investigated in two types of supersymmetric $SO(10)$ scenarios taking
into account of the effects of neutrino physics, i.e.\ the observed
large generation mixing and tiny mass scale. The first scenario is 
minimal, including right-handed neutrinos at intermediate scale
with the unification of third-generation Yukawa couplings. Another is
the case that the large mixing of atmospheric neutrinos originates
from the charged-lepton sector. Under the $SO(10)$-motivated boundary
conditions for supersymmetry-breaking parameters, typical low-energy
particle spectrum is discussed and the parameter space is identified
which satisfies the conditions for successful radiative electroweak
symmetry breaking and the experimental mass bounds of
superparticles. In particular, the predictions of the bottom quark
mass and the $b\to s\gamma$ branching ratio are fully analyzed. In
both two scenarios, new types of radiative electroweak symmetry
breaking are achieved with the effects of neutrino couplings. The
Yukawa unification becomes compatible with the bottom quark mass and
the experimental constraints from flavor-violating rare processes, and
the hierarchical superparticle mass spectrum is obtained.
\end{abstract}
\end{titlepage}

\newpage

%%%%%%%%%%%%%%%%%%%%%%%%%%%%%%%%%%%%%%%%%%%%%%%%%%%%%%%%%%%%%%%%%%%%%%
\section{Introduction}
%%%%%%%%%%%%%%%%%%%%%%%%%%%%%%%%%%%%%%%%%%%%%%%%%%%%%%%%%%%%%%%%%%%%%%
For the last decades, a lot of articles have been devoted to trying to
understand the underlying particle theory beyond the standard 
model (SM). Among them, the minimal supersymmetric standard 
model (MSSM) is conceived to be one of the most promising candidates
for its brevity and various attractive features. A well-known example
is the unification of SM gauge coupling constants at some high-energy
scale~\cite{GUT}. The exploration of supersymmetric grand unified
theory (GUT) has been therefore one of the most important subjects in
particle physics. In addition to the gauge coupling unification, the
GUT framework often leads to the unification of Yukawa coupling 
constants. For example, in the minimal $SU(5)$ model\cite{GG}, a
right-handed down-type quark and a lepton doublet belong to a single
quintuplet representation of $SU(5)$ group, and especially gives the
bottom-tau Yukawa unification at high-energy scale. It is found in the
MSSM that the bottom-tau unification is preferred in light of the
experimentally measured values of the bottom quark and tau lepton
masses. Moreover the detailed numerical analyses of
renormalization-group (RG) running of gauge and Yukawa couplings have
brought up a possibility of top-bottom-tau Yukawa unification with a
large ratio between two vacuum expectation values (VEVs) of the MSSM
Higgs doublets~\cite{btau}. Hence one is naturally led to the scenario
where all matter fermions of one generation are unified in a single
representation of GUT group. The simplest candidate for such
unified gauge group is $SO(10)$ where the SM fermions are included in
16-dimensional representations. Various phenomenological studies 
of $SO(10)$-type Yukawa unification have been performed in the
literature~\cite{Yunify,HRS,COPW,RS,BOP,Tata,BDR,TW}.

In the phenomenological study of Yukawa unification, one of the main
subjects is the masses of third-generation fermions. It is known that
the bottom quark mass is largely affected by low-energy threshold
corrections which are induced via decoupling of superpartners of SM
particles and bring a change of several tens of percents to the bottom
quark mass estimation~\cite{Hempfling,HRS,COPW,RS}. The magnitude of
the corrections depends on low-energy values of supersymmetry (SUSY)
breaking parameters and supersymmetric Higgs mass
parameter. Consequently the low-energy estimation of fermion masses 
also has strong dependence on the superparticle spectrum. It is argued 
from detailed analysis that the threshold correction to the bottom
quark mass must be suppressed than its naively expected value so that
the Yukawa unification leads to the top, bottom, and tau masses within
the experimentally allowed ranges~\cite{HRS,COPW,RS,TW}.

In addition to the successful prediction of charged fermion masses,
the recent experimental results for neutrino physics also seem to
prefer the unified theory. An important property of neutrinos is their
tiny mass scale compared to the other SM fermions, the smallness of
which scale is naturally realized by introducing right-handed
neutrinos and their large Majorana masses, called the seesaw
mechanism~\cite{seesaw}. The right-handed neutrinos are unified into
16-dimensional representations of $SO(10)$, combined with the other SM
fermions. Moreover the recent study of the solar and atmospheric
neutrinos~\cite{neuexp} has revealed that there is large generation
mixing in the lepton sector, while the corresponding mixing angles in
the quark sector are known to be small. Such contrastive generation
structure between quarks and leptons is naively difficult to be
obtained in unified theory, since as stated above the matter fermions
are combined into GUT multiplets. A way to ameliorate this problem is
proposed in GUT framework to naturally accommodate the large lepton
mixing with asymmetric forms of Yukawa matrices, the so-called
lopsided forms~\cite{ABB,SY,Ramond,NY,BK,others}.

In GUT models with the above-mentioned neutrino property, the RG
evolution of coupling constants is expected to be altered from the
naive top-bottom-tau Yukawa unification by including the effects of
neutrino couplings and/or the generation structure. Assuming the
seesaw mechanism and taking no account of lepton large mixing, the
phenomenological study of neutrino Yukawa effects on gauge and Yukawa
RG evolution has been performed, e.g.\ in
Refs.~\cite{VS,AK,CEIN}. The analysis has shown that the evaluation 
of gauge couplings and third-generation fermion masses is slightly
changed; for example, the prediction of top quark mass is made up
to 3~GeV~\cite{AK} and the low-energy value of the strong gauge
coupling constant is decreased by a few percents~\cite{CEIN}. A
relevance of leptonic generation mixing for the bottom-tau Yukawa
unification has also been discussed~\cite{btau-atmos}. These RG
studies however only deal with the gauge and Yukawa coupling
evolutions. As stated above, the low-energy threshold corrections to
the bottom quark mass often play a significant role in the analysis of
fermion masses. Since the corrections are determined by superparticle
mass spectrum, the inclusion of dimensionful parameters into the RG
analysis is an important issue for a complete study of neutrino effects.

Concerning the RG evolution of mass parameters, the radiative
electroweak symmetry breaking (EWSB)~\cite{EWSB} should be carefully
examined. The successful radiative EWSB generally restricts
SUSY-breaking parameters at high-energy scale and also low-energy
superparticle spectrum in the top-bottom-tau Yukawa unification. The
resulting spectrum often leads to a large magnitude of low-energy
threshold corrections, following which the prediction of
third-generation fermion masses are difficult to be consistent with
the experimentally allowed ranges~\cite{HRS,COPW,RS,TW}.

In the present work, we perform detailed analysis of radiative EWSB
and third-generation fermion masses in grand unified models, taking
into account the neutrino property indicated by the recent experimental
results. In addition, the $b\to s\gamma$ decay rate is also evaluated
since the experimental constraint on the decay rate tends to severely
restrict low-energy superparticle spectrum~\cite{BOP,bsg}. For
comparison and completeness, we first review the top-bottom-tau Yukawa
unification without neutrino effects, following which we will include
the effects of neutrino physics in two cases; the first is to consider
the Yukawa unification of top, bottom, tau, and third-generation
neutrino where no other (small) elements in Yukawa matrices are
involved, and in the second case, the large generation mixing in the
lepton sector is included in the analysis by assuming the lopsided
form of Yukawa matrices. In both cases, we introduce large Majorana
masses for right-handed neutrinos, and tiny neutrino masses are
induced through the seesaw mechanism. Since low-energy threshold
correction to the bottom quark mass is rather sensitive to
SUSY-breaking parameters, the running bottom quark mass is treated as
an output parameter in order to reveal the behavior of threshold
corrections. On the other hand, the physical top quark mass is used as
an input quantity. In this paper we adopt the 
value $m_t^{\rm pole}\!=178$~GeV~\cite{PDG} and give some comments on
the case of the most recent report on the top quark 
mass $m_t^{\rm pole}\!=172.7$~GeV~\cite{Tevatron}.

The paper is organized as follows. In Section~\ref{sec:general} we
present an overview of the third-generation fermion 
masses, the $b\to s\gamma$ constraint, and radiative EWSB in Yukawa
unification scenario where low-energy SUSY-breaking quantities are
treated as free parameters. This treatment reveals preferred types
of low-energy superparticle spectrum and provides an useful reference
for later discussions of radiative EWSB in specific models. In
Section~\ref{sec:so10} the phenomenology of the top-bottom-tau Yukawa
unification is discussed. Sections~\ref{sec:wRHnu} and \ref{sec:lop}
include the neutrino effects into the analysis of EWSB; with the
Yukawa unification of top-bottom-tau and third-generation neutrino in 
Section~\ref{sec:wRHnu} and with the lopsided form of mass textures
in Section~\ref{sec:lop}. We summarize the results in
Section~\ref{sec:sum}.

%%%%%%%%%%%%%%%%%%%%%%%%%%%%%%%%%%%%%%%%%%%%%%%%%%%%%%%%%%%%%%%%%%%%%%
\bigskip
\section{General Aspects in Yukawa Unification}
\label{sec:general}
%%%%%%%%%%%%%%%%%%%%%%%%%%%%%%%%%%%%%%%%%%%%%%%%%%%%%%%%%%%%%%%%%%%%%%
In this section, we present an overview of low-energy phenomenology in
general Yukawa unification scenario; the third-generation fermion
masses, the $b\to s\gamma $ decay, and the EWSB are discussed. We
assume the gauge coupling unification and also Yukawa coupling
unification at the GUT scale. The low-energy values of these
dimensionless couplings are determined by solving RG equations. On the 
other hand, there is no assumption about GUT models and dimensionful
parameters at high-energy regime. Accordingly, low-energy
SUSY-breaking parameters are treated as independent variables. The
general analysis given here reveals preferred types of low-energy
spectrum and would be useful for later discussion in specific
high-energy models with neutrino effects.

Below the GUT scale, the theory is assumed to be the MSSM whose
superpotential is given by
\begin{equation}
  W_{\rm MSSM} \,=\, Q_i(Y_u)_{ij}\bar u_j H_u 
  +Q_i(Y_d)_{ij}\bar d_j H_d +L_i(Y_e)_{ij}\bar e_j H_d +\mu H_u H_d,
  \label{MSSMspot}
\end{equation}
where $Q_i$, $\bar u_i$, $\bar d_i$, $L_i$, $\bar e_i$ and $H_u$,
$H_d$ are three-generation matter superfields and Higgs superfields,
respectively. The $3\times3$ complex matrices $Y_u$, $Y_d$, $Y_e$ with
generation indices mean Yukawa coupling constants and $\mu$ is the
supersymmetric Higgs mass parameter. We assume the Yukawa coupling
unification in that the 3-3 elements of Yukawa matrices, i.e.\ top,
bottom, and tau Yukawa couplings, $y_t$, $y_b$ and $y_\tau$, are
unified at high-energy scale:
\begin{equation}
  y_t(M_G) \,=\, y_b(M_G) \,=\, y_\tau(M_G) \;\equiv\, y_{{}_G},
\end{equation}
where $M_G\simeq10^{16}$~GeV is the GUT scale. The precise definition 
of $M_G$ will be given in the next section.

%%%%%%%%%%%%%%%%%%%%%%%%%%%%%%%%%%%%%%%%%%%%%%%%%%%%%%%%%%%%%%%%%%%%%%
\subsection{Top, Bottom, Tau Masses in Yukawa Unification}
\label{sec:tbtau}
%%%%%%%%%%%%%%%%%%%%%%%%%%%%%%%%%%%%%%%%%%%%%%%%%%%%%%%%%%%%%%%%%%%%%%
We first review the RG evolution of third-generation Yukawa couplings
in Yukawa unification scenario. It is found that the Yukawa
unification is compatible with the observed values of heavy fermion
masses, while the low-energy prediction of mass eigenvalues is rather
sensitive to superparticle spectrum~\cite{HRS,RS,TW}. In the following
analysis, the exact top-bottom-tau Yukawa coupling unification is
assumed at the GUT scale and other small entries in Yukawa matrices
are safely neglected. Once the unified gauge 
coupling $g_{{}_G}=g_1(M_G)=g_2(M_G)=g_3(M_G)$ and the unified Yukawa
coupling $y_{{}_G}$ are fixed, one can evaluate with the MSSM RG
equations the $\overline{\text{DR}}$ running gauge and Yukawa
couplings at the $Z$-boson mass scale $M_Z$. The Yukawa couplings and
the third-generation $\overline{\text{DR}}$ running fermion masses are
related as
{\allowdisplaybreaks%
\begin{eqnarray}
  m_t^{\overline{\text{DR}}}(M_Z) &\,=\,& 
  v_H\sin\beta\, y_t^{\overline{\text{DR}}}(M_Z)\,
  (1+\Delta_t), \nonumber \\
  m_b^{\overline{\text{DR}}}(M_Z) &\,=\,& 
  v_H\cos\beta\, y_b^{\overline{\text{DR}}}(M_Z)\,
  (1+\Delta_b), \nonumber \\
  m_\tau^{\overline{\text{DR}}}(M_Z) &\,=\,& 
  v_H\cos\beta\, y_\tau^{\overline{\text{DR}}}(M_Z)\,
  (1+\Delta_\tau),
\end{eqnarray}}%
where $v_H$ parametrizes the $\overline{\text{DR}}$ Higgs VEV at the
electroweak scale which is taken as $v_H=174.6$~GeV\@. 
The angle $\beta$ is defined by the VEV ratio of two Higgs doublets
as $\tan\beta=\langle H_u^0\rangle/\langle H_d^0\rangle$. The
low-energy threshold corrections from heavy
superparticles~\cite{PBMZ}, $\Delta_t$, $\Delta_b$ and $\Delta_\tau$,
have been included. The Yukawa unification at high-energy scale and
the MSSM RG evolution generally lead to almost the same size of top
and bottom Yukawa couplings, $y_t^{\overline{\text{DR}}}(M_Z)\simeq
y_b^{\overline{\text{DR}}}(M_Z)$, and consequently the 
ratio $\tan\beta$ should be very large to attain the observed mass
hierarchy between the top and bottom quarks. The required value 
of $\tan\beta$ is roughly estimated as
\begin{equation}
\tan\beta \;\simeq\; 
\frac{m_t^{\overline{\text{DR}}}(M_Z)}{m_b^{\overline{\text{DR}}}(M_Z)} 
\;\simeq\; {\cal O}(50-60),
\end{equation}
if one neglects threshold corrections and small differences between
the top and bottom Yukawa couplings generated through the RG evolution
down to low-energy regime.

In such a large $\tan\beta$ case, low-energy threshold corrections
from heavy superparticles become important~\cite{HRS,COPW,PBMZ}. The
sizes of corrections to gauge couplings and top quark 
mass $\Delta_t$ mainly depend on the overall scale of superparticle
masses. On the other hand, the correction to bottom quark 
mass $\Delta_b$ is controlled by a ratio between SUSY-breaking masses
and the $\mu$ parameter. This is a consequence of the fact 
that $\Delta_b$ is dominated by finite parts of threshold corrections
which are approximately constituted of the following two
contributions~\cite{HRS,COPW}:
\begin{eqnarray}
\qquad \Delta_ b^{\tilde g} \;&\,\simeq\;& \frac{2\tan\beta}{3\pi}
  \bigg(\frac{g_3^2}{4\pi}\bigg)\,\mu M_3\,
  I(M_3^2,m_{\tilde b_1}^2,m_{\tilde b_2}^2), \label{mbthre1} \\
  \Delta_ b^{\tilde\chi^+} \!&\,\simeq\;& \frac{\tan\beta}{4\pi}
  \bigg(\frac{y_t^2}{4\pi}\bigg)\,\mu A_t\,
  I(\mu^2,m_{\tilde t_1}^2,m_{\tilde t_2}^2), \label{mbthre2}
\end{eqnarray}
where
\begin{equation}
  I(x,y,z) \,=\, 
  \frac{xy\ln(y/x)+yz\ln(z/y)+zx\ln(x/z)}{(x-y)(y-z)(z-x)}.
\end{equation}
The corrections $\Delta_ b^{\tilde g}$ 
and $\Delta_ b^{\tilde \chi^+}$ denote the contributions 
to $\Delta_b$ from gluino-scalar bottom and chargino-scalar top loop
diagrams. The SUSY-breaking couplings $M_3$, $A_t$,
$m_{\tilde t_{1,2}}$, $m_{\tilde b_{1,2}}$ are the gluino mass, the
trilinear coupling of scalar top quark, and the mass eigenvalues of
top and bottom scalar quarks, respectively. The complete form of
corrections including generation mixing is found, 
e.g.\ in~\cite{BRP}. The loop function $I(x,y,z)$ behaves 
as ${\cal O}(1/M)$ with $M$ being the maximum value 
among $x$, $y$, and $z$. If there is no large hierarchy among
SUSY-breaking mass parameters and $\mu$, the 
corrections $|\Delta_b^{\tilde g}|$ and 
$|\Delta_ b^{\tilde \chi^+}|$ are expected to become ${\cal O}(1)$ for
a large value of $\tan\beta$. Also $\Delta_\tau$ is dominated by
finite parts in the large $\tan\beta$ case, but it is generally
smaller than $\Delta_b$ since the large gauge and Yukawa couplings are
absent in the expression of $\Delta_\tau$. In the Yukawa unification 
scenario, $\Delta_t$ and $\Delta_\tau$ roughly 
become $|\Delta_{t}|,\,|\Delta_\tau|\lesssim{\cal O}(0.1)$ when
superparticles are lighter than a few TeV~\cite{PBMZ}.

In this section we take SUSY-breaking parameters as free variables and
let the threshold corrections $\Delta_i$ represent low-energy
superparticle spectrum. In other words, once a preferred range 
of $\Delta_i$ is found, that in turn gives a constraint on low-energy
SUSY-breaking parameters. For fixed values of low-energy gauge
couplings, top-quark and tau-lepton masses, and the threshold
corrections $\Delta_{t,\tau}$, the Yukawa unification hypothesis
predicts the low-energy bottom quark mass as a function 
of $\Delta_b$. The allowed range of $\Delta_b$ is evaluated by the
following numerical procedure. First the input parameters are 
the $\overline{\text{MS}}$ gauge 
couplings $g_1^{\overline{\text{MS}}}(M_Z)$ 
and $g_2^{\overline{\text{MS}}}(M_Z)$ given in~\cite{ACKMPRW},
$\alpha_3^{\overline{\text{MS}}}(M_Z)=0.1187\,$ 
($\alpha_3=g_3^2/4\pi$), $M_Z=91.1876$~GeV and
$m_\tau^{\rm pole}=1776.99$~MeV~\cite{PDG}. The prediction of
bottom-quark mass is correlated to the input value of top-quark mass,
which is varied in the 
range $165\text{~GeV}<m_t^{\rm pole}<185$~GeV\@. At the 
scale $M_Z$, the $\overline{\text{MS}}$ gauge couplings are converted
to the $\overline{\text{DR}}$ ones including low-energy corrections of
suitable range. The two-loop MSSM RG equations~\cite{2loopRGE} are
traced from $M_Z$ up to the scale where 
the $\overline{\text{DR}}$ $SU(2)_L$ and $U(1)_Y$ (in GUT
normalization) gauge couplings meet. That defines the GUT 
scale $M_G$ and the unified gauge coupling $g_{{}_G}=
g_1^{\overline{\text{DR}}}(M_G)=
g_2^{\overline{\text{DR}}}(M_G)$. At the GUT scale, the exact
top-bottom-tau Yukawa unification is imposed. The Yukawa couplings for
the first and second generations are neglected:
\begin{equation}
  Y_u^{\overline{\text{DR}}}(M_G) \,=\,
  Y_d^{\overline{\text{DR}}}(M_G) \,=\,
  Y_e^{\overline{\text{DR}}}(M_G) \,=\,
  \left(\begin{array}{ccc}
    ~~&~~& \\
    & & \\
    & & y_{{}_G}\! \\
  \end{array}\right).
\end{equation}
For fixed values of $\Delta_t$ and $\Delta_\tau$, the unified Yukawa
coupling $y_{{}_G}$ has a one-to-one correspondence 
to $m_t^{\rm pole}$ and can be determined. The procedure is as
follows. The two-loop MSSM RG evolution from $M_G$ down 
to $M_Z$ determines $y_t^{\overline{\text{DR}}}(M_Z)$, 
$y_b^{\overline{\text{DR}}}(M_Z)$ and 
$y_\tau^{\overline{\text{DR}}}(M_Z)$, and then with an input value of
tau-lepton mass and assumed $\Delta_\tau$, a required value 
of $\tan\beta$ is found, where $m_\tau^{\overline{\text{DR}}}(M_Z)$ is
extracted from $m_\tau^{\rm pole}$ using the SM 
corrections~\cite{ACKMPRW}. Once $\tan\beta$ is 
fixed, $m_t^{\rm pole}$ is evaluated 
by $y_t^{\overline{\text{DR}}}(M_Z)$, $v_H$, assumed $\Delta_t$, and
the SM QCD contribution~\cite{PBMZ}. A series of the above
calculations is reiterated by varying $y_{{}_G}$ until to achieve the
input value of $m_t^{\rm pole}$. Finally, the prediction 
of $m_b^{\overline{\text{DR}}}(M_Z)$ is derived as a function 
of $m_t^{\rm pole}$ and $\Delta_b$, and is convert 
to $m_b^{\overline{\text{MS}}}(m_b)$ taken into account the two-loop
SM QCD corrections~\cite{HRS}. The prediction 
of $m_b^{\overline{\text{MS}}}$ depends on other input parameters
besides the top-quark mass and $\Delta_b$. We will mention this later.

\begin{figure}[t]
\begin{center}
\includegraphics[width=5cm,clip]{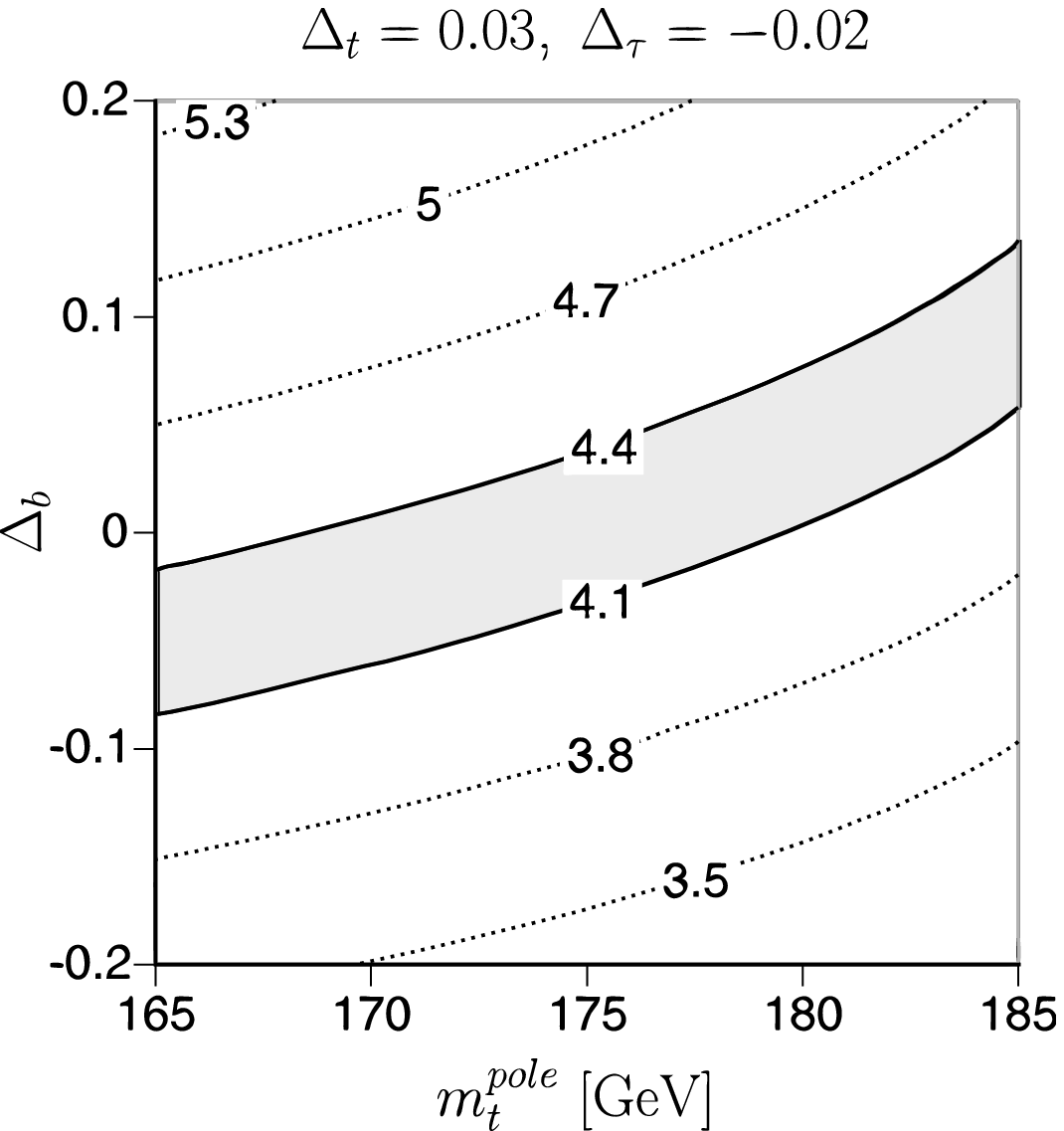}\hspace*{1cm}
\includegraphics[width=5cm,clip]{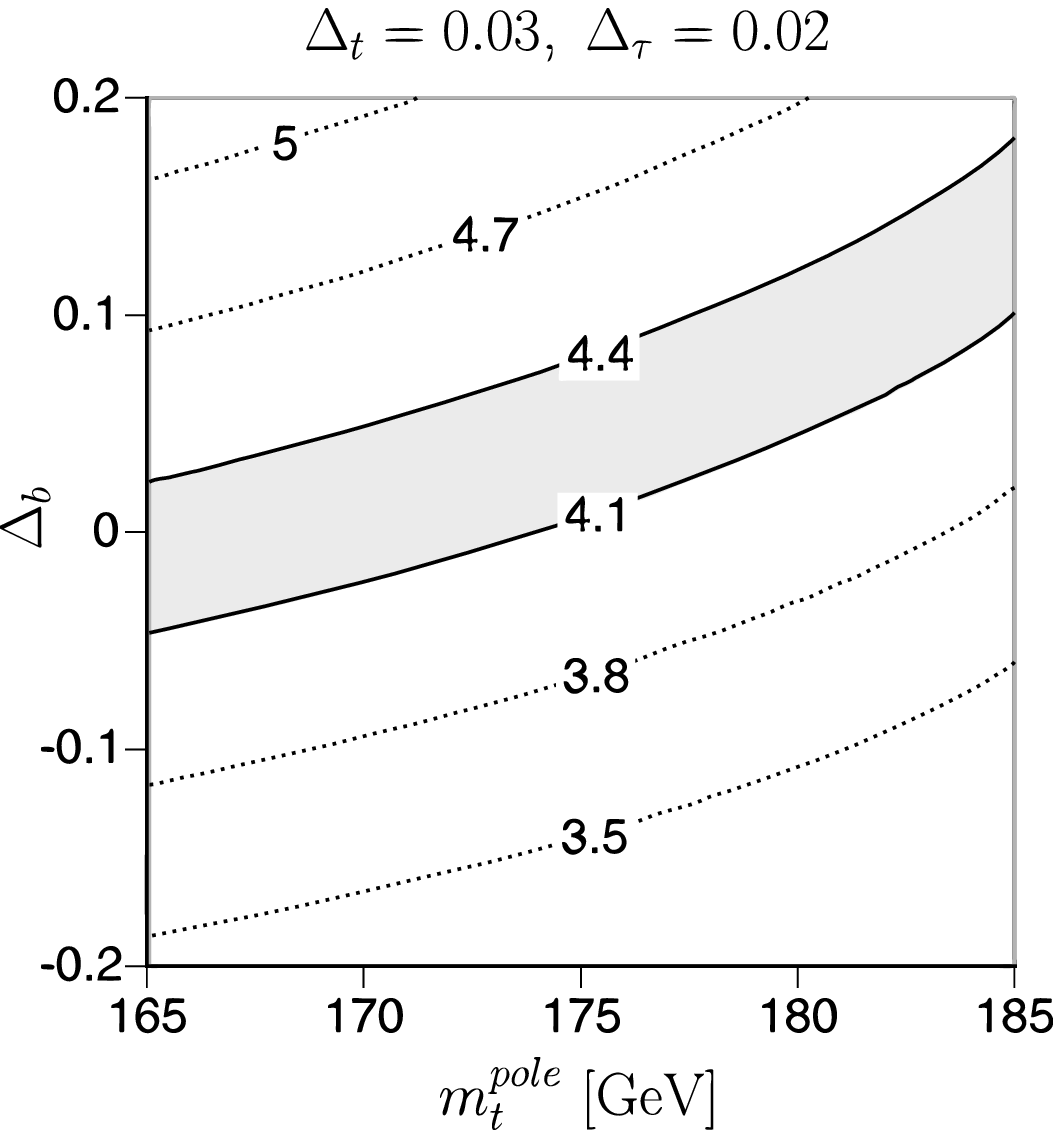} \\[5mm]
\includegraphics[width=5cm,clip]{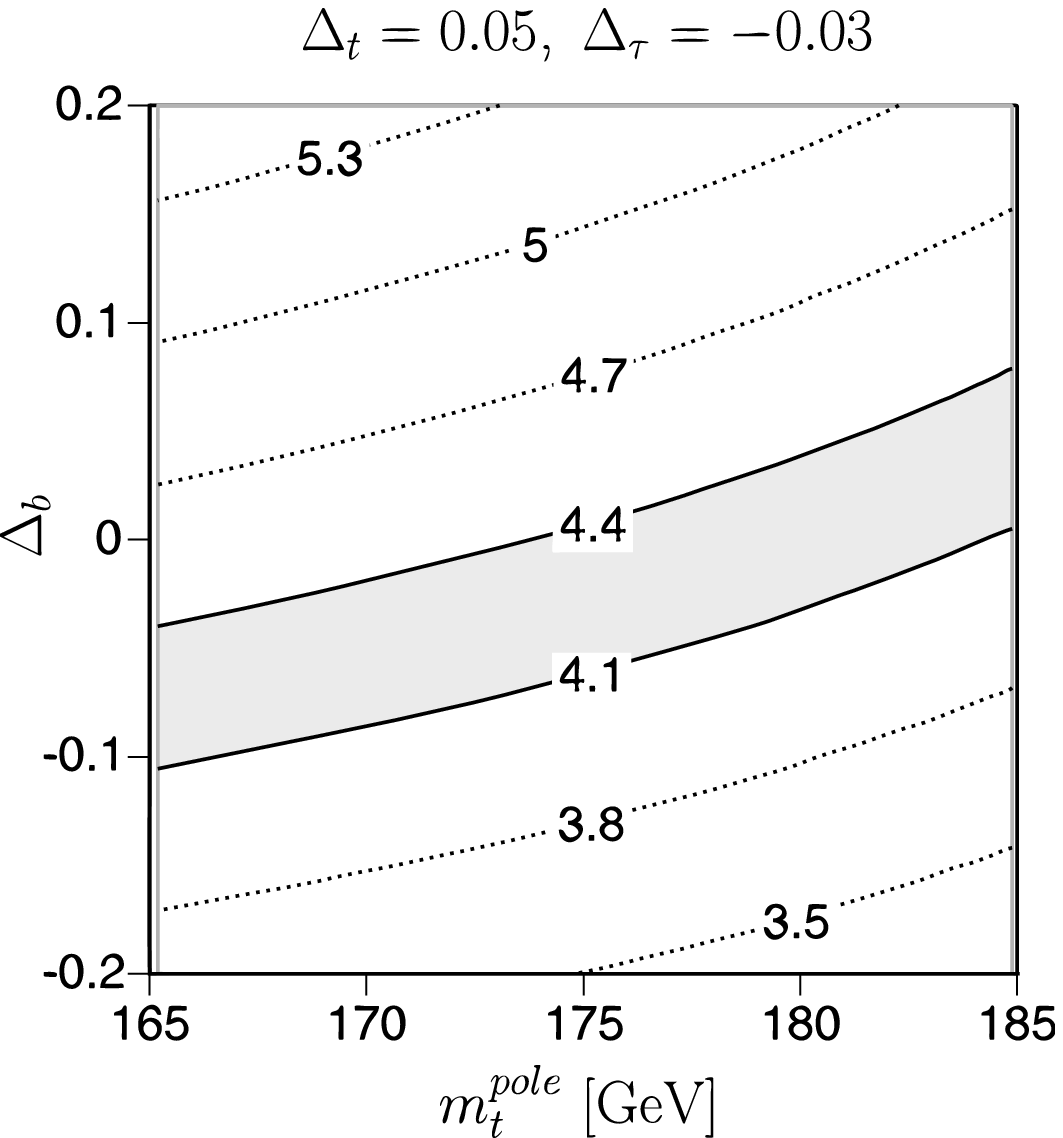}\hspace*{1cm}
\includegraphics[width=5cm,clip]{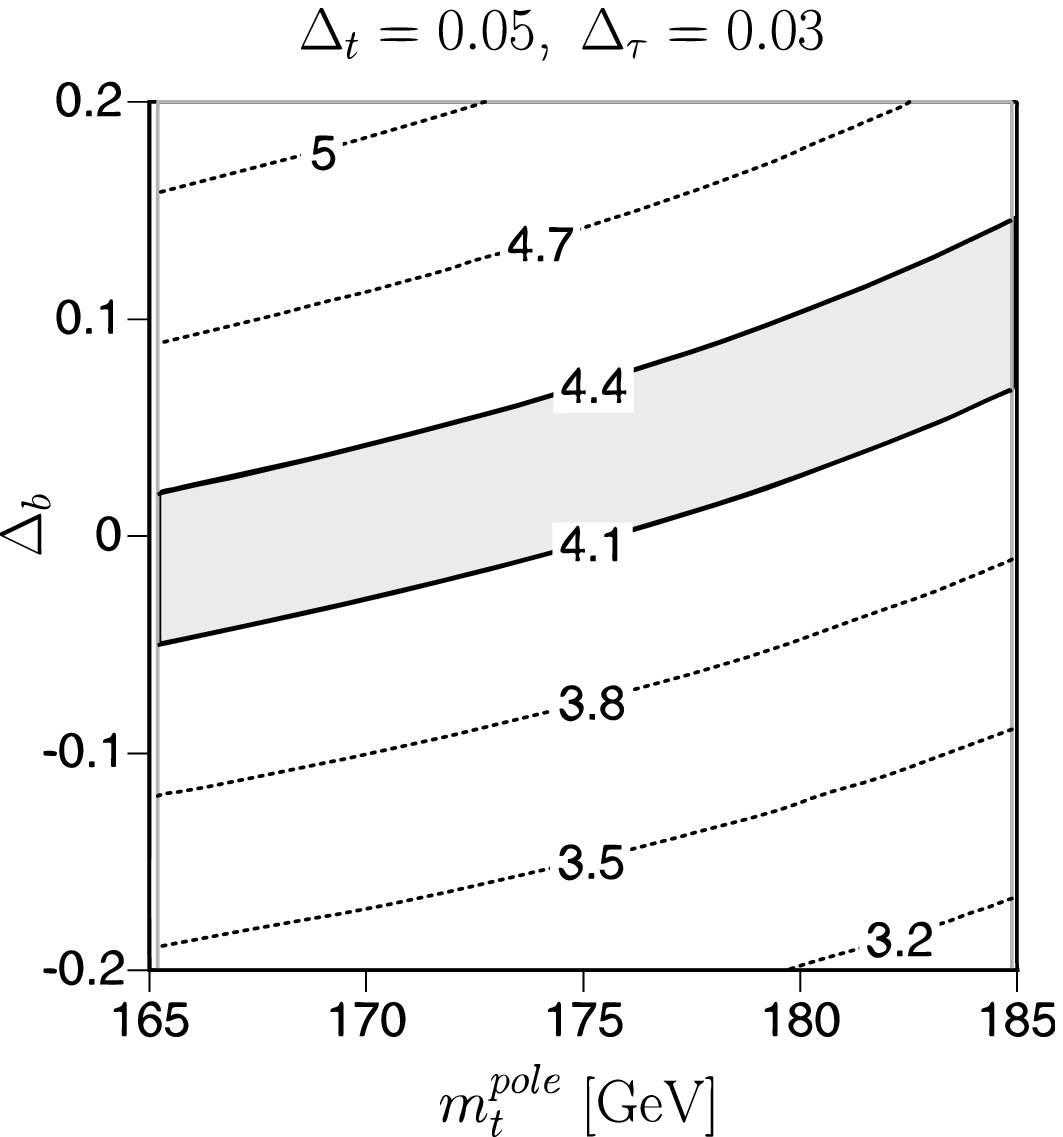}
\end{center}
\caption{Typical predictions of bottom-quark mass in the Yukawa
unification. The low-energy value $m_b^{\overline{\text{MS}}}(m_b)$ is
shown as the function of top-quark mass and threshold 
correction $\Delta_b$. The input parameters are taken
as the $\overline{\text{MS}}$ gauge couplings 
$\alpha_1^{\overline{\text{MS}}}(M_Z)=0.01698$ and
$\alpha_2^{\overline{\text{MS}}}(M_Z)=0.03364$,
$\alpha^{\overline{\text{MS}}}_3(M_Z)=0.1187$ ($\alpha_i=g_i^2/4\pi$),
$M_Z=91.1876$~GeV and $m_\tau^{\rm pole}=1776.99$~MeV\@. 
The threshold corrections to top-quark and tau-lepton masses are, as
examples, set to be $\Delta_t=0.03$ and $\Delta_\tau=-0.02$ in the
left-top, $\Delta_t=0.03$ and $\Delta_\tau=0.02$ in the right 
top, $\Delta_t=0.05$ and $\Delta_\tau=-0.03$ in the left 
bottom, and $\Delta_t=0.05$ and $\Delta_\tau=0.03$ in the right-bottom
figures, respectively.\bigskip}
\label{mbfig}
\end{figure}

Fig.~\ref{mbfig} shows $m_b^{\overline{\text{MS}}}(m_b)$ as the
function of $m_t^{\rm pole}$ and $\Delta_b$. In the figures, we 
set $\Delta_t$ and $\Delta_\tau$ to typical values which roughly
correspond to the case that superparticle masses are less than a few
TeV\@. The observed value of bottom-quark mass is given 
by $m_b^{\overline{\text{MS}}}(m_b)=4.1-4.4$~GeV~\cite{PDG} and shown
as shaded regions in the figures. It is immediately found that the
threshold correction $|\Delta_b|$ must be 
small; $|\Delta_b|\lesssim 0.1$ in a wide range of top-quark mass,
which result is consistent with the previous analysis~\cite{RS,TW}.
The result is slightly changed by varying input parameters. For
example, $m_b^{\overline{\text{MS}}}(m_b)$ is altered $\pm0.2$~GeV
with $\alpha_3^{\overline{\text{MS}}}(M_Z)$ in the 
range $0.116-0.121$ for fixed values 
of $m_t^{\rm pole}$ and $\Delta_b$. We also checked the dependence
of $m_b^{\overline{\text{MS}}}(m_b)$ on unknown GUT-scale threshold
corrections to Yukawa couplings. If a five percents deviation 
of $y_b^{\overline{\text{DR}}}(M_G)$ 
or $y_\tau^{\overline{\text{DR}}}(M_G)$ from $y_{{}_G}$ is 
included, $|m_b^{\overline{\text{MS}}}(m_b)|$ typically has an
ambiguity less than $0.3$~GeV\@. The GUT-scale correction 
to $y_t^{\overline{\text{DR}}}(M_G)$ is turned out to be
irrelevant.

Such a small value of experimentally preferred $|\Delta_b|$ indicates
that there must be some hierarchical structure among SUSY-breaking
mass parameters and $\mu$ parameter, otherwise, two types of large
threshold corrections (\ref{mbthre1}) and (\ref{mbthre2}) must cancel
out to each other. The latter case largely depends on the detail of
superparticle spectrum and is therefore a model-dependent option. As
for the former case, however, an interesting resolution is found. It
can be seen from the expression (\ref{mbthre1}) and (\ref{mbthre2})
that, a small correction $|\Delta_b|$ is obtained if scalar quark
masses are relatively larger than the other mass parameters; the
gluino mass, scalar top trilinear coupling, and $\mu$ parameter. It
was pointed out in~\cite{HRS} that such a hierarchical mass pattern
does follow from symmetry argument: the 
Peccei-Quinn (PQ) symmetry~\cite{PQ} which rotates Higgs particles
suppresses the $\mu$ parameter, and the R symmetry which acts on the
fermionic coordinate forbids holomorphic SUSY-breaking parameters in
the symmetric limit. Therefore if approximate PQ and R symmetries are
realized in the theory, the Yukawa unification hypothesis predicts
experimentally-allowed masses for the third-generation fermions (top,
bottom quarks and tau lepton)~\cite{HRS,RS,TW}.

%%%%%%%%%%%%%%%%%%%%%%%%%%%%%%%%%%%%%%%%%%%%%%%%%%%%%%%%%%%%%%%%%%%%%%
\subsection{$\boldsymbol{b\to s\gamma}$ Decay}
%%%%%%%%%%%%%%%%%%%%%%%%%%%%%%%%%%%%%%%%%%%%%%%%%%%%%%%%%%%%%%%%%%%%%%
As stated previously, the Yukawa unification generally requires a
large value of $\tan\beta$. It is well known that in the 
large $\tan\beta$ case the experimental observation 
of $b\to s\gamma$ decay provides severe constraints on low-energy
superparticle spectrum~\cite{COPW,RS,BOP,bsg}. In this subsection we
shortly review the characteristic feature of $b\to s\gamma$ constraint
in the large $\tan\beta$ case.

The experimentally-measured branching ratio of 
the $b\to s\gamma$ decay is ${\cal B}(b\to s\gamma)=
(3.55\pm0.24^{+0.09}_{-0.10}\pm0.03)\times10^{-4}$~\cite{bsgexp}, and
the SM prediction for the branching ratio is theoretically in
excellent agreement with the experimental observation~\cite{bsgSM}. In
the MSSM, the $b\to s\gamma$ decay amplitude consists of several loop
diagrams: up-type quark--W boson ($A_{\rm SM}$), up-type
quark--charged Higgs boson ($A_{H^+}$), up-type scalar 
quark--chargino ($A_{\tilde\chi^+}$), down-type scalar 
quark--neutralino ($A_{\tilde\chi^0}$), and down-type scalar
quark--gluino ($A_{\tilde g}$). The magnitudes of these amplitudes
depend on masses of fields running in the internal loops. The 
amplitude $A_{H^+}$ always gives a constructive contribution to 
the $W$-boson loop $A_{\rm SM}$, which together make up the standard
model contribution, while the other contributions $A_{\tilde\chi^+}$, 
$A_{\tilde\chi^0}$, and $A_{\tilde g}$ are constructive or
destructive, depending on the signs of scalar trilinear couplings 
and $\mu$ parameter. It is known 
that $A_{\tilde\chi^+}$, $A_{\tilde\chi^0}$, and $A_{\tilde g}$ scale
as $\tan\beta$ and are amplified with Yukawa
unification~\cite{BOP}. In most cases, since $A_{\tilde\chi^+}$ is
largest among the superparticle-induced contributions, we focus 
on $A_{\tilde\chi^+}$ in the following qualitative discussion, for
simplicity. In numerical evaluation, all the decay modes are included
according to the formulas in~\cite{bsgformula}.

In the Yukawa unification scenario, $\tan\beta$ is large and the
chargino contribution $A_{\tilde\chi^+}$ becomes important. The decay
rate of $b\to s\gamma$ is therefore sensitive to the sign 
of $A_{\tilde\chi^+}$. It is found that the sign 
of $A_{\tilde\chi^+}$ is determined by that of $\mu$ in the minimal
supergravity scenario where scalar trilinear coupling is negative at
low-energy regime. For a 
positive $\mu$ parameter, $A_{\tilde\chi^+}$ gives a destructive
contribution to the standard model prediction. If a cancellation
occurs between $A_{\tilde\chi^+}$ and $A_{H^+}$, a relatively light
superparticle spectrum is permitted~\cite{BOP,bsg}. The degree of such
cancellation is roughly controlled by mass ratio among the charged
Higgs boson, charginos, and up-type scalar quarks. On the other hand,
for a negative $\mu$ parameter, $A_{\tilde\chi^+}$ gives a
constructive contribution to the standard model
prediction. Considering the fact that the standard model contribution
is well fitted to the experimental result, new physics contribution
must be suppressed. In particular, superparticle spectrum with a
negative $\mu$ parameter is highly constrained from a viewpoint 
of $b\to s\gamma$ process.

\begin{figure}[t]
\begin{center}
\includegraphics[width=6cm,clip]{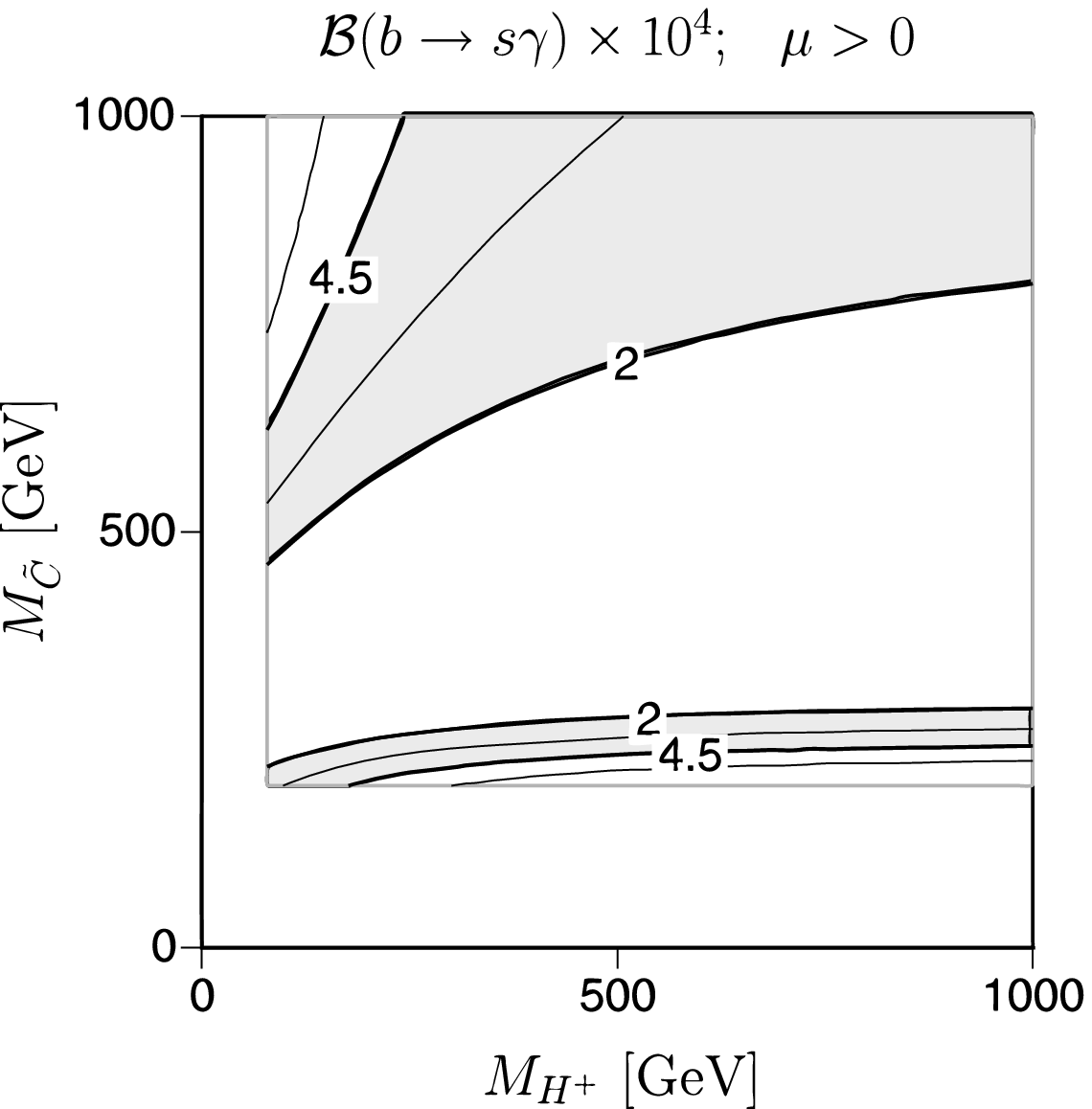}\hspace*{1cm}
\includegraphics[width=6cm,clip]{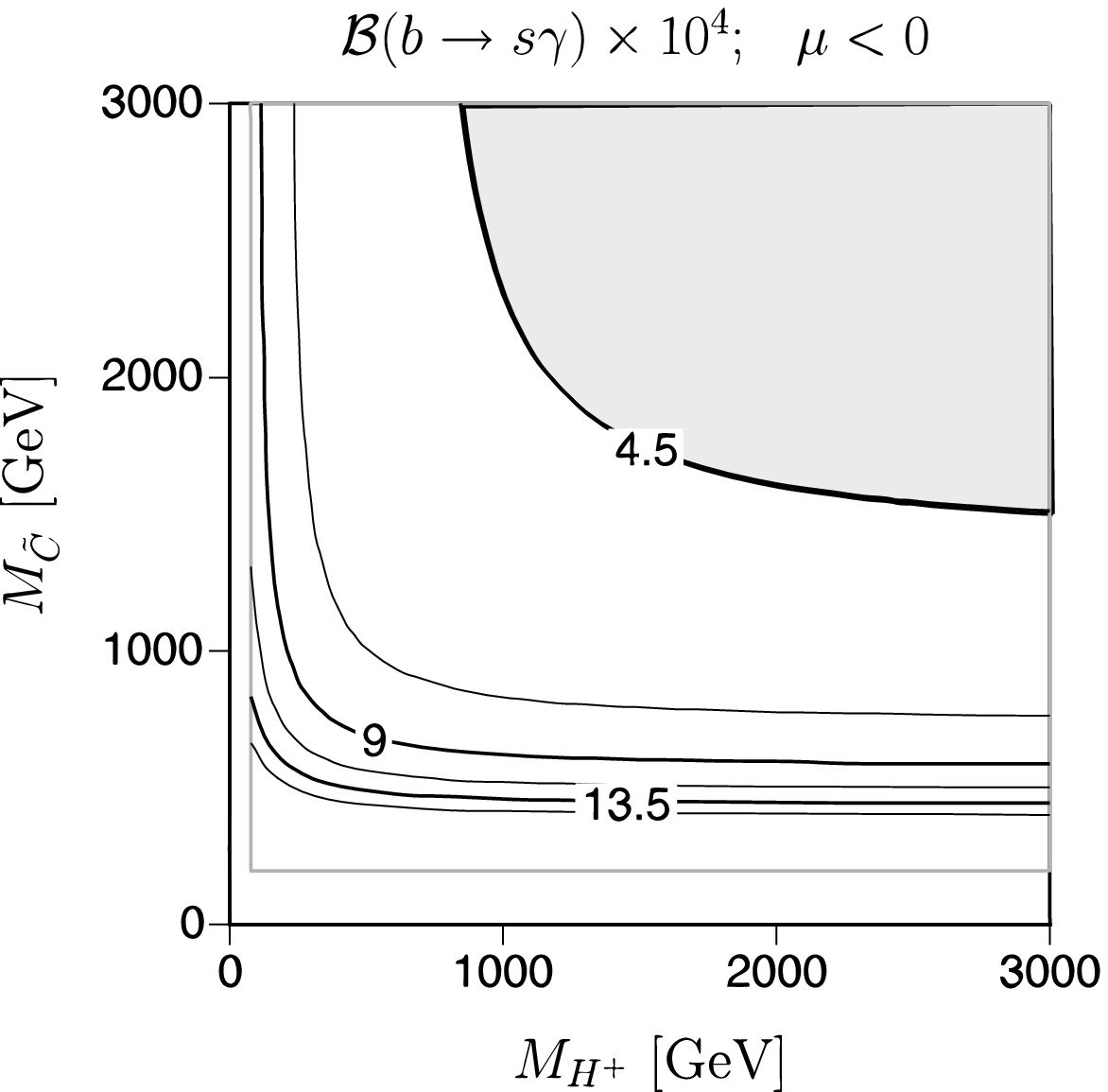}
\end{center}
\caption{Typical results for $b\to s\gamma$ branching ratio in the
large $\tan\beta$ case ($\tan\beta=50$). The two axes denote the
masses of charged Higgs boson and colored superparticles. The sign 
of $\mu$ is set to be positive (negative) in the left (right)
figure. The input parameters besides those in Fig.~\ref{mbfig} are the
W-boson mass $M_W^{\rm pole}=80.425$~GeV, 
$m_b^{\overline{\text{MS}}}(m_b)=4.25$~GeV, 
$m_t^{\rm pole}=178$~GeV, 
$m_u^{\overline{\text{MS}}}(M_Z)/m_t^{\overline{\text{MS}}}(M_Z)=
8.6\times10^{-7}$, 
$m_c^{\overline{\text{MS}}}(M_Z)/m_t^{\overline{\text{MS}}}(M_Z)=
3.7\times10^{-3}$,
$m_d^{\overline{\text{MS}}}(M_Z)/m_b^{\overline{\text{MS}}}(M_Z)=
1.0\times10^{-3}$,
$m_s^{\overline{\text{MS}}}(M_Z)/m_b^{\overline{\text{MS}}}(M_Z)=
2.2\times10^{-2}$, and the observed values of the generation mixing
matrix elements.\bigskip}
\label{bsgfig}
\end{figure}

Fig.~\ref{bsgfig} shows typical results for 
the $b\to s\gamma$ branching ratio in the large $\tan\beta$ case. The
vertical axis means the mass of charged Higgs boson and the horizontal
one $M_{\tilde C}$ which characterizes superparticle masses (note
that, in this section, SUSY-breaking parameters are free
variables). As an example to discuss qualitative feature, we set in
the figures the masses of colored superparticles
and $|\mu|$ to be $M_{\tilde C}$, and those of uncolored ones,
trilinear couplings of scalar top and bottom to 
be $M_{\tilde C}/2$. The lower limits 
of $M_{H^+}$ and $M_{\tilde C}$ come from the experimental mass bounds
on charged Higgs boson and gluino: $M_{H^+}>79.3$~GeV 
and $M_3>195$~GeV (95\% CL)~\cite{PDG}. The set of other input
parameters is shown in the figure caption. Since there may be some
theoretical ambiguities in the estimation, we take in this paper a
rather conservative constraint on the $b\to s\gamma$ branching ratio
\begin{equation}
  2.0\times10^{-4} \,<\, {\cal B}(b\to s\gamma) \,<\, 4.5\times10^{-4}.
\end{equation}
The region which satisfies this constraint is shaded in the figures.

It is found from Fig.~\ref{bsgfig} that the $b\to s\gamma$ branching
ratio shows quite different behavior between the positive and negative
values of $\mu$ parameter. In the $\mu>0$ case, there exist large
parameter regions which satisfy the experimental constraint. The
branching ratio becomes larger in the 
regions $M_{H^+}\ll M_{\tilde C}$ and $M_{H^+}\gg M_{\tilde C}$. That
is a consequence of the fact that $A_{\tilde\chi^+}$ gives a
destructive contribution to $A_{\rm SM}$ and $A_{H^+}$. For
$300\text{~GeV}\lesssim M_{\tilde C}\lesssim 500$~GeV, 
$A_{\tilde\chi^+}$ has a similar magnitude (and opposite sign) of the
sum of $A_{SM}$ and $A_{H^+}$, and the branching ratio almost
vanishes. In the region where $M_{\tilde C}$ is closer to its
experimental lower bound, the branching ratio takes a larger value,
which is dominated by superparticle loop diagrams. In this way, in
order to satisfy the $b\to s\gamma$ constraint with a 
positive $\mu$ parameter, the masses of charged Higgs boson and
superparticles must lie in almost the same order to realize a
cancellation among the partial amplitudes, or both the charged Higgs
boson and superparticles are heavy as much as 1~TeV\@. On the other 
hand, for $\mu<0$, the $b\to s\gamma$ process severely restricts
the charged Higgs boson and superparticle 
masses. Neither $M_{H^+}$ nor $M_{\tilde C}$ is allowed to be smaller
than a few TeV.

To summarize this subsection, in the large $\tan\beta$ case as in the
Yukawa unification, the observation of $b\to s\gamma$ decay provides
severe constraints on low-energy superparticle 
spectrum. For $\mu>0$, since a cancellation among different
contributions is possible, a lighter superparticle spectrum is allowed
when the masses of charged Higgs boson and superparticles lie in a
similar order. For $\mu<0$, various contributions are additive and
make up a large branching ratio, and hence light superparticles are
disfavored; to have branching ratio within experimentally-allowed
range, both the charged Higgs boson and superparticles must be heavier
than a few TeV.

%%%%%%%%%%%%%%%%%%%%%%%%%%%%%%%%%%%%%%%%%%%%%%%%%%%%%%%%%%%%%%%%%%%%%%
\subsection{MSSM Higgs Potential}
%%%%%%%%%%%%%%%%%%%%%%%%%%%%%%%%%%%%%%%%%%%%%%%%%%%%%%%%%%%%%%%%%%%%%%
In the large $\tan\beta$ case, successful EWSB requires several
conditions among mass parameters in the theory. In particular, 
the $\mu$ parameter and CP-odd neutral Higgs boson mass are related to
SUSY-breaking mass parameters of up- and down-type Higgs fields. Such
conditions should be satisfied at the electroweak scale and are
promoted to those for GUT-scale parameters. The detailed discussions
of radiative EWSB for specific high-energy models will be given in
later sections.

In the MSSM, the tree-level Higgs potential takes the following form:
{\allowdisplaybreaks%
\begin{eqnarray}
  V_{\text{Higgs}} &=& (|\mu|^2+m^2_{H_u})(|H_u^0|^2+|H_u^+|^2)
  +(|\mu|^2+m^2_{H_d})(|H_d^0|^2+|H_d^-|^2)  \nonumber \\[1mm]
  && \quad +\big[B\mu(H_u^+H_d^--H_u^0H_d^0)+{\rm c.c.}\big] 
  \nonumber \\ 
  && \quad +\frac{g_2^2 + g^{\prime 2}}{8}
  \big(|H_u^0|^2+|H_u^+|^2-|H_d^0|^2-|H_d^-|^2\big)^2
  +\frac{g_2^2}{2}\big|H_u^+H_d^{0*}+H_u^0H_d^{-*}\big|^2 ,\quad
\end{eqnarray}}%
where $H_u=(H_u^+,H_u^0)^{\rm T},H_d=(H_d^0,H_d^-)^{\rm T}$ are the
lowest components of Higgs superfields, $m_{H_{u,d}}^2$ are the
SUSY-breaking mass parameters for up- and down-type Higgs 
scalars, and $B$ is the SUSY-breaking Higgs mixing parameter. The 
coupling $g^\prime$ is $U(1)_Y$ gauge coupling which is related 
to $g_1$ in GUT normalization as $g^\prime=\sqrt{3/5}\,g_1$. All the
couplings are running parameters defined in 
the $\overline{\text{DR}}$ scheme, but for notational simplicity, the
subscripts ``$\overline{\text{DR}}$'' will not be explicitly shown in
the below.

When the Higgs scalars develop non-vanishing VEVs, one of VEVs of the
charged scalars is always set to be zero with $SU(2)_L$ gauge
transformation, and another charged one becomes zero due to the
minimization condition. Therefore the Higgs 
potential $V_{\rm Higgs}$ induces the correct pattern 
of EWSB; $SU(2)_L\times U(1)_Y\to U(1)_{\rm EM}$. Also using 
the $U(1)_Y$ gauge transformation and a phase rotation 
of $H_u$ and $H_d$, the VEVs of neutral components can be made real
and positive. Thus the Higgs VEVs are parametrized by real, positive
parameters as
\begin{equation}
  \langle H_u\rangle \,=\,
  \left(\begin{array}{c} 0 \\ v_H\sin\beta \end{array}\right), \qquad 
  \langle H_d\rangle \,=\, 
  \left(\begin{array}{c} v_H\cos\beta \\ 0 \end{array}\right),
\end{equation}
where $0\leq\beta\leq\pi/2$. In the vacuum, the Z-boson mass $M_Z$ is
given by $M_Z^2=\frac{1}{2}(g_2^2+g^{\prime 2})v_H^2$. The nonzero and
finite VEVs are obtained when the mass parameters in the potential
satisfy the following inequalities~\cite{EWSB}:
\begin{equation}
  (m^2_{H_u}+|\mu|^2)(m^2_{H_d}+|\mu|^2)-|B\mu|^2 \,<\, 0
\end{equation}
and
\begin{equation}
  (m^2_{H_u}+|\mu|^2)+(m^2_{H_d}+|\mu|^2)-2|B\mu| \,>\, 0
\end{equation}
at the renormalization scale $Q\sim M_{\rm SUSY}$ which is a typical
mass scale of SUSY-breaking parameters. The former condition implies
the origin of field space is made unstable to have broken electroweak
symmetry, and the latter one lifts up the flat direction 
with $\tan\beta=1$, otherwise the potential is unbounded from below in
that direction. At the vacuum of potential, the stationary conditions
are written as
{\allowdisplaybreaks%
\begin{eqnarray}
  \frac{M_Z^2}{2} &\;=\;& 
  \frac{m_{H_d}^2-m_{H_u}^2\tan^2\beta}{\tan^2\beta-1}-|\mu|^2, \\ 
  \tan\beta+\cot\beta &\;=\;& 
  \frac{m_{H_u}^2+m_{H_d}^2+2|\mu|^2}{|B\mu|},
\end{eqnarray}}%
at the classical level. The MSSM RG equations allow us to freely
choose low-energy $\mu$ and $B$ with the freedom of 
GUT-scale $\mu$ and $B$ without affecting any other SUSY-breaking
parameters. Thus the above stationary conditions are solved 
for $\mu$ and $B$. If the Z-boson mass and a required value 
of $\tan\beta$ cannot be fitted by $\mu$ and $B$, a desired pattern of
EWSB does not occur.

In the large $\tan\beta$ limit, the stationary conditions are
approximately rewritten in the following simple form: 
{\allowdisplaybreaks%
\begin{eqnarray}
  M_Z^2 &\;\simeq\;& -2(m_{H_u}^2+|\mu|^2), \\[1mm]
  \tan\beta &\;\simeq\;& \frac{m_{H_d}^2+m_{H_u}^2+2|\mu|^2}{|B\mu|}.
\end{eqnarray}}%
Then the constraints on the Higgs mass parameters read
{\allowdisplaybreaks%
\begin{eqnarray}
  |\mu|^2 &\;\simeq\;& -m_{H_u}^2-\frac{M_Z^2}{2} \;>\; 0,
  \label{con1} \\[1mm]
  |B\mu|\tan\beta &\;\simeq\;& m_{H_d}^2-m_{H_u}^2-M_Z^2 \;>\; 0.
  \label{con2}
\end{eqnarray}}%
The Higgs SUSY-breaking mass parameters must satisfy the
inequalities (\ref{con1}) and (\ref{con2}) so that the 
observed $Z$-boson mass and a positively definite value 
of $\tan\beta$ are realized by consistently 
choosing $\mu$ and $B$. It is interesting to note that the left-handed
sides of these inequalities are related to mass squareds of physical
particles in the symmetry broken phase. In (\ref{con1}), $|\mu|^2$ is
relevant to tree-level masses of charginos and neutralinos. Also the
inequality (\ref{con2}) is concerned with the tree-level mass
eigenvalue of CP-odd neutral Higgs boson:
\begin{equation}
  M_A^2 \;=\; m_{H_u}^2+m_{H_d}^2 +2|\mu{}|^2 \;\simeq\;
  m_{H_d}^2-m_{H_u}^2-M_Z^2.
\end{equation}
The current experimental mass bounds of the lightest 
neutralino $\tilde \chi_1^0$, the lighter 
chargino mass eigenstate $\tilde \chi_1^+$, and the CP-odd neutral
Higgs boson 
are $m_{\tilde \chi_1^0}>46$~GeV, $m_{\tilde\chi_1^+}>67.7$~GeV, 
and $M_A>90.4$~GeV (95\% CL), respectively~\cite{PDG}. Therefore
(\ref{con1}) and (\ref{con2}) actually impose severer constraints on
the Higgs mass parameters. Moreover, in the large $\tan\beta$ case,
one-loop corrections to $M_A$ and $\mu$ generally tend to be large,
which is mainly due to the tadpole contribution to one-loop effective
potential. In the numerical analysis in later sections, we take into
account of these mass bounds and one-loop corrections.

The mass bound on $M_A^2$ (\ref{con2}) implies that, at low-energy
regime, two Higgs SUSY-breaking 
masses $m_{H_u}^2$ and $m_{H_d}^2$ must be separated and the down-type
Higgs mass squared must be larger than the up-type Higgs mass
squared. If $\tan\beta$ is small, a larger value of the top Yukawa
coupling than the bottom (and tau) ensures such a mass separation
via the RG evolution down to low energy, and successful EWSB is
radiatively achieved~\cite{EWSB}. However for the 
large $\tan\beta$ case as in the Yukawa unification, the top and
bottom Yukawa couplings are of almost same order of magnitude
throughout the RG evolution, and the splitting of two Higgs masses is
not guaranteed. As will be seen in the following sections, 
if $SO(10)$-like GUT models are assumed, this mass bound of CP-odd
neutral Higgs boson excludes large regions of high-energy
SUSY-breaking parameter space.

%%%%%%%%%%%%%%%%%%%%%%%%%%%%%%%%%%%%%%%%%%%%%%%%%%%%%%%%%%%%%%%%%%%%%%
\bigskip
\section{$\boldsymbol{SO(10)}$ Unification}
\label{sec:so10}
%%%%%%%%%%%%%%%%%%%%%%%%%%%%%%%%%%%%%%%%%%%%%%%%%%%%%%%%%%%%%%%%%%%%%%
In this section we discuss the minimal $SO(10)$-type GUT scenario
which is a naive high-energy realization of the Yukawa
unification. Contrary to the analysis in the previous section,
SUSY-breaking parameters at the GUT scale are assumed to be unified,
that is, a top-down approach to low-energy SUSY phenomenology. The
radiative EWSB and various other aspects in this class of unification
scenario have been widely investigated in the
literature~\cite{COPW,RS,BOP,OP,MOP,Tata,BDR}. We focus here on the 
study of radiative EWSB, third-generation fermion masses, 
and $b\to s\gamma$ process for comparison with later
discussions. 

The model we now consider is specified by the following assumptions:
(i) $SO(10)$ gauge symmetry is broken down to the SM gauge group at
the GUT scale $M_G$, below which the theory is just the MSSM\@. 
(ii) The MSSM matter and Higgs superfields are included in $SO(10)$
multiplets $16_i$ (matter) and $10_H$ (Higgs), respectively.
(iii) The MSSM Yukawa terms (\ref{MSSMspot}) come from a GUT-scale
superpotential $W=16_iY_{ij}16_j10_H$. The Yukawa matrix $Y$ has a
large ${\cal O}(1)$ component $Y_{33}\equiv y_{{}_G}$, and therefore
the top, bottom, and tau Yukawa couplings are unified at the GUT
scale. (iv) SUSY-breaking terms also respect 
the $SO(10)$ symmetry. Thus the independent SUSY-breaking parameters
at the GUT scale are
\begin{equation}
  m_{16}^2,\; m_{10}^2,\; M_{1/2},\; A_0,\; B_0,
\end{equation}
where $m_{16}^2$ ($m_{10}^2$) denotes the matter (Higgs) scalar
masses, $M_{1/2}$ the universal gaugino mass parameter, 
and $A_0$ the universal scalar trilinear coupling. As mentioned
before, the $B$ parameter is determined at the electroweak scale by
solving the EWSB conditions and hence the high-energy boundary 
value $B_0$ is irrelevant to the analysis in this paper. The boundary
values of SUSY-breaking parameters in the MSSM are matched to the
GUT-scale independent parameters as
{\allowdisplaybreaks%
\begin{gather}
  m_{\tilde Q}^2(M_G)_{ij} \,=\, m_{\tilde u}^2(M_G)_{ij} \,=\, 
  m_{\tilde d}^2(M_G)_{ij} \,=\, m_{\tilde L}^2(M_G)_{ij} \,=\, 
  m_{\tilde e}^2(M_G)_{ij} \,=\, m_{16}^2\,\delta_{ij}, \\
  m_{H_u}^2(M_G) \,=\, m_{H_d}^2(M_G) \,=\, m_{10}^2, \\[1mm]
  M_1(M_G) \,=\, M_2(M_G) \,=\, M_3(M_G) \,=\, M_{1/2}, \\[1mm]
  A_u(M_G)_{ij} \,=\, A_d(M_G)_{ij} \,=\, A_e(M_G)_{ij} \,=\, 
  A_0\,\delta_{ij},
\end{gather}}%
where $m_{\tilde Q}^2$, $m_{\tilde u}^2$, $m_{\tilde d}^2$, 
$m_{\tilde L}^2$, $m_{\tilde e}^2$, $A_u$, $A_d$, $A_e$ are the MSSM
matter scalar masses and trilinear couplings in the generation space,
respectively, and $M_1$, $M_2$, $M_3$ denote the SM gaugino
masses. Note that even if the universal scalar masses are assumed at a
cutoff scale, RG effects above $M_G$ might induce some difference
between $m_{10}^2$ and $m_{16}^2$ at $M_G$~\cite{PP}, whose size of
splitting is however rather model dependent, e.g.\ sensitive to
GUT-breaking Higgs sector. Thus $m_{10}^2$ and $m_{16}^2$ can be taken
as independent variables, and throughout this paper we equivalently
use the following notation:
\begin{equation}
  \tilde m_0^2 \,=\, \frac{m_{10}^2+m_{16}^2}{2}, \qquad
  \xi \,=\, \frac{m_{10}^2-m_{16}^2}{m_{10}^2+m_{16}^2},
  \label{xi}
\end{equation} 
in other words,
\begin{equation}
  m_{10}^2 \,=\, \tilde m_0^2(1+\xi), \qquad 
  m_{16}^2 \,=\, \tilde m_0^2(1-\xi).
\end{equation}
  
In the following, we first discuss about the radiative EWSB and then
study the parameter space allowed by third-generation fermion masses
and the $b\to s\gamma$ observation. As seen in the previous section,
the Yukawa unification hypothesis leads to third-generation fermion
masses strongly correlated to the threshold 
correction $\Delta_b$, which has large dependence on SUSY-breaking
mass and $\mu$ parameters. Since these parameters are severely
restricted by the EWSB conditions and the $b\to s\gamma$ constraint,
it is a non-trivial issue whether the above type of $SO(10)$
unification is phenomenologically viable. We will present a detailed
analysis on this subject in the below.

%%%%%%%%%%%%%%%%%%%%%%%%%%%%%%%%%%%%%%%%%%%%%%%%%%%%%%%%%%%%%%%%%%%%%%
\subsection{Radiative EWSB in $\boldsymbol{SO(10)}$ Unification}
\begin{figure}[t]
\begin{center}
\includegraphics[width=5cm]{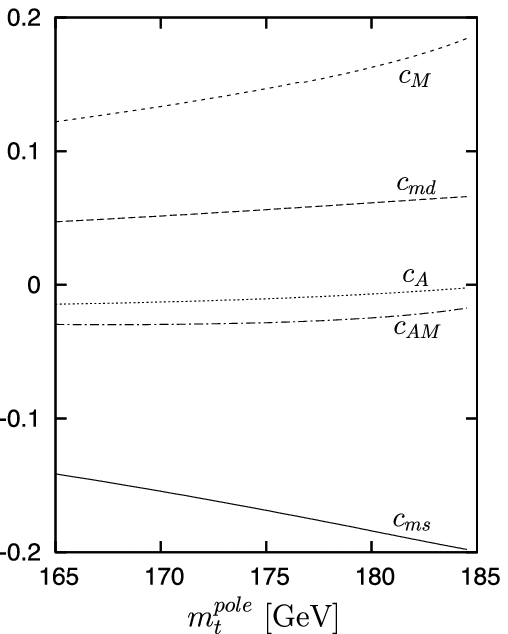}\hspace*{2cm}
\includegraphics[width=5cm]{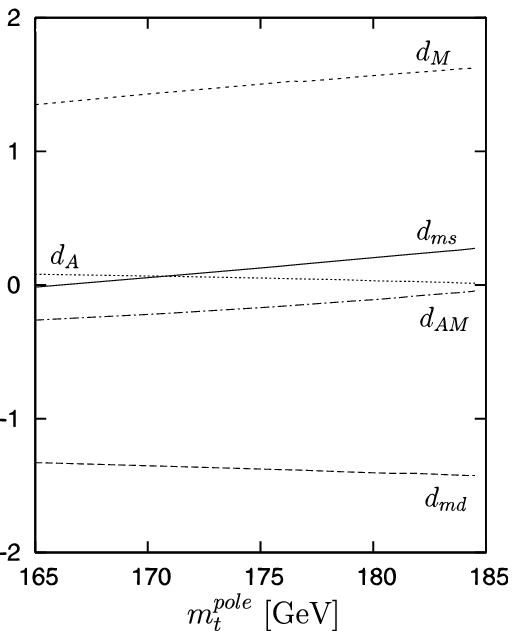}
\end{center}
\caption{Typical behaviors of $c$'s and $d$'s in the MSSM RG 
solutions (\ref{mafit1}) and (\ref{mufit1}) evaluated at the
renormalization scale $Q=1$~TeV\@. In the figures, the input
parameters are the same as in Fig.~\ref{mbfig} and the threshold
corrections to top and tau-lepton masses are set 
to $\Delta_t=0.03$ and $\Delta_\tau=-0.02$.\bigskip}
\label{fit1}
\end{figure}

%%%%%%%%%%%%%%%%%%%%%%%%%%%%%%%%%%%%%%%%%%%%%%%%%%%%%%%%%%%%%%%%%%%%%%
To have an insight about low-energy superparticle spectrum in 
the $SO(10)$-type unification, we start to discuss about the radiative
EWSB\@. Since the Yukawa coupling unification leads to a large value
of $\tan\beta$, the EWSB conditions are now approximately given by the
simple form (\ref{con1}) and (\ref{con2}). In this case, as discussed
before, the experimental lower bounds on $M_A$ and $|\mu|$ strongly
restrict low-energy values of mass parameters in the Higgs potential,
which are promoted to the constraints on GUT-scale parameters through
the RG evaluation. The numerical solutions of the MSSM RG equations
with large $\tan\beta$ are written in the following form by solving
the EWSB conditions about $M_A$ and $\mu$:
{\allowdisplaybreaks%
\begin{eqnarray}
  M_A^2 &\;=\;& (c_{ms}+c_{md}\xi)\tilde m_0^2 +c_M M^2_{1/2} 
  +c_{AM}A_0M_{1/2} +c_A A_0^2 -M_Z^2, 
  \label{mafit1} \\
  |\mu|^2 &\;=\;& (d_{ms}+d_{md}\xi)\tilde m_0^2 +d_M M_{1/2}^2 
  +d_{AM}A_0M_{1/2} +d_A A_0^2 -\frac{M_Z^2}{2}.
  \label{mufit1}
\end{eqnarray}}%
The numerical coefficients $c$'s and $d$'s are dimensionless
quantities which are determined by the gauge and Yukawa
couplings. Typical behaviors of $c$'s and $d$'s at 1 TeV are shown in
Fig.~\ref{fit1} as the function of $m_t^{\rm pole}$, where the
input parameters are the same as in Fig.~\ref{mbfig} and the threshold
corrections to top and tau-lepton masses are set 
to $\Delta_t=0.03$ and $\Delta_\tau=-0.02$ as an example. Several
implications of the above RG solutions are investigated below in order.

First, let us study the CP-odd neutral Higgs 
mass $M_A$~(\ref{mafit1}). As discussed previously, the positiveness 
of $M_A^2$, i.e.\ a successful EWSB, requires the separation of Higgs
SUSY-breaking mass parameters, $m_{H_u}^2<m_{H_d}^2$, at the electroweak
scale. In the present $SO(10)$-type unification where the two Higgs
masses are unified at the GUT scale, the EWSB must be triggered purely
by radiative corrections. It is found from Fig.~\ref{fit1} that
the contributions of gaugino mass $c_M$ and of scalar 
mass $c_{ms}$ have similar magnitudes with opposite signs, while the
other effects are relatively small. This suggests that a difference
between up- and down-type Higgs masses is not generated in the RG
evolution due to the structure of Yukawa couplings: the GUT-scale
Yukawa unification means not only the top but also the bottom and tau
Yukawa couplings are as large as ${\cal O}(1)$ which induce
significant RG effects on the down-type Higgs mass to make it equal to
the up-type Higgs mass in low-energy Higgs potential. Consequently the
condition for positive $M_A^2$ naively seems difficult to be satisfied.

In the present model we now consider, the separation of two Higgs mass
parameters is produced by $y_\tau$ and $g_1$~\cite{COPW}. Notice that,
if one took a limit $y_\tau,\,g_1\to0$, the theory has 
an $SU(2)$ symmetry which is identified to the global version 
of $SU(2)_R$ in the Pati-Salam unification 
group $SU(4)\times SU(2)_L\times SU(2)_R$~\cite{PS}. In the symmetric
limit, two Higgs SUSY-breaking masses are identical and the radiative
EWSB does not occur. The positive value of $c_M$ (gaugino mass effect)
reflects the fact that the $SU(2)_R$ breaking ($y_\tau,\,g_1\neq0$)
induces $y_t>y_b$ in the RG evolution, and then 
lowers $m_{H_u}^2$ than $m_{H_d}^2$ at low energy, which difference is
enhanced in the case of large gaugino mass. On the other hand, the
negative value of $c_{ms}$ (scalar mass effect) is a result that 
the $SU(2)_R$ breaking ($y_\tau\neq0$ and the absence of neutrino
Yukawa coupling) induces $m_{H_u}^2>m_{H_d}^2$ at low energy, which is
enhanced by larger scalar masses in the RG evolution. As a result, if
smaller terms of $A_0$ and $\xi$ are neglected, the experimental lower
bound on CP-odd neutral Higgs mass $M_A\gtrsim M_Z$ gives the
following restriction between gaugino and scalar mass parameters:
\begin{equation}
  M_{1/2}^2 \;\gtrsim\; \frac{-c_{ms}\tilde m_0^2+2M_Z^2}{c_M}.
\end{equation}
It is found from explicit numerical values in Fig.~\ref{fit1} that
the right-hand side is larger than $\tilde m_0^2$, and hence the above
restriction is conservatively rewritten as
\begin{equation}
  M_{1/2}^2 \;\gtrsim\; \tilde m_0^2.
  \label{MAbound}
\end{equation}
This inequality is an important and strong constraint on the GUT-scale
SUSY-breaking parameters; a half of parameter space is ruled out. It
is also noticed that $M_A$ is bounded from above by gaugino mass parameter
\begin{equation}
  M_A^2 \;\lesssim\; c_M M^2_{1/2} -M_Z^2.
\end{equation}
In the $SO(10)$-type unification, therefore, the CP-odd neutral Higgs
boson is generally predicted to be light~\cite{COPW}. The 
constraint (\ref{MAbound}) is not sensitive to the other 
parameters $A_0$ and $\xi$. The $A_0$ dependent terms have only tiny
effects because the coefficients $c_A$ and $c_{AM}$ are very 
small, $|c_A|,|c_{AM}|\lesssim{\cal O}(0.01)$. An extremely large
value of $|A_0|$ tends to make the EWSB vacuum unstable~\cite{CCB} and
is disfavored. As for the $\xi$ dependence, the CP-odd neutral Higgs
mass is a bit affected, depending on the sign 
of $\xi$. However $\xi$ is constrained by other superparticle mass
bounds; in particular, a large value of $|\xi|$ leads to scalar tau
being the lightest supersymmetric particle (LSP), and hence 
the $\xi$ term cannot be so large.

Let us turn to studying the second condition (\ref{mufit1}) concerning
the higgsino masses. The numerical solution of the RG equations ($d$'s
in Fig.~\ref{fit1}) indicates that the dominant positive
contribution is the gaugino mass effect ($d_M$). The only possible
correction comes from the scalar mass effect, especially 
the $\xi$ term ($d_{md}$), which can lower $|\mu|^2$ 
when $\xi\gtrsim\frac{-d_{ms}}{d_{md}}$, 
e.g.\ $m_{10}^2\gtrsim1.3m_{16}^2$ for $m_t^{\rm pole}=178$~GeV\@.
The constraint (\ref{MAbound}) however means that such scalar mass
contribution cannot be larger than that of gaugino, even if $\xi$ has
its maximal value $1$. Furthermore a large value of $\xi$ leads to the
LSP scalar tau lepton and is disfavored. The other effects from 
the $A_0$ and $M_Z$ terms are negligibly small. Thus the gaugino mass
effect becomes dominant in large region of SUSY-breaking parameters;
the low-energy $|\mu|$ is approximately given by the gaugino mass
\begin{equation}
  |\mu|^2 \;\sim\; d_M M_{1/2}^2.
  \label{mu}
\end{equation}
In this case, the lighter neutralinos and chargino become 
gaugino-like, since the unified gaugino mass at the GUT scale leads 
to $M_1\simeq0.4M_{1/2}$ and $M_2\simeq 0.8M_{1/2}$ at the electroweak
scale, which are generally smaller than $|\mu|$.

In these ways the radiative EWSB in the $SO(10)$-type unification
requires a restricted type of low-energy superparticle spectrum. This
is mainly due to the constraint (\ref{MAbound}) on the GUT-scale
parameters which generally predicts (i) scalar quark masses are
correlated with the gluino mass through the RG evolution and cannot be
much larger than it, (ii) scalar leptons also cannot be much heavier
than the $SU(2)_L$ gaugino, (iii) the gaugino components are dominant
in the lighter neutralinos and chargino, and (iv) a relatively light
CP-odd neutral Higgs boson is expected. It is stressed here that
Eq.~(\ref{mu}) and mass eigenvalues related to the $\mu$ parameter
depend on the magnitude of $\xi$ term in the RG solution
(\ref{mufit1}), which is restricted by the inequality (\ref{MAbound})
and the requirement that the LSP is charge neutral. The relevance 
of $\xi$-dependent contribution will be investigated in later sections
and found to sometimes play an important role in establishing the
successful radiative EWSB in other scenarios.

In the next subsection a detailed study will be given for the
experimental constraints in the minimal $SO(10)$-type scenario. Before 
proceeding to numerical analysis, we here present a summary of the
results, referring to the mass spectrum naively expected from the
above discussion. First, the threshold correction to bottom quark mass
is generally not suppressed. That depends on the relative strength of
scalar quark mass, gluino mass, and $\mu$ parameters. The EWSB 
constraints (\ref{MAbound}) and (\ref{mu}) mean that the PQ and R
symmetries are largely violated, and the threshold correction becomes
large as discussed in Section~\ref{sec:tbtau}. Secondly,
the $b\to s\gamma$ amplitude becomes large since the charged
Higgs contribution $A_{H^+}$ is enhanced by a light charged Higgs
boson $M_{H^+}^2=M_A^2+M_W^2$. In particular, for $\mu<0$, 
the observation of $b\to s\gamma$ rare decay severely restricts the
parameter space of the model. The enhancements of the threshold 
correction $\Delta_b$ and the $b\to s\gamma$ amplitude generally make
the minimal $SO(10)$-type unification difficult to be consistent with
the observation.

%%%%%%%%%%%%%%%%%%%%%%%%%%%%%%%%%%%%%%%%%%%%%%%%%%%%%%%%%%%%%%%%%%%%%%
\subsection{Parameter Space Analysis}
%%%%%%%%%%%%%%%%%%%%%%%%%%%%%%%%%%%%%%%%%%%%%%%%%%%%%%%%%%%%%%%%%%%%%%
We perform the numerical analysis of parameter space in the 
minimal $SO(10)$-type unification which space is allowed by the
experimental constraints from the bottom quark mass and 
the $b\to s\gamma$ decay rate. The input parameters are the same as
those in Section~\ref{sec:general}, and the top quark mass 
is $m_t^{\rm pole}=178$~GeV\@. The unified Yukawa 
coupling $y_{{}_G}$ is evaluated by using the two-loop MSSM RG
equations and one-loop SUSY threshold corrections to top and tau
Yukawa couplings, $\Delta_t$ and $\Delta_\tau$, which are controlled
by SUSY-breaking mass parameters. The low-energy threshold corrections
to gauge couplings are also taken into account. Once $y_{{}_G}$ is
determined, one can solve the EWSB conditions, the mass bound of
CP-odd neutral Higgs boson, and the requirement of neutral LSP at the
scale $Q=M_{\rm SUSY}$ which is typically defined by scalar quark mass
parameters as $M_{\rm SUSY}=(m^2_{\tilde Q 33}m^2_{\tilde u 33})^{1/4}$. 
The current experimental lower bounds on gaugino masses are included
as in the previous section, and the mass bounds on the scalar top,
bottom, tau are $m_{\tilde t_1}>95.7$~GeV, $m_{\tilde b_1}>89$~GeV, 
and $m_{\tilde \tau_1}>81.9$~GeV (95\% CL),
respectively~\cite{PDG}. Finally the bottom quark mass is estimated
with one-loop SUSY threshold correction and two-loop SM QCD
correction~\cite{PBMZ,BRP,HRS}, and the $b\to s\gamma$ branching ratio
is calculated according to the formulas in~\cite{bsgformula}.

\begin{figure}[t]
\begin{center}
\includegraphics[width=5cm]{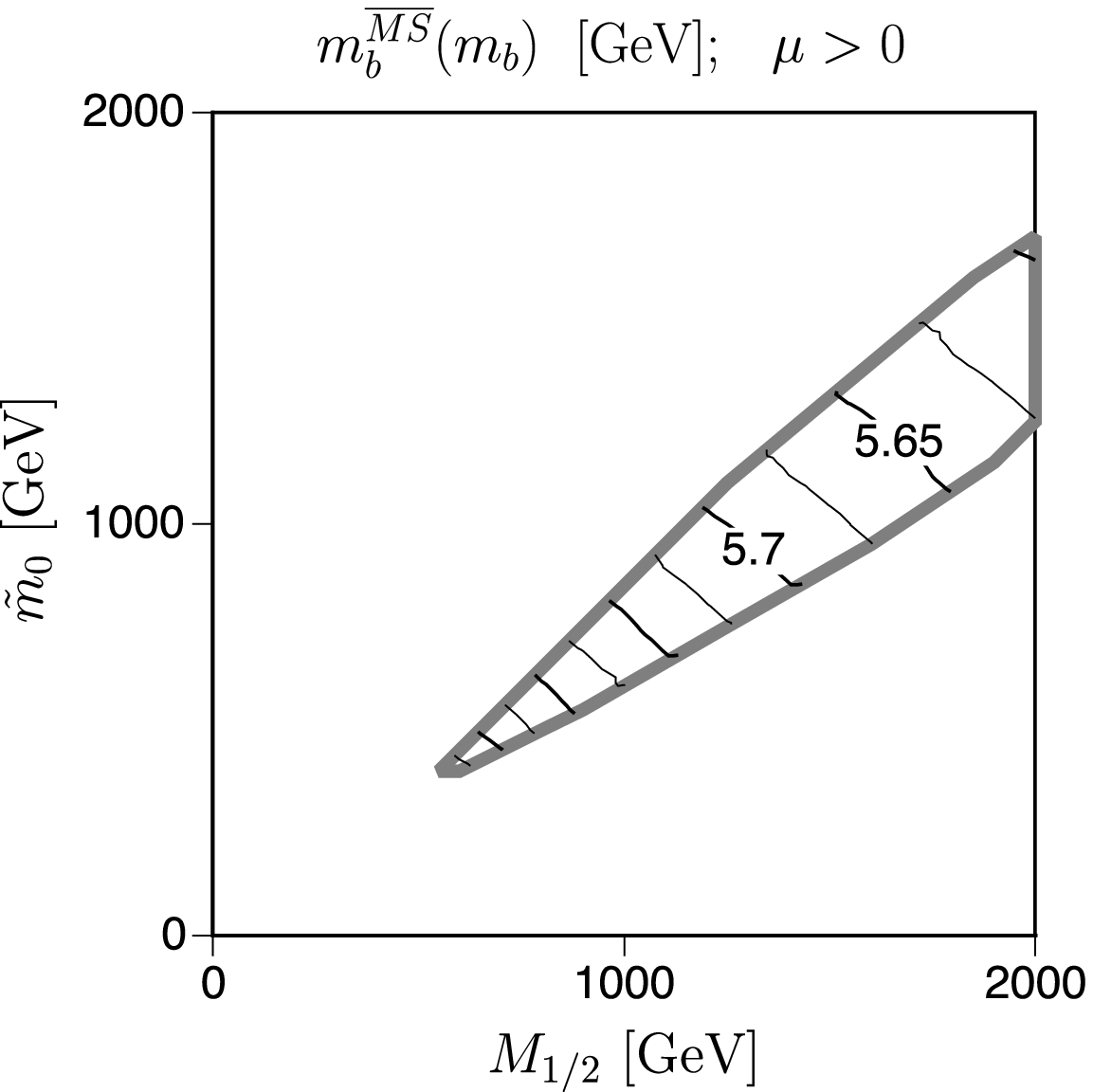}\hspace*{1cm}
\includegraphics[width=5cm]{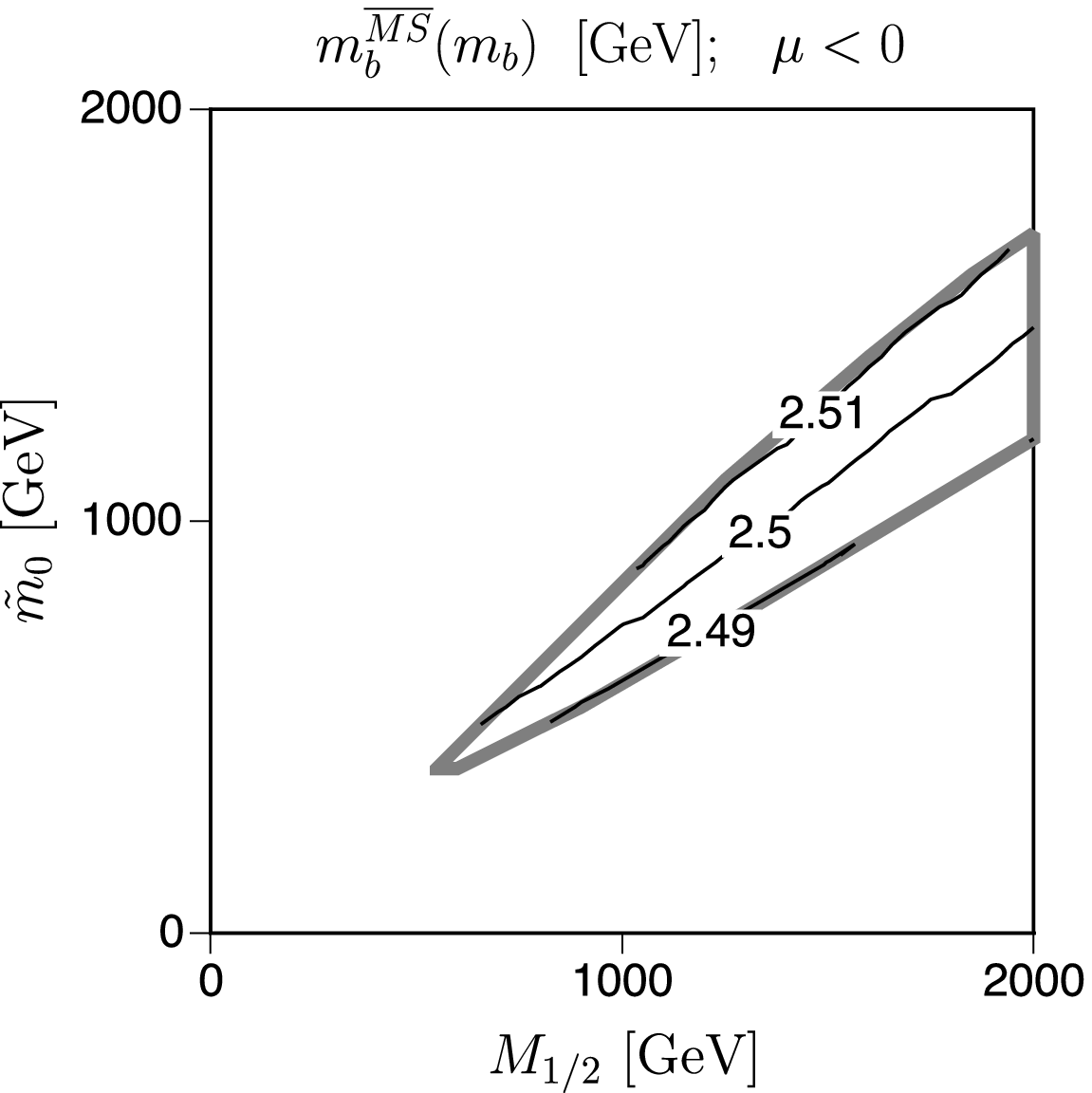}\\[5mm]
\includegraphics[width=5cm]{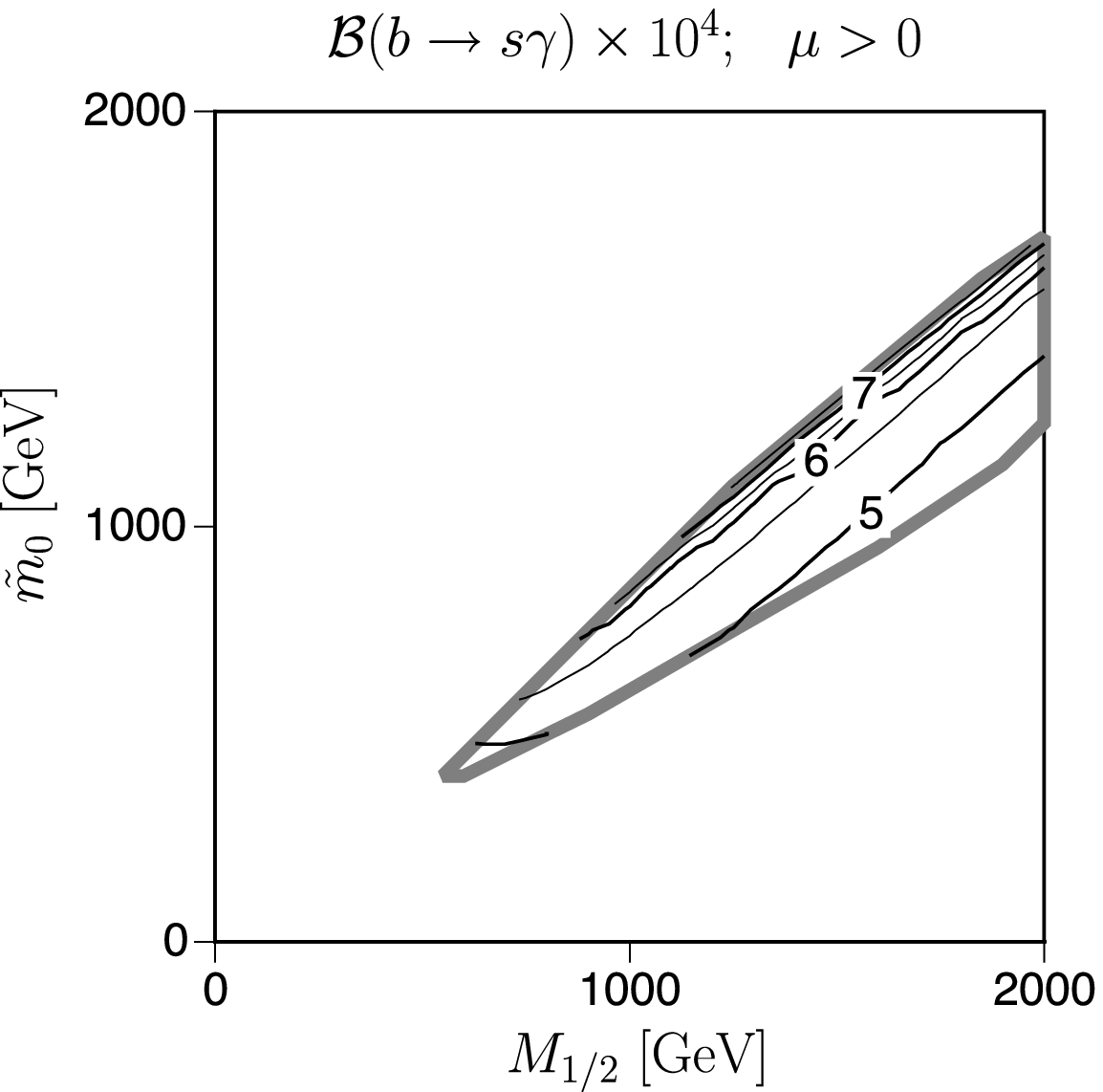}\hspace*{1cm}
\includegraphics[width=5cm]{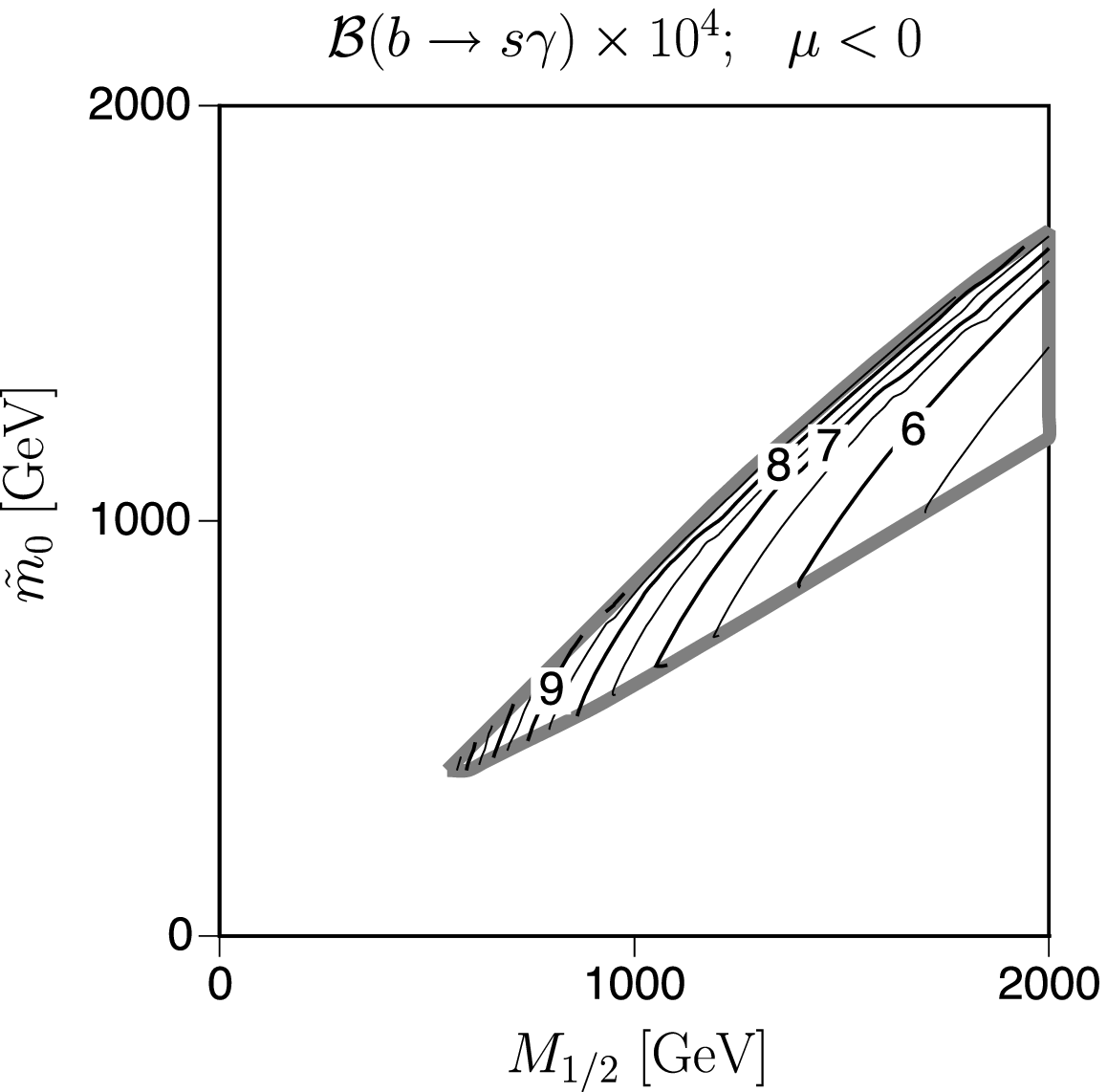}
\end{center}
\caption{The parameter space consistent with the radiative EWSB, the
experimental mass bounds of superparticles, and the requirement of
neutral LSP in the minimal $SO(10)$-type unification. The bottom quark
mass $m_b^{\overline{\text{MS}}}(m_b)$ and 
the $b\to s\gamma$ branching ratio are also shown in the figures. The
GUT-scale scalar mass parameters are set as $A_0=\xi=0$. The two-loop
MSSM RG equations for gauge and Yukawa couplings and the one-loop ones
for dimensionful parameters are used. The one-loop SUSY threshold
corrections to gauge and Yukawa couplings are included. The radiative
EWSB conditions are solved by use of the one-loop effective potential
at the scale $Q=M_{\rm SUSY}$ which is defined by the scalar quark
masses as $M_{\rm SUSY}=(m^2_{\tilde Q 33}m^2_{\tilde u 33})^{1/4}$. 
In each figure, the left-top region is excluded by the mass bound of
CP-odd neutral Higgs boson and the right-bottom region is ruled out
from the fact that the scalar tau lepton becomes the LSP.\bigskip}
\label{Mgm0}
\end{figure}
Fig.~\ref{Mgm0} shows that the parameter space consistent with
the radiative EWSB, the experimental mass bounds of superparticles,
and the requirement of neutral LSP\@. In the figures, the predictions
of bottom quark mass $m_b^{\overline{\text{MS}}}(m_b)$ and 
the $b\to s\gamma$ branching ratio are shown in the allowed parameter
regions. For simplicity, vanishing scalar trilinear couplings and the
universal scalar masses ($A_0=\xi=0$) have been assumed in the
figures. The numerical result here shows that the GUT-scale gaugino
mass $M_{1/2}$ must be larger than the universal scalar 
mass $\tilde m_0$, which confirms the previous 
analysis (\ref{MAbound}). Too a large value of gaugino 
mass $M_{1/2}>2\tilde m_0$ is excluded by the requirement of neutral
LSP, in which region the scalar tau lepton becomes the LSP\@. The
unification of Yukawa couplings generally makes scalar lepton mass
eigenvalues small due to the $y_\tau$ contribution to scalar lepton
masses in the RG evolution and large left-right mixing elements in the
scalar tau mass matrix which is proportional 
to $\tan\beta$. In the region allowed by the experimental
constraints, SUSY spectrum is severely constrained from the 
inequality (\ref{MAbound}). For example, in the present model,
the $\mu$ parameter at low energy has a strong correlation with the
gaugino mass; $|\mu|^2\simeq1.4M_{1/2}^2$. Therefore the lightest
neutralino and chargino become gaugino-like. Also scalar quark masses
are correlated with $M_{1/2}$ and have few dependence on the initial 
value $\tilde m_0$; the mass of light scalar top is given 
by $m_{\tilde t_1}^2\simeq3.1M_{1/2}^2$ or 
equivalently $m_{\tilde t_1}^2\simeq0.5M_3^2$. The CP-odd neutral
Higgs boson is generally expected to be light in the 
minimal $SO(10)$-type unification; $M_A^2\lesssim0.06M_{1/2}^2$ in the
parameter region in Fig.~\ref{Mgm0}.

The minimal $SO(10)$-type unification leads to the constrained mass
spectrum largely violating the PQ and R symmetries. That predicts a
large value of the threshold correction $|\Delta_b|$, that 
is, $0.3\lesssim|\Delta_b|\lesssim 0.4$ in the allowed parameter
region of Fig.~\ref{Mgm0}. As examined in
Section~\ref{sec:tbtau}, the Yukawa unification requires a small value
of $|\Delta_b|$ in order to obtain the bottom quark mass within the
experimentally allowed range. In the present case, too large a
magnitude of the finite threshold correction $|\Delta_b|$ spoils the
successful bottom mass prediction. It is observed from
Fig.~\ref{Mgm0} that $m_b$ is too large for $\mu>0$ and too
small for $\mu<0$. The bottom mass prediction has little sensitivity
to the overall SUSY-breaking scale. Further 
the $b\to s\gamma$ branching ratio generally becomes large in the
minimal $SO(10)$-type unification. This is due to the large 
amplitude $|A_{H^+}|$ enhanced by a small value of the charged Higgs
boson mass. As discussed before, a cancellation between the 
amplitudes $A_{H^+}$ and $A_{\tilde \chi^+}$ is possible with a
positive $\mu$ parameter. Now we have a hierarchical spectrum that
scalar quarks are much heavier than the charged Higgs boson, and 
then $|A_{H^+}|$ becomes 
large; $|A_{H^+}|\gtrsim1.3|A_{\tilde \chi^+}|$ in the parameter
region of Fig.~\ref{Mgm0}. Therefore the cancellation among
the amplitudes is not enough to suppress the non-SM contributions 
to $b\to s\gamma$ transition even for $\mu>0$, and the experimental
constraint is serious. For the $\mu<0$ case, the constraint becomes
severer than the $\mu>0$ case.

\begin{figure}[t]
\begin{center}
\includegraphics[width=5cm,clip]{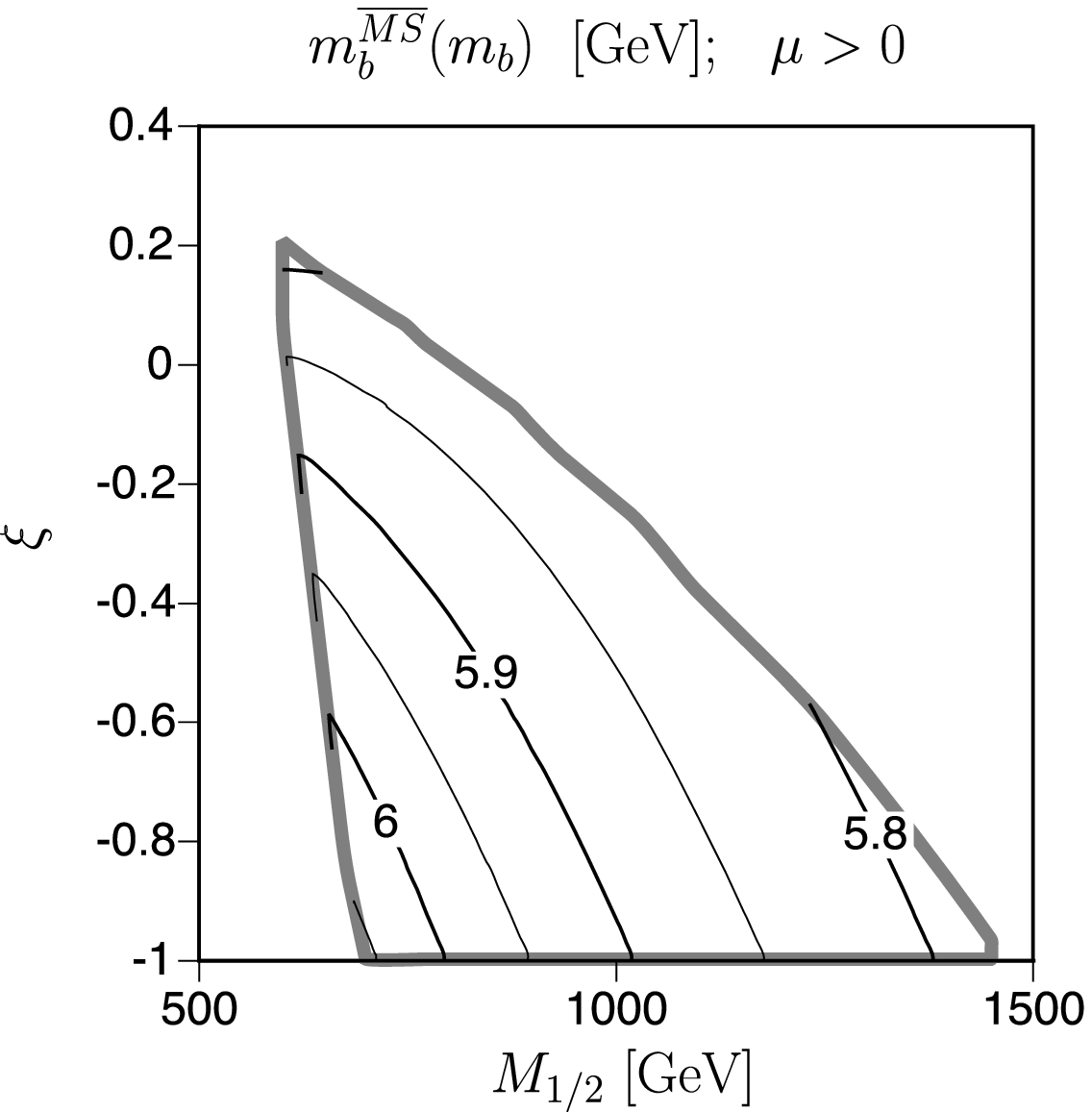}\hspace*{1cm}
\includegraphics[width=5cm,clip]{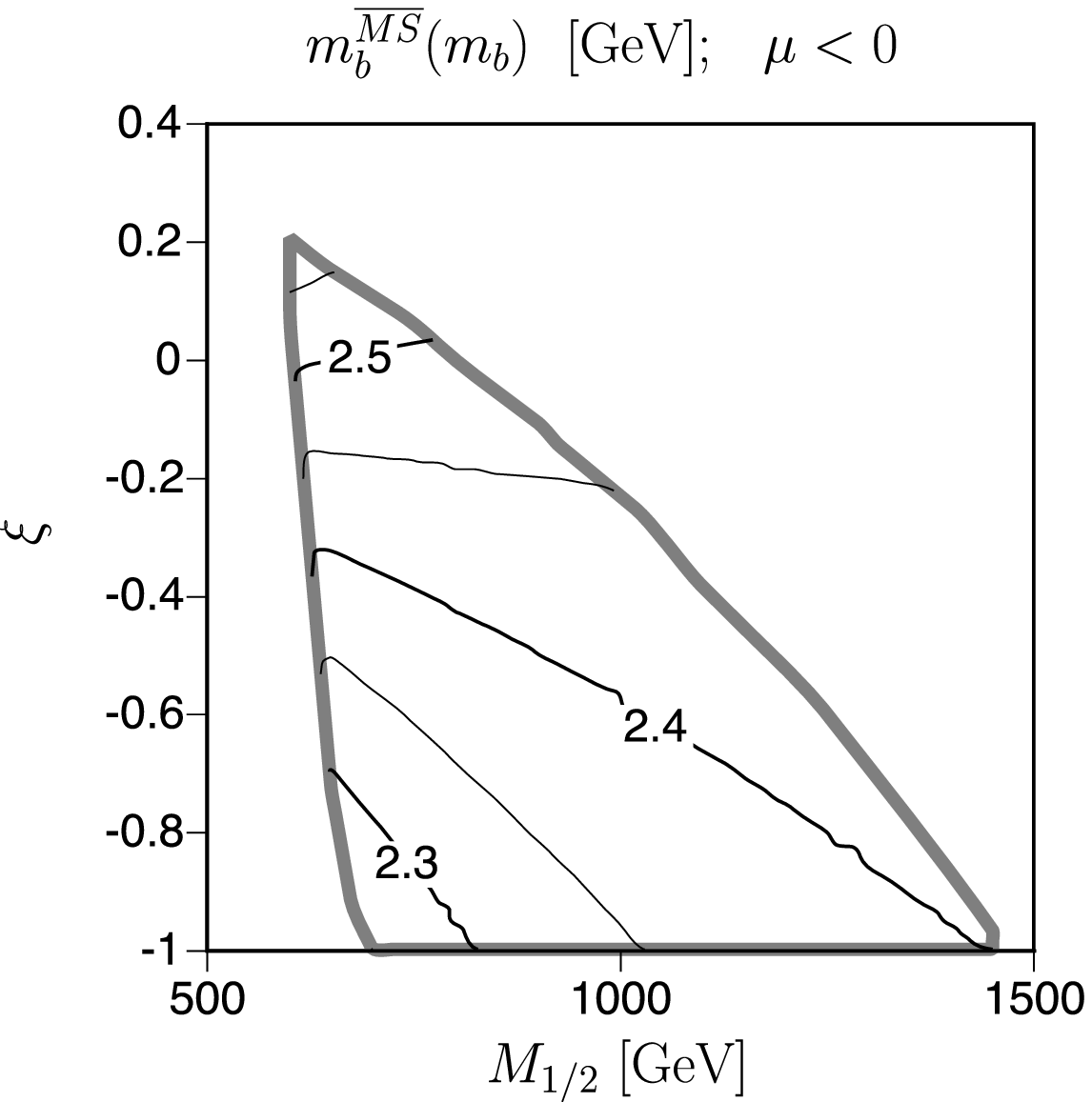}\\[5mm]
\includegraphics[width=5cm,clip]{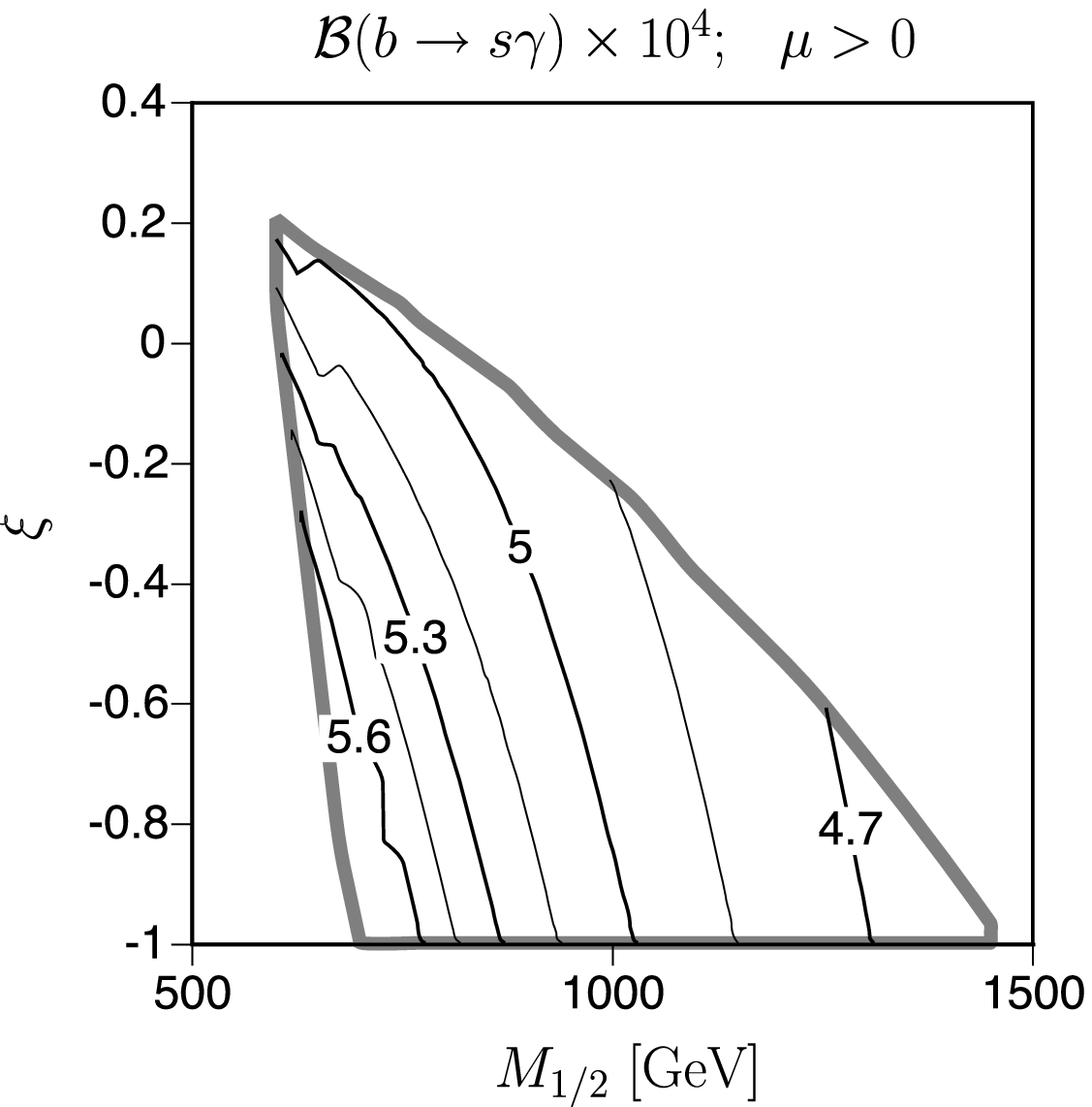}\hspace*{1cm}
\includegraphics[width=5cm,clip]{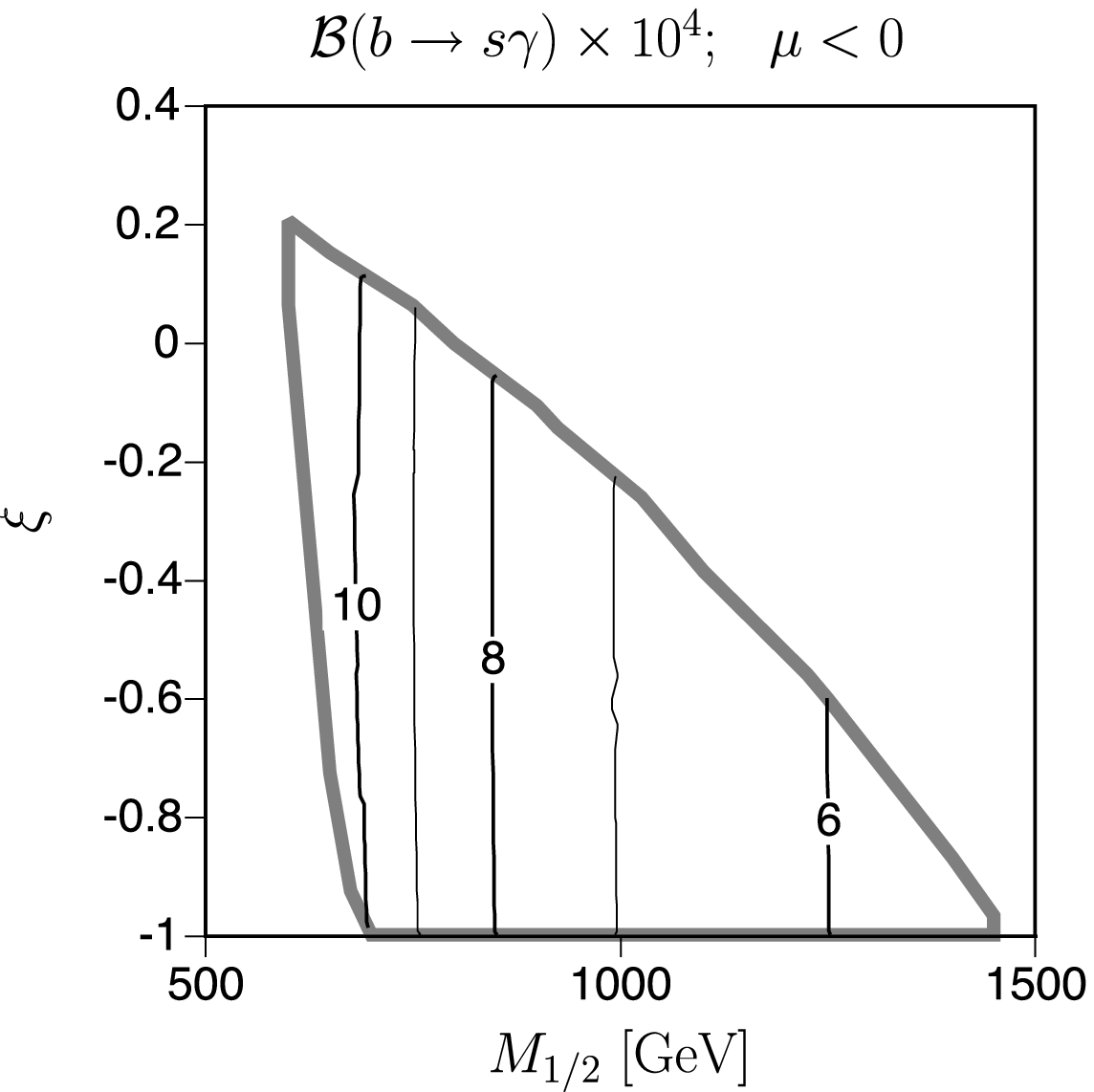}
\end{center}
\caption{The same as Fig.~\ref{Mgm0} but 
for $M_{1/2}$--$\xi$ parameter space. The GUT-scale scalar mass
parameters are set as $\tilde m_0=500$~GeV and $A_0=0$. In each
figure, the narrow left side is excluded by the experimental mass
bound on CP-odd neutral Higgs boson and the right-top region is ruled
out from the fact that the scalar tau lepton becomes the LSP.\bigskip}
\label{Mgxi}
\end{figure}

\begin{figure}[t]
\begin{center}
\includegraphics[width=5cm,clip]{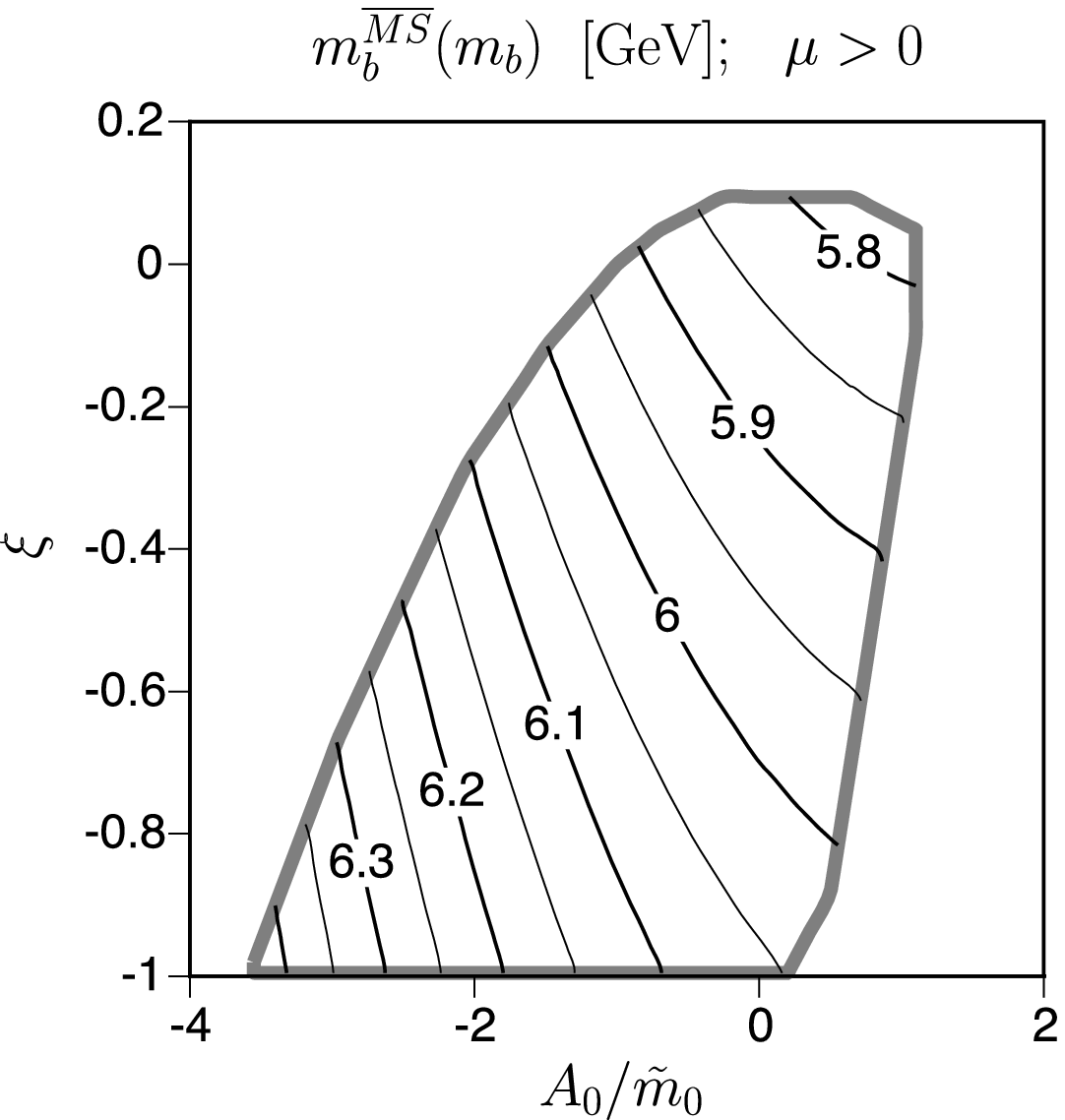}\hspace*{1cm}
\includegraphics[width=5cm,clip]{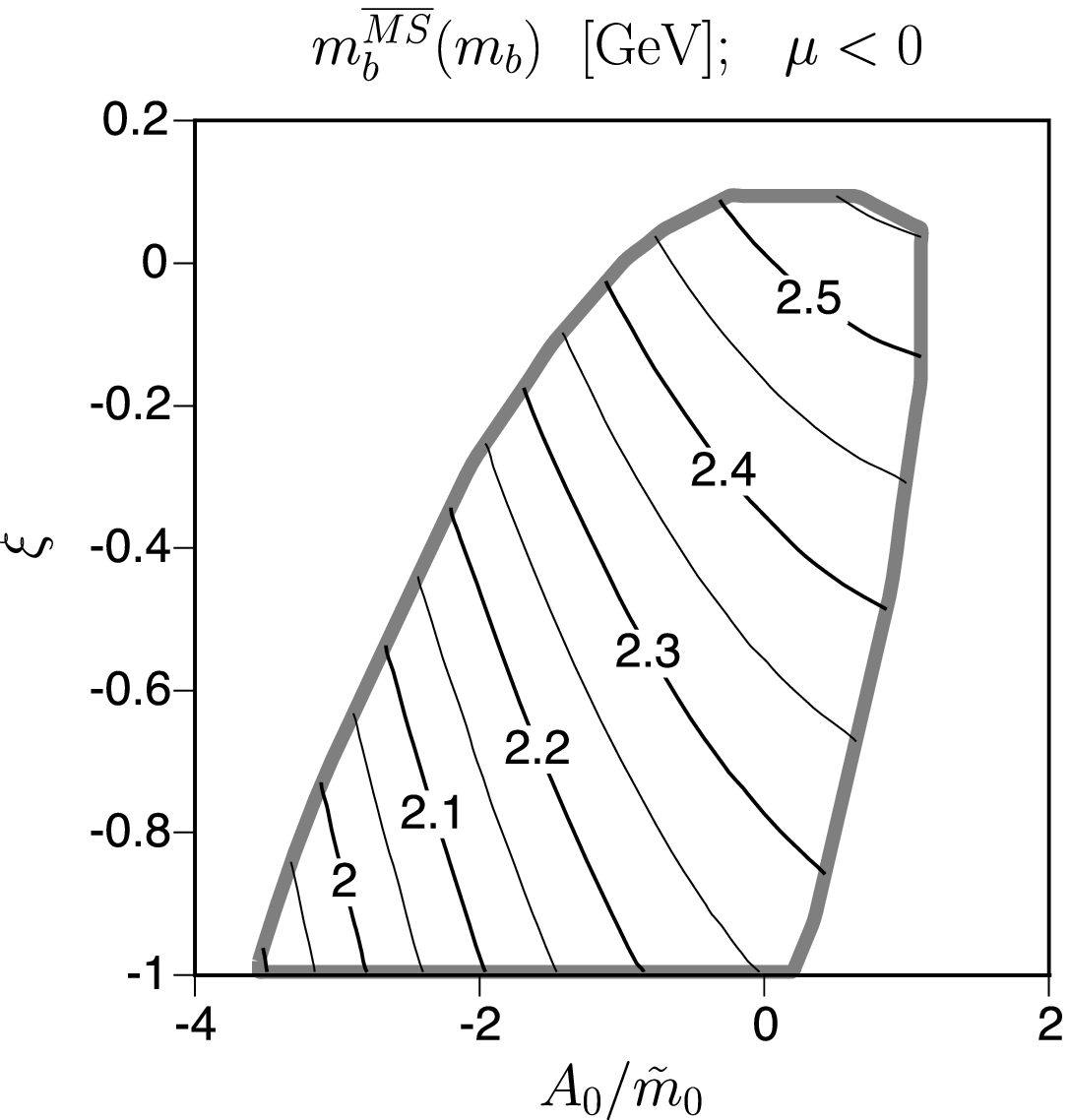}\\[5mm]
\includegraphics[width=5cm,clip]{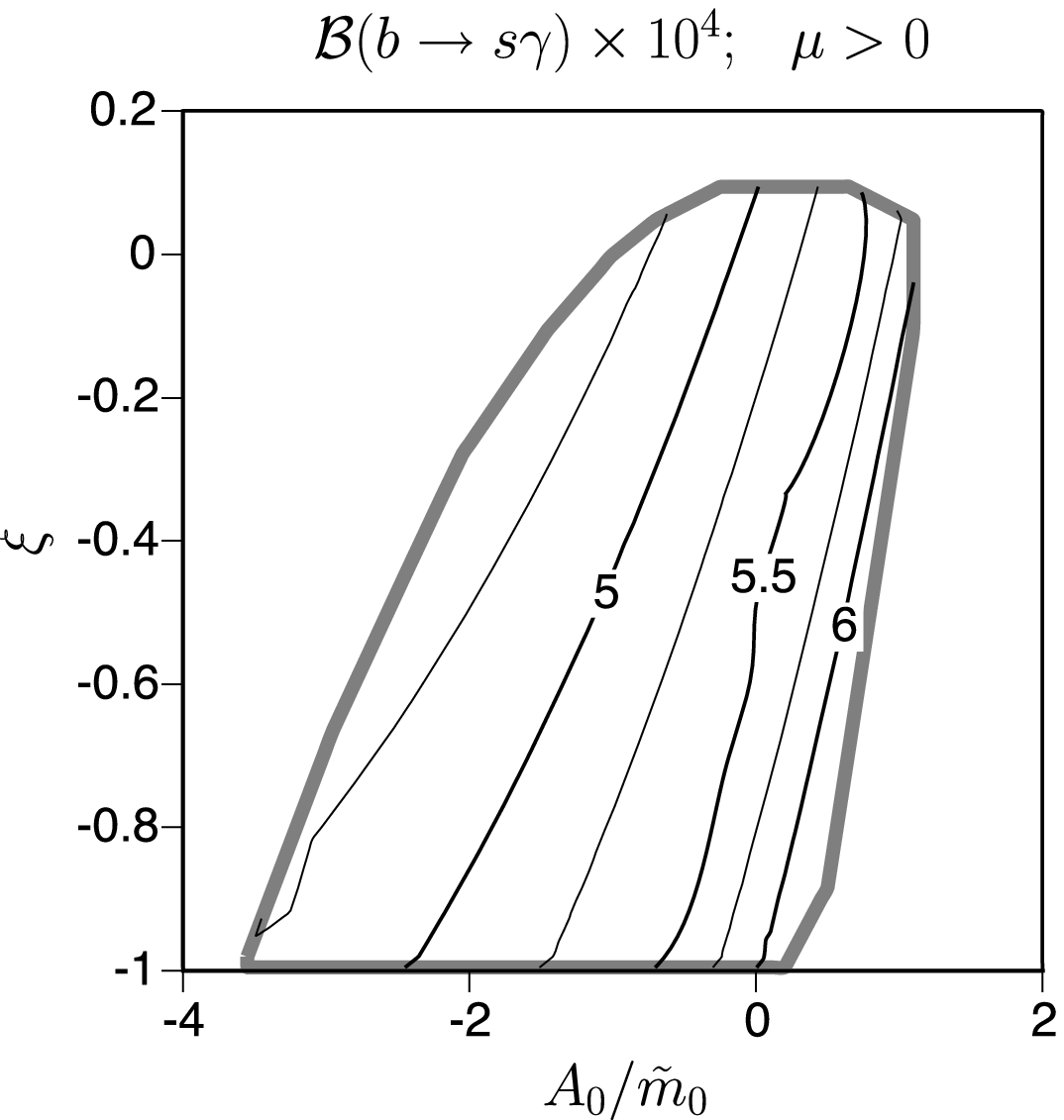}\hspace*{1cm}
\includegraphics[width=5cm,clip]{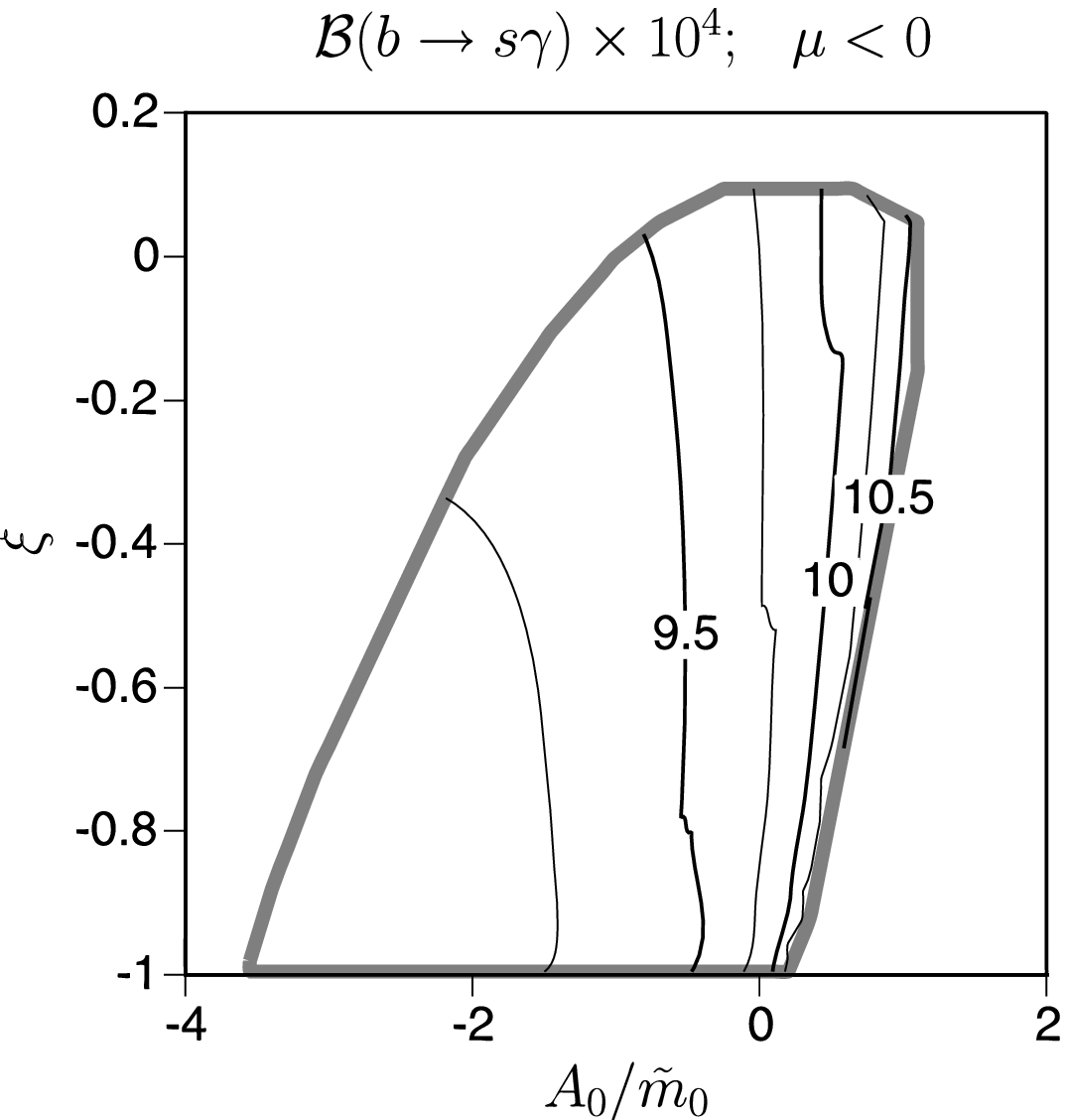}
\end{center}
\caption{The same as Fig.~\ref{Mgm0} but 
for $A_0$--$\xi$ parameter space. The GUT-scale scalar mass parameters
are set as $\tilde m_0=500$~GeV and $M_{1/2}=700$~GeV. In each figure,
the right side is excluded by the experimental mass bound on CP-odd
neutral Higgs boson and the left and top regions are ruled out from
the fact that the scalar tau lepton becomes the LSP.\bigskip}
\label{A0xi}
\end{figure}

Next, let us examine the parameter dependences 
on $A_0$ and $\xi$, that is, non-vanishing scalar trilinear couplings
and the difference of SUSY-breaking masses between matter and Higgs
fields. Figs.~\ref{Mgxi} and~\ref{A0xi} are the same 
as Fig.~\ref{Mgm0} but for $M_{1/2}$--$\xi$ and $A_0$--$\xi$
parameter spaces, respectively. The experimental mass bound on CP-odd
neutral Higgs boson excludes the left (right) side of parameter space
in Fig.~\ref{Mgxi} (Fig.~\ref{A0xi}). The charged LSP regions
correspond to the right-top (left and top) region 
in Fig.~\ref{Mgxi} (Fig.~\ref{A0xi}).

The CP-odd neutral Higgs mass has little dependence on $\xi$. This
behavior can be understood from the tiny value of $c_{md}$ in
(\ref{mafit1}). As mentioned before, the CP-odd neutral Higgs mass is
scaled with the difference between the Higgs mass parameters. A
non-zero $\xi$ generates a separation between Higgs and matter scalar
masses at the GUT scale, but does not directly contribute to the
separation inside Higgs masses. Thus the experimental bound on CP-odd
neutral Higgs mass is not relaxed with the freedom 
of $\xi$. The $\xi$ dependence of the $b\to s\gamma$ branching ratio
is also small. This is a consequence of small $\xi$ dependence on the
charged Higgs mass. In contrast to the CP-odd neutral Higgs mass, the
scalar tau mass has larger dependence on $\xi$; a larger 
negative $\xi$ raises the mass of scalar tau and the constraint from
the LSP is relaxed. The more $\xi$ increases, the lighter scalar tau
lepton is. The roughly upper half of the $M_{1/2}$--$\xi$ plane is
excluded by the LSP condition. This in turn implies that 
the $\xi$-dependent term ($d_{md}$) in the RG solution (\ref{mufit1})
is much smaller than the gaugino mass 
effect ($d_M$), and $|\mu|$ cannot be smaller than $M_{1/2}$; for 
example, $1.2\lesssim |\mu|^2/M_{1/2}^2\lesssim 2.0$ in the allowed
parameter region in Fig.~\ref{Mgxi}. For a 
negative $\xi$, the $\mu$ parameter is increased which enhances the
magnitude of threshold correction $\Delta_b$, and consequently, the
bottom quark mass becomes large.

The $A_0$ dependence of CP-odd neutral Higgs mass is more relevant
than that on $\xi$. It is found from the RG solution (\ref{mufit1})
that, for a positive (negative) $A_0$, the CP-odd neutral Higgs mass
generally becomes smaller (larger) than the $A_0=0$ case. In
Fig.~\ref{A0xi}, we 
have $90\text{~GeV}\lesssim M_A\lesssim 180$~GeV\@. Note that a rather
large value of $|A_0|$ lowers $M_A$ even for a negative value 
of $A_0$ because such a large $|A_0|$ enhances the tau Yukawa effect
in the RG evolution which lowers $m_{H_d}^2$ rather 
than $m_{H_u}^2$ in low-energy regime. In a similar way, the masses of
scalar tau lepton and scalar quarks have sizable dependences 
on $A_0$. A large $|A_0|$ lowers these masses through the RG evolution
down to low energy, and in particular, the large $|A_0|$ region is
excluded since the LSP is scalar tau lepton. As seen from
(\ref{mufit1}), a negative $A_0$ raises $|\mu|$ and then the bottom
mass prediction is enhanced and becomes worse. Contrary to this 
behavior, the $b\to s\gamma$ branching ratio is decreased by a
negative $A_0$. This is a consequence of the $A_0$ dependence 
of $M_{H^+}$; a negative $A_0$ raises $M_{H^+}$ and 
lowers $A_{H^+}$ whose absolute value is generally larger 
than $A_{\tilde \chi^+}$ in the minimal $SO(10)$-type
unification. Therefore the experimental constraint 
from the $b\to s\gamma$ process is relaxed.

These $\xi$ and $A_0$ dependences slightly modify 
the $M_{1/2}$--$\tilde m_0$ parameter space which is consistent with
the mass bound of CP-odd neutral Higgs boson and the requirement of
neutral LSP (Fig.~\ref{Mgm0}). However such change is small
and does not allow the EWSB with $M_{1/2},\,|\mu|\ll\tilde m_0$, with
which spectrum the threshold correction to bottom quark mass is
suppressed in a technically natural way. The 
minimal $SO(10)$-type unification is therefore difficult to reproduce
the observed value of bottom quark mass. The $b\to s\gamma$ rare decay
process also restricts a light superparticle spectrum at low-energy
regime.

%%%%%%%%%%%%%%%%%%%%%%%%%%%%%%%%%%%%%%%%%%%%%%%%%%%%%%%%%%%%%%%%%%%%%%
\subsection{Discussions}
%%%%%%%%%%%%%%%%%%%%%%%%%%%%%%%%%%%%%%%%%%%%%%%%%%%%%%%%%%%%%%%%%%%%%%
As shown in Section~\ref{sec:tbtau}, the enhanced threshold
corrections to bottom quark mass are proportion 
to $\mu M_3$ and $\mu A_t$, which are controlled by the PQ and R
symmetries. These symmetries are useful in the $SO(10)$-type
unification since the suppression of the threshold correction is
required to attain the experimentally allowed fermion masses in the
Yukawa unification~\cite{HRS,RS,TW}. It is however found that 
the $SO(10)$-type unification does not allow to impose the PQ and R
symmetries because a successful EWSB, in particular, the separation of
two Higgs scalar masses, requires a symmetry-violating 
condition $M_{1/2}\gtrsim\tilde m_0$ for the GUT-scale SUSY-breaking
parameters. This behavior is a consequence that the RG 
solution (\ref{mafit1}) receives a sizable negative contribution from
scalar masses. If the coefficient $c_{ms}$ in (\ref{mafit1}) turns to
be positive or there exist some additional positive contributions, the
radiative EWSB with a small gaugino mass $M_{1/2}<\tilde m_0$ is
possible, and accordingly one could
take $M_{1/2},\,A_0,\,B_0\ll\tilde m_0$, that is, the R symmetric
radiative EWSB is viable. Moreover, in such a case, the PQ symmetry is
also realized since the gaugino mass effect in the 
solution (\ref{mufit1}) could be canceled by the scalar mass
contribution.

It is pointed out~\cite{D} that the $D$-term effect of 
additional $U(1)$ symmetry contributes to the CP-odd neutral Higgs
mass without disturbing the PQ and R symmetries. Then the radiative
EWSB with $M_{1/2}<\tilde m_0$ is available with an 
appropriate $D$-term contribution included. Along this line, there are
various studies about phenomenological aspects 
in $SO(10)$ models (see, e.g.\ \cite{RS,BDR,Tata}). On the other hand,
a similar type of EWSB is also viable with some specific types of
non-universal scalar masses which do not respect $SO(10)$ unified
gauge symmetry. In these cases, approximate PQ and R symmetries are
realized in low-energy mass spectrum.

In the following sections, we investigate alternative possibilities to
attain PQ and/or R symmetric spectrum which include the effects
inspired by neutrino physics in $SO(10)$ models, and will find new
types of radiative EWSB scenarios.

%%%%%%%%%%%%%%%%%%%%%%%%%%%%%%%%%%%%%%%%%%%%%%%%%%%%%%%%%%%%%%%%%%%%%%
\bigskip
\section{$\boldsymbol{SO(10)}$ Unification with Neutrino Couplings}
\label{sec:wRHnu}
%%%%%%%%%%%%%%%%%%%%%%%%%%%%%%%%%%%%%%%%%%%%%%%%%%%%%%%%%%%%%%%%%%%%%%
It has been indicated by various recent experiments that neutrinos
have tiny mass scale less than a few eV, which is extremely smaller
than the other SM fermion masses. In the framework 
of $SO(10)$ unification, a 16-plet contains a single field under the
SM gauge group which may be naturally identified to a right-handed
neutrino. The $SO(10)$-invariant superpotential 
term $16\,16\,10_H$ generates neutrino Yukawa couplings among the
left- and right-handed neutrinos and the up-type Higgs boson. Thus
neutrinos obtain a similar size of Dirac masses to the other SM
fermions and the observed tiny mass scale seems unnatural. A promising
way to cure this problem is to introduce large Majorana masses for
right-handed neutrinos. Integrated out the heavy right-handed
neutrinos, tiny Majorana masses are generated for left-handed
neutrinos~\cite{seesaw}. Thus the superpotential terms below the GUT
scale in this scenario is given by
\begin{equation}
  W \;=\; W_{\rm MSSM} + L_i(Y_\nu)_{ij}\bar\nu_j H_u 
  +\frac{1}{2}\bar\nu_i({\cal M}_\nu)_{ij}\bar\nu_j.
\end{equation}
The last two terms are introduced in addition to the MSSM
superpotential (\ref{MSSMspot}) where $Y_\nu$ is the neutrino Yukawa
matrix and ${\cal M}_\nu$ denotes large-scale Majorana masses for
right-handed neutrino superfields $\bar\nu_i$ ($i=1,2,3$). As in the
previous Yukawa unified scenarios, we naturally have a hierarchical
order of neutrino Yukawa couplings and then only the third diagonal
element is large; $(Y_\nu)_{33}\equiv y_\nu\sim{\cal O}(1)$, which is
expected to be of the same order of the top Yukawa coupling 
in $SO(10)$ unification. Such a hierarchy assumption might also be
applied to the Majorana mass matrix of right-handed neutrinos. In the
following analysis we simply have the 3-3 element of the Majorana mass
matrix as $({\cal M}_\nu)_{33}\equiv M_\nu=10^{14}$~GeV\@. The 
minimal $SO(10)$-type unification discussed in the previous section
corresponds to the case that $M_\nu$ is equal to or larger than the
GUT-breaking scale. The other matrix elements of neutrino couplings
are smaller than these dominant 3-3 elements, and might be responsible
for explaining the observed large generation mixing of light
neutrinos~\cite{enhance}. The details of these small matrix elements
are completely irrelevant to the following RG analysis and can be
dropped. If there were other Yukawa matrix elements 
than $(Y_\nu)_{33}$ which take ${\cal O}(1)$ values, the effects of
neutrino couplings are enhanced accordingly. However in this paper we
assume a conservative case that only the tau-neutrino Yukawa coupling
is large.

In this section, we study the unification scenario 
with $M_\nu<M_G$. As a result, the effects of neutrino couplings
become important in the RG evolution between $M_G$ and $M_\nu$. The
influences of neutrino Yukawa couplings have been studied concerning
on the gauge coupling unification~\cite{CEIN} and the third-generation
fermion masses~\cite{VS,AK}. Also in Ref.~\cite{KO}, the radiative
EWSB is examined in the case of large violation of PQ and R
symmetries. It was found that such neutrino Yukawa effect is less than
a few percents and the qualitative discussion, e.g.\ that given in
Section~\ref{sec:tbtau}, is also applied to our present case. In
particular, a successful prediction of bottom quark mass still
requires that the threshold correction $\Delta_b$ at SUSY-breaking
scale must be suppressed. On the other hand, as will be shown below,
neutrino couplings play a significant role in the RG evolution of
SUSY-breaking parameters. Thus the picture of radiative EWSB and
superparticle mass spectrum are found to be considerably altered from
the minimal $SO(10)$-type unification. We will show that neutrino
coupling effects make it possible to attain PQ and R symmetric
radiative EWSB which is preferred by the bottom quark mass prediction
in the large $\tan\beta$ case.

%%%%%%%%%%%%%%%%%%%%%%%%%%%%%%%%%%%%%%%%%%%%%%%%%%%%%%%%%%%%%%%%%%%%%%
\subsection{Radiative EWSB with Large Neutrino Couplings}
%%%%%%%%%%%%%%%%%%%%%%%%%%%%%%%%%%%%%%%%%%%%%%%%%%%%%%%%%%%%%%%%%%%%%%
Let us first see how the radiative EWSB scenario is altered by
introducing neutrino couplings. The right-handed neutrinos contribute
the one-loop RG equations for Higgs SUSY-breaking mass parameters
which are given by
{\allowdisplaybreaks%
\begin{eqnarray}
  \frac{dm_{H_u}^2}{d\ln Q} &\;=\;&
  \frac{dm_{H_u}^2}{d\ln Q}\bigg|_{\rm MSSM} \!+\,
  \frac{y_\nu^2}{8\pi^2}\big(m_{H_u}^2+m_{\tilde L}^2+m_{\tilde\nu}^2
  +A_\nu^2\big), \label{mhuRGE} \\
  \frac{dm_{H_d}^2}{d\ln Q} &\;=\;&
  \frac{dm_{H_d}^2}{d\ln Q}\bigg|_{\rm MSSM}
\end{eqnarray}}%
above the decoupling scale of (the third-generation) right-handed
neutrino $M_\nu$. The mass 
parameters $m_{\tilde L}^2$, $m_{\tilde\nu}^2$, and $A_\nu$ are the
abbreviations of the third diagonal elements of left-, right-handed
scalar neutrino mass matrices, and trilinear coupling of scalar
neutrinos, respectively. The wavefunction renormalization of up-type
Higgs scalar is affected by propagating right-handed neutrinos but
that of down-type Higgs scalar is not. As a result, only the RG
equation of up-type Higgs mass (\ref{mhuRGE}) receives the additional
terms ($\propto y_\nu^2$) from the neutrino sector. These terms are
naturally positive in the RG evolution down to $M_\nu$ and 
lowers $m_{H_u}^2$ in the infrared region, compared to the MSSM
prediction. The down-type Higgs mass is not altered at one-loop order.

As seen in the previous section, the $SO(10)$-type Yukawa unification
generally leads to only a tiny difference between up- and down-type
Higgs masses at the electroweak scale and excludes a large portion of
parameter space by the experimental bound of CP-odd neutral Higgs
mass which is proportional to that mass difference. This fact reflects
the $SU(2)_R$ symmetry which is violated only by the 
small $U(1)_Y$ gauge coupling and the absence of right-handed
neutrinos. In particular, the latter decreases the down-type Higgs
mass by tau Yukawa effect in the RG evolution. In the present
scenario, including the neutrino couplings in the RG equation 
of $m_{H_u}^2$ tends to cancel the tau Yukawa effect and provides a
positive contribution to the CP-odd neutral Higgs mass
squared. Consequently the radiative EWSB is expected to be made more
natural than the minimal $SO(10)$-type unification.

The one-loop RG equations for scalar lepton masses are also affected
by the neutrino couplings. In particular the scalar tau mass
eigenvalue is decreased and the requirement of charge-neutral LSP
becomes severer than the minimal $SO(10)$-type unification. If
considered the R symmetric radiative EWSB, the scalar tau lepton could
readily be made heavier than the lightest neutralino. That will be
checked in the numerical analysis below.

In the following we simply take $y_\nu=y_{{}_G}$ at the GUT scale for
comparison to the analysis of Yukawa unification in the previous
sections. Evaluating the RG evolution of Higgs mass parameters, we
obtain the RG solutions in the EWSB vacuum:
{\allowdisplaybreaks%
\begin{eqnarray}
  M_A^2 &\;=\;&  (e_{ms}+e_{md}\xi)\tilde m_0^2 
  +e_{m\nu}m_{\tilde\nu0}^2 +e_M M_{1/2}^2 
  +e_{AM}A_0M_{1/2} +e_{A_\nu M}A_{\nu0}M_{1/2} \nonumber \\[1mm] 
  && \qquad +e_AA_0^2 +e_{A_\nu}A_{\nu0}^2 +e_{AA_\nu}A_0A_{\nu0}
  -M_Z^2,  \label{mafit2} \\[2mm]
  |\mu|^2 &\;=\;& (f_{ms}+f_{md}\xi)\tilde m_0^2 
  +f_{m\nu}m_{\tilde\nu0}^2 +f_M M_{1/2}^2 
  +f_{AM}A_0M_{1/2} +f_{A_\nu M}A_{\nu0}M_{1/2} \nonumber \\
  && \qquad +f_AA_0^2 +f_{A_\nu}A_{\nu0}^2 +f_{AA_\nu}A_0A_{\nu0}
  -\frac{M_Z^2}{2},  \label{mufit2}
\end{eqnarray}}%
where $m_{\tilde \nu 0}^2$ and $A_{\nu 0}$ are the boundary values 
of $m_{\tilde \nu}^2$ and $A_\nu$ at the GUT-breaking scale. In order
to clarify the effects of neutrino couplings, we have separated them
from the other parameters of charged fields, the latter of which are
simply assumed to be unified at the GUT scale for comparison to the
previous $SO(10)$-type unified scenario. The separation of
neutrino SUSY-breaking parameters while keeping (approximate) Yukawa
unification is dynamically corroborated, e.g.\ in the framework that
right-handed neutrinos contain low-energy remnants of gauge-singlet
superfields. When the mixture is tiny between such extra singlets and
the third-generation $1_3$ in the $16_3$ multiplet, the Yukawa
unification is almost preserved. On the other hand, SUSY-breaking
parameters of neutrinos are significantly modified if the extra
singlets receive larger breaking effects than the ordinary
matter. That could be easily realized if the extra fields directly
couple to SUSY-breaking sector and the others are not. A simpler and
alternative mechanism for separating the neutrino effect is to assume
that neutrino Yukawa couplings for the first and second generations
have ${\cal O}(1)$ values, which turn out to enhance RG running
effects of neutrino couplings. That is possible since the neutrino
mass spectrum has not been experimentally determined unlike the other
charged fermions. Anyway the neutrino couplings are regarded as free
parameters and can be large. It may be interesting to note that large
values of neutrino couplings do not cause phenomenological problems
such as the destabilization of the EWSB vacuum because of the
existence of huge supersymmetric Majorana masses of right-handed 
neutrinos.

\begin{figure}[t]
\begin{center}
\includegraphics[width=5cm,clip]{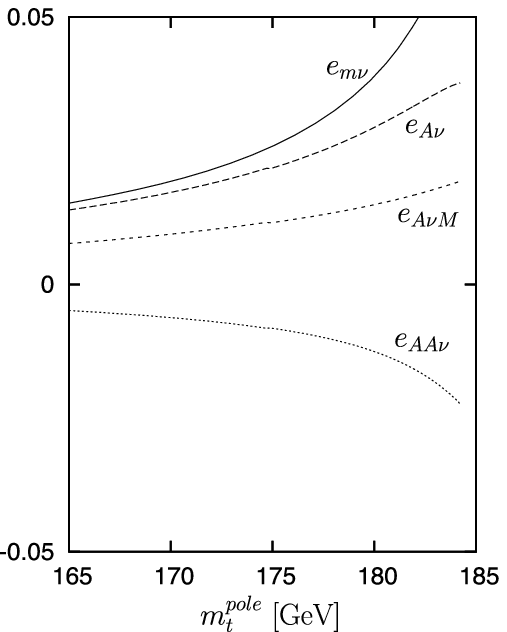}\hspace*{2cm}
\includegraphics[width=5cm,clip]{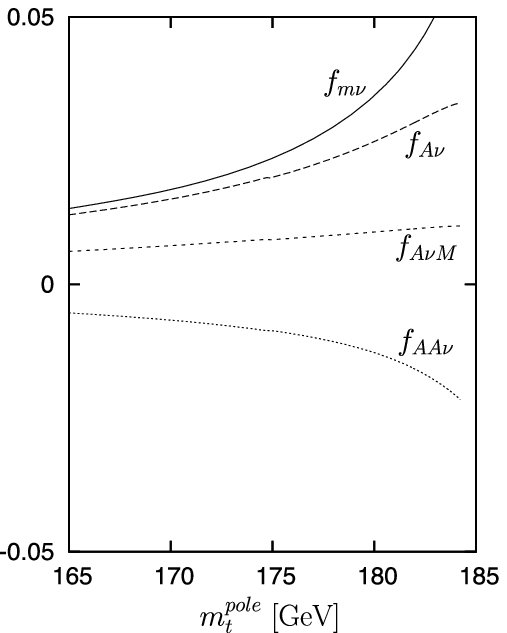}
\end{center}
\caption{Typical behaviors of the coefficients $e$'s and $f$'s in the
RG solutions (\ref{mafit2}) and (\ref{mufit2}) evaluated at the
renormalization scale $Q=1$~TeV\@. Here we take input parameters the
same as in the previous figure~\ref{fit1} in addition to the
decoupling scale of the third-generation right-handed 
neutrino $M_\nu=10^{14}$~GeV\@.\bigskip}
\label{fit2}
\end{figure}
The coefficients in the solutions (\ref{mafit2}) and (\ref{mufit2})
are roughly estimated from the RG equation (\ref{mhuRGE}) of the
up-type Higgs mass parameter. The RG running 
between $M_G$ and $M_\nu$ generates a 
departure $\delta m_{H_u}^2$ from the MSSM prediction of up-type Higgs
mass. At one-loop order, the neutrino coupling effect is given by
\begin{equation}
  \delta m_{H_u}^2 \;=\; 
  -(2\tilde m_0^2+m_{\tilde\nu0}^2 +A_{\nu0}^2)\,\epsilon 
  \,+{\cal O}(\epsilon^2),
\end{equation}
where the positive parameter $\epsilon$ is
\begin{equation}
  \epsilon \;=\; \frac{y_\nu^2}{8\pi^2}
  \ln\bigg(\frac{M_G}{M_\nu}\bigg) \;\simeq\; 0.06\,y_\nu^2. 
\end{equation}
We find from this expression that the coefficients 
satisfy $e_{ms}\simeq c_{ms}+2\epsilon$ and $e_{m\nu}
\simeq e_{A_\nu}\simeq\epsilon$. The 
others, $e_{md}$, $e_M$, $e_A$ and $e_{AM}$, are almost the same as
the corresponding coefficients $c$'s in the MSSM case, and also the
neutrino part $e_{A_\nu M}$ and $e_{AA_\nu}$ are the next-to-leading
order of $\epsilon$. The similar behaviors are expected to hold for
the solution (\ref{mufit2}) since the leading-order neutrino RG effect
comes from the modification of $m_{H_u}^2$ in low-energy regime. We
checked these behaviors by numerically solving the RG
equations. Fig.~\ref{fit2} represents typical values 
of $e$'s and $f$'s as the functions of the top quark mass, especially
concerning the neutrino 
parameters $m_{\tilde\nu0}^2$ and $A_{\nu0}$. In the calculation, the
input parameters are taken as the same as in Fig.~\ref{fit1}, and
the EWSB conditions are solved at the renormalization 
scale $Q=1$~TeV\@. It is found from the figures that the above
analytic estimation is consistent with the exact numerical one. Thus
the neutrino coupling effects are characterized by $\epsilon$.

While $\epsilon$ is not so large, it provides significant effect on the
radiative EWSB\@. An important point here is that, in addition to the
usual gaugino mass effect, there are new sources of positive
contribution to the CP-odd neutral Higgs mass squared $M_A^2$, that
is, the neutrino 
effects $e_{m\nu}m_{\tilde\nu0}^2+e_{A_\nu}A_{\nu0}^2$ in
(\ref{mafit2}). These terms can raise $M_A$ without leading to a large
gaugino mass, and the experimental bound on $M_A$ is made consistent
with the R symmetric low-energy superparticle spectrum which allows
the prediction of bottom quark mass well within the experimental
range. In the limit $M_{1/2},A_0\to0$, the RG
solution (\ref{mafit2}) becomes
\begin{equation}
  M_A^2 \;=\; \big[\,e_{ms}+e_{md}\xi
  +(e_{m\nu}N^2+e_{A\nu}N_A^2)(1-\xi)\,\big]\,\tilde m_0^2 -M_Z^2,
\end{equation}
where $N$ and $N_A$ represent the effects of neutrino couplings
compared to other matter SUSY-breaking parameters; $N^2=
m_{\tilde\nu0}^2/m_{16}^2$ and $N_A^2=A_{\nu0}^2/m_{16}^2$. If one
neglects $M_Z$ and $\xi$, the positive $M_A^2$ implies
\begin{equation}
  N^2+N_A^2 \;\gtrsim\; \frac{-e_{ms}}{\epsilon} \;\sim\; 4.
  \label{NNA}
\end{equation}
Therefore only a few times larger values of neutrino couplings are
needed to obtain phenomenologically preferred mass spectrum. A
non-vanishing $\xi$ has little dependence on this lower bound as will
be shown in the following numerical analysis.

The PQ symmetric spectrum is also available if one considers the above
R symmetric radiative EWSB and a suitable value of $\xi$ parameter. In
the solution for $|\mu|^2$~(\ref{mufit2}), the positive contribution
from gaugino masses can be cancelled by the $\xi$ term
($f_{md}$). Note that there also exists the neutrino coupling effect
in the evaluation of $\mu$ parameter which raises $|\mu|$ compared
with the minimal $SO(10)$-type unification. However this effect is not
important since the characteristic size of neutrino coupling 
effect ($\epsilon$) is much smaller than the dominant contributions
from scalar masses ($f_{md}$) and gaugino masses ($f_M$).

The neutrino coupling effects in the RG evolution of mass parameters
lead to low-energy superparticle spectrum quite different from that in
the minimal $SO(10)$-type unification. In particular, the PQ and R
symmetries are found to appear in the mass spectrum by introducing
natural, sizable effects of neutrino couplings. Such a spectrum is
known to be favorable to low-energy phenomenology in the 
large $\tan\beta$ case. In the next subsection, we will examine the
parameter space consistent with the PQ and R symmetric radiative EWSB,
the experimental mass bounds of superparticles, and the requirement of
neutral LSP and also discuss the predictions of bottom quark mass and
the $b\to s\gamma$ rare decay.

%%%%%%%%%%%%%%%%%%%%%%%%%%%%%%%%%%%%%%%%%%%%%%%%%%%%%%%%%%%%%%%%%%%%%%
\subsection{Parameter Space Analysis}
%%%%%%%%%%%%%%%%%%%%%%%%%%%%%%%%%%%%%%%%%%%%%%%%%%%%%%%%%%%%%%%%%%%%%%
Let us perform the parameter space analysis of 
the $SO(10)$ unification with neutrino couplings. We particularly
focus on the realization of the PQ and/or R symmetric radiative EWSB
and its low-energy phenomenology. This type of radiative EWSB is
triggered by moderate values of mass and/or trilinear coupling of the
third-generation right-handed neutrino [see Eq.~(\ref{NNA})]. In
the RG evolution, scalar masses and trilinear couplings generally have
similar effects since they appear together in beta functions. A
difference might be generated in the evolution of neutrino trilinear
coupling between $M_G$ and $M_\nu$, which results in, e.g.\ the
difference between $e_{m\nu}$ and $e_{A_\nu}$ in the 
solution (\ref{mafit2}). Such differences are however generally small,
and the two cases with a large $N$ and with a large $N_A$ lead to
almost the same low-energy superparticle spectrum. In what follows we
will show, as an illustration, the parameter space analysis for the
case of varying scalar neutrino mass ($N$) with the vanishing neutrino
trilinear couplings ($N_A=0$).

\begin{figure}[t]
\begin{center}
\includegraphics[width=5cm,clip]{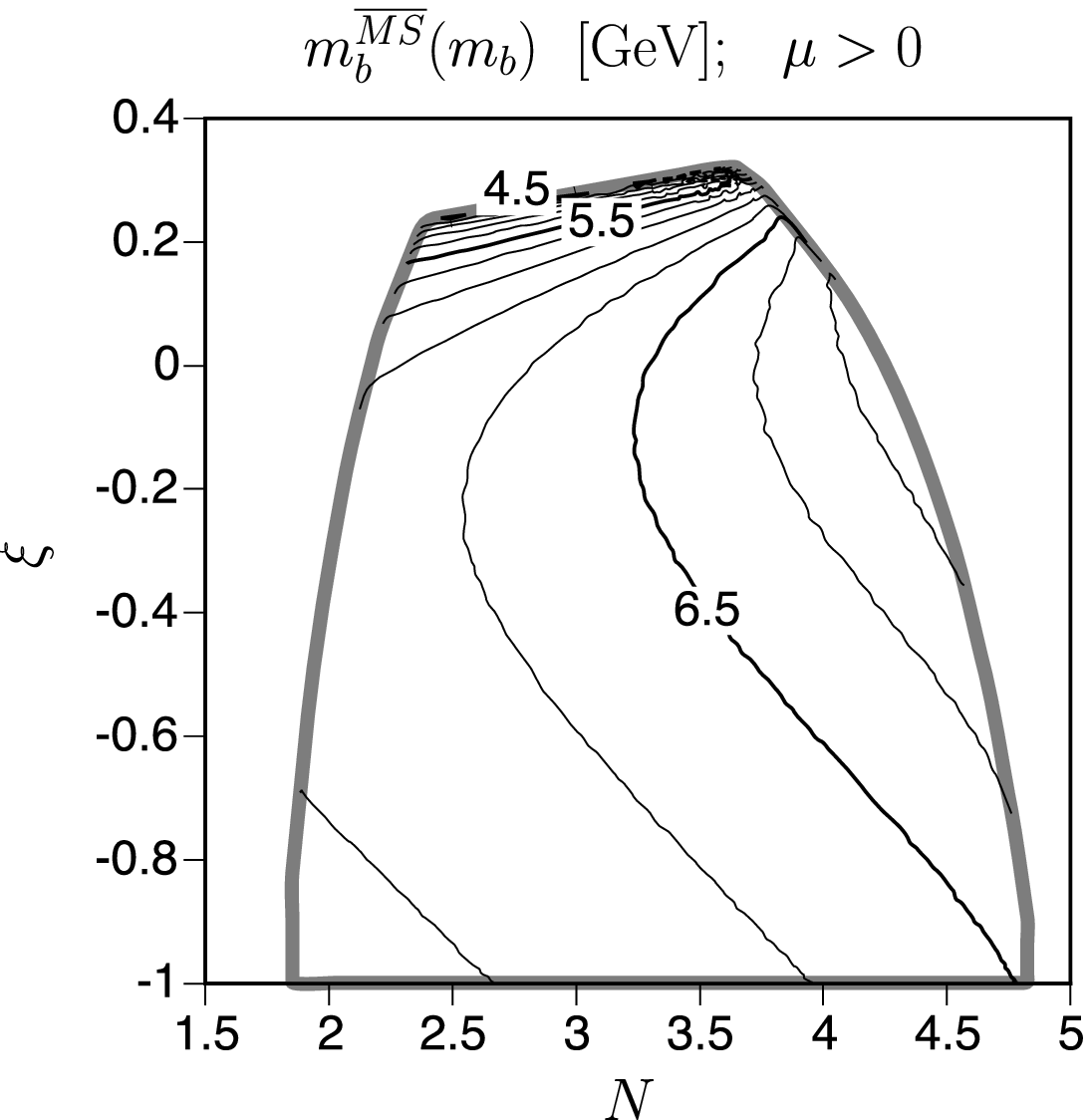}\hspace*{1cm}
\includegraphics[width=5cm,clip]{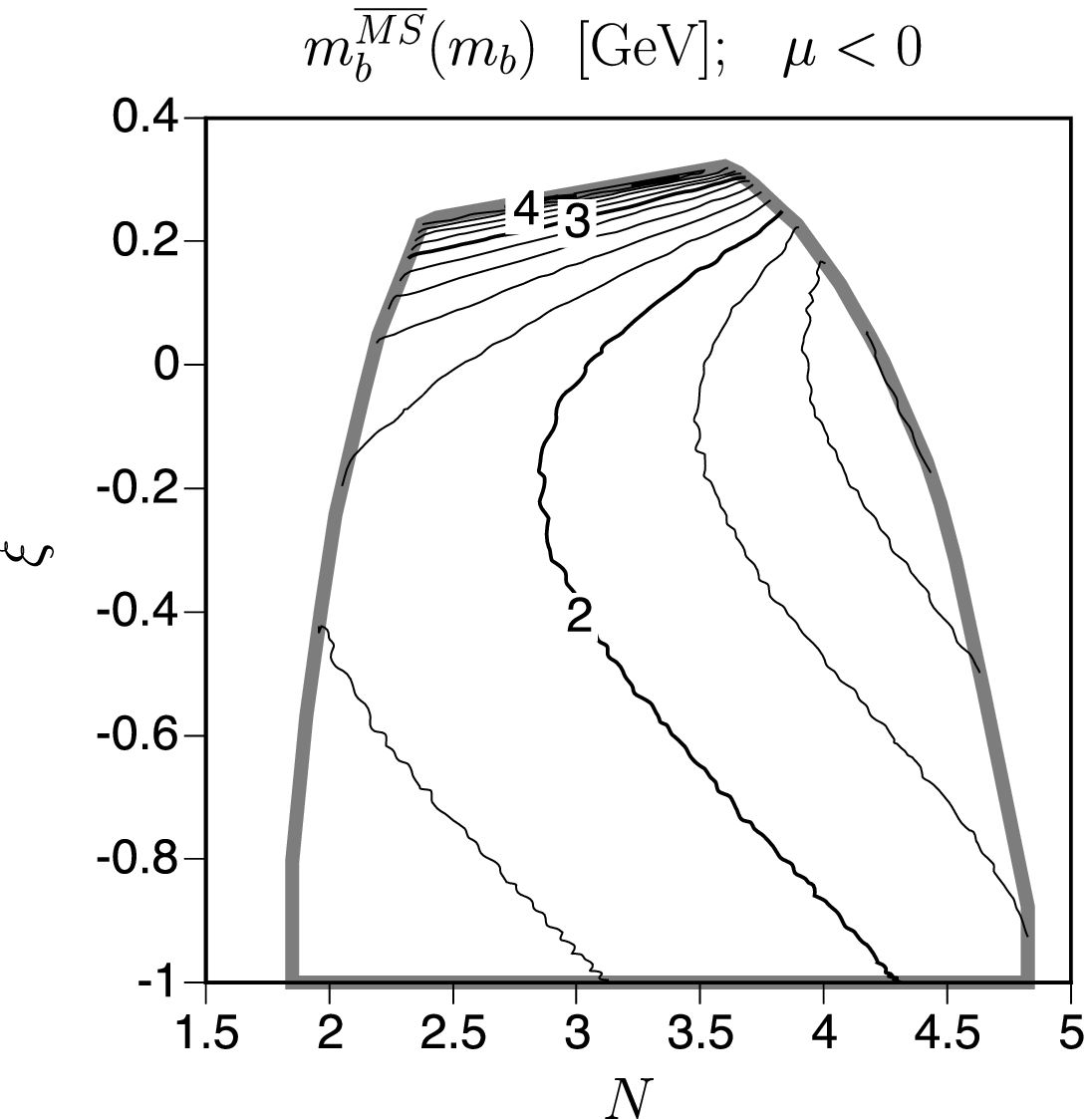}\\[5mm]
\includegraphics[width=5cm,clip]{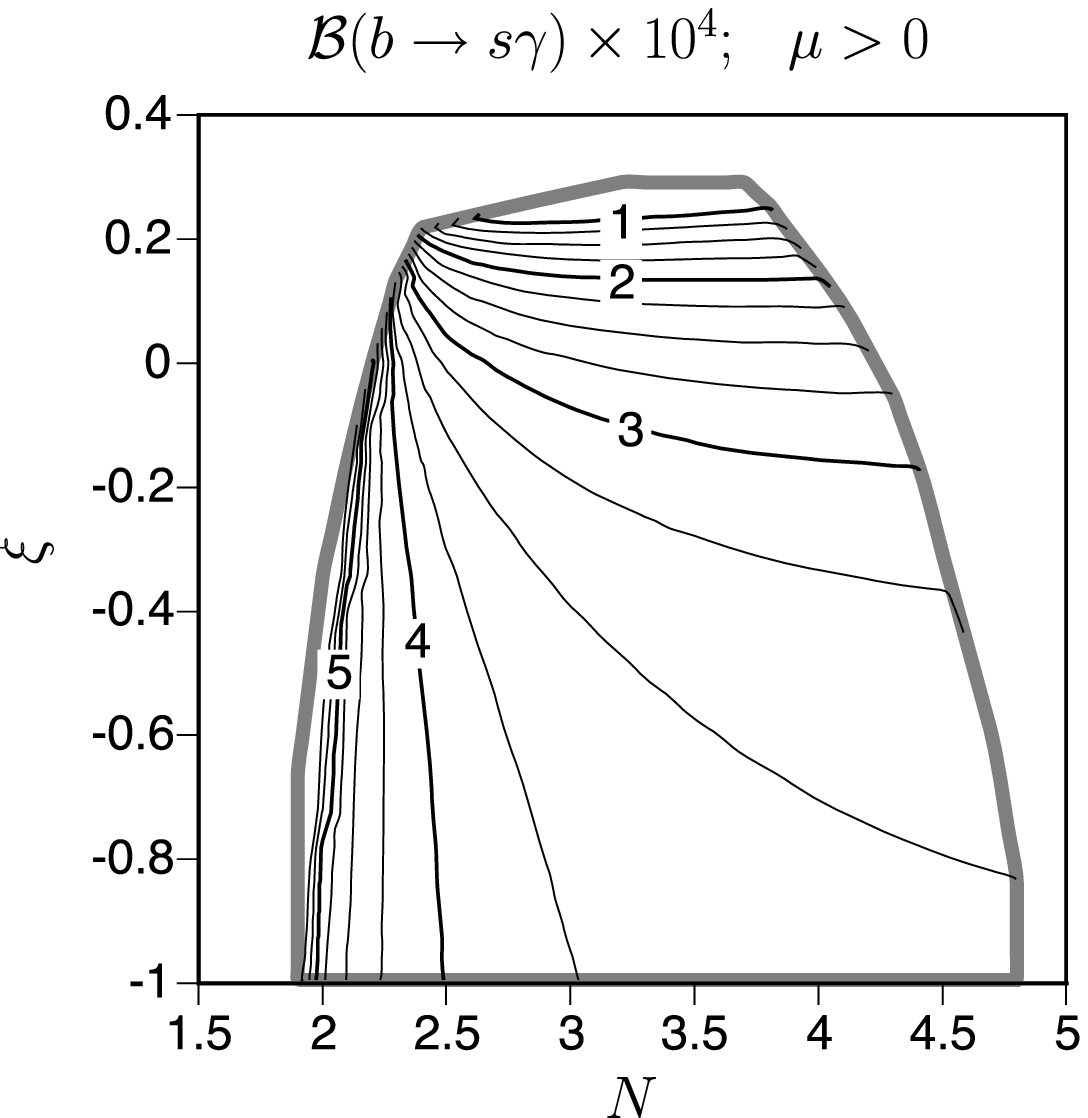}\hspace*{1cm}
\includegraphics[width=5cm,clip]{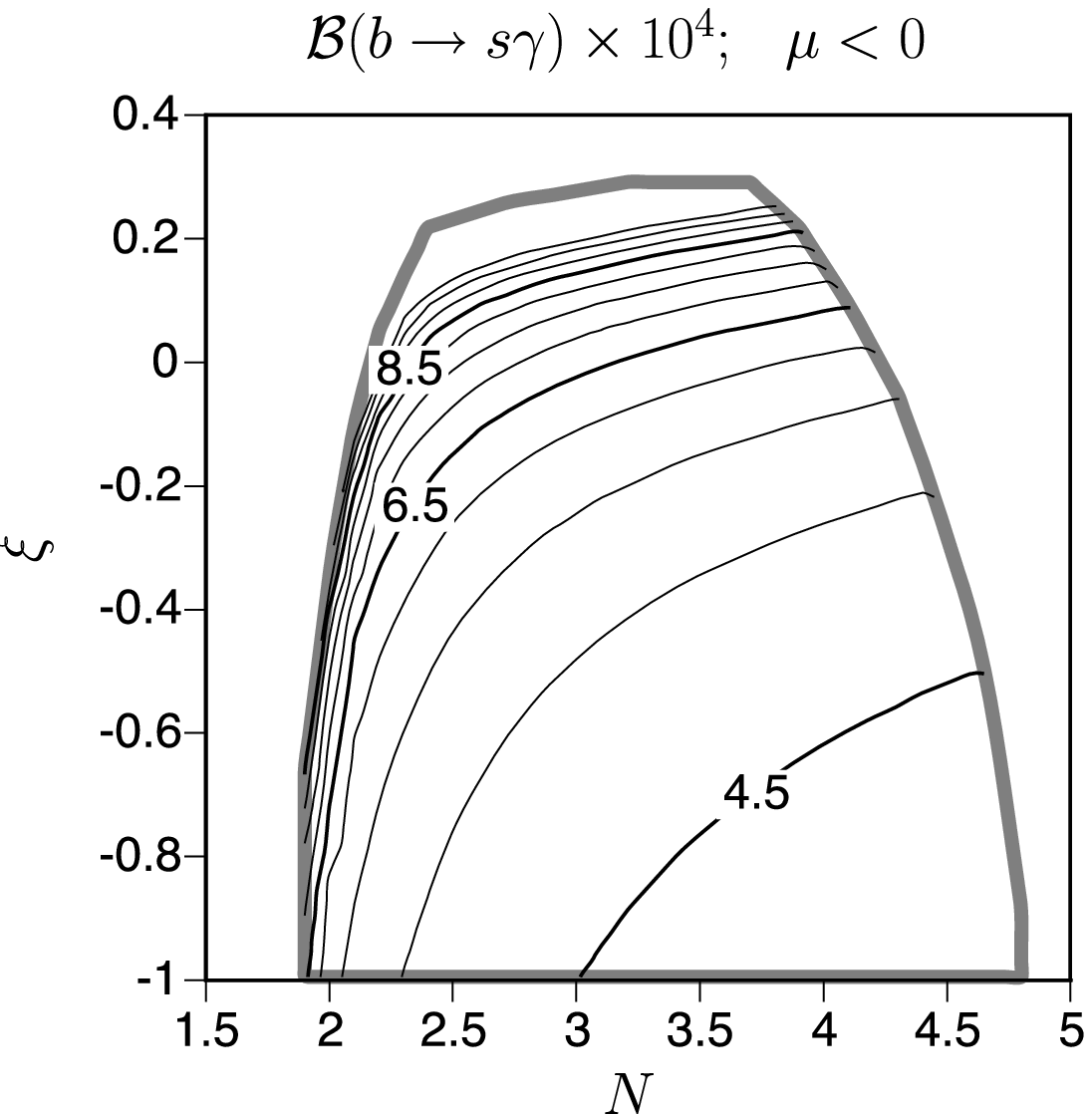}
\end{center}
\caption{The parameter space consistent with the radiative EWSB, the
experimental mass bounds of superparticles, and the requirement of
neutral LSP in the $SO(10)$ unification with neutrino couplings. The
horizontal and vertical axes denote the SUSY-breaking mass of
third-generation scalar neutrino and the matter/Higgs discrepancy
parameter, respectively. The bottom quark 
mass $m_b^{\overline{\text{MS}}}(m_b)$ and 
the $b\to s\gamma$ branching ratio are also shown in the figures. The
GUT-scale mass parameters are set 
as $M_{1/2}=300$~GeV, $\tilde m_0=2$~TeV, and $A_0=0$. The other input
parameters are taken as the same as in the previous
figure~\ref{Mgm0}. The radiative EWSB conditions, the superparticle
mass bounds, the SUSY threshold corrections and 
the $b\to s\gamma$ branching ratio are calculated in the same way as
Fig.~\ref{Mgm0}. In each figure, the left, top, and right
regions are excluded, respectively, by the mass bound of CP-odd
neutral Higgs boson, the lower bounds of chargino and neutralino
masses, and the requirement that the scalar tau lepton should not be
the LSP.\bigskip}
\label{Nxi}
\end{figure}

The two important model parameters for successful radiative EWSB are
the scalar neutrino mass denoted by $N$ and the matter/Higgs mass
discrepancy parameter $\xi$ defined 
in (\ref{xi}). Fig.~\ref{Nxi} shows the parameter regions 
of ($N,\,\xi$) consistent with the radiative EWSB conditions, the
experimental bounds on superparticle masses, and the requirement of
neutral LSP\@. In the figures, the predictions of bottom quark 
mass $m_b^{\overline{\text{MS}}}(m_b)$ and 
the $b\to s\gamma$ branching ratio are shown in the allowed parameter
regions. The boundary values of SUSY-breaking variables at the GUT
scale are $M_{1/2}=300$~GeV, $\tilde m_0=2$~TeV, and $A_0=0$. The
other input parameters are taken as the same as in the previous
figure~\ref{Mgm0}.
In Fig.~\ref{Nxi}, the left side of the parameter space is
excluded by the experimental mass bound of CP-odd neutral Higgs
boson. This suggests that a large SUSY-breaking mass of right-handed
scalar neutrino raises the Higgs mass. It is also found that the
analytic estimation~(\ref{NNA}) is well satisfied and 
its $\xi$ dependence is small. On the other hand, the $\xi$ dependence
of $\mu$ parameter is relatively large: a positive, larger value 
of $\xi$ rapidly lowers the prediction 
of $|\mu|$ for $M_{1/2}\ll\tilde m_0$. That rules out the top region
in each figure by the lower mass bounds of charginos and
neutralinos. In the parameter space of Fig.~\ref{Nxi}, we
obtain the CP-odd neutral Higgs mass $M_A\lesssim3M_{1/2}$ and 
the $\mu$ parameter as small as 50~GeV\@. Too large values 
of $\xi$ and right-handed scalar neutrino mass reduce the mass
eigenvalues of scalar lepton doublets through the RG evolution down to
low-energy regime. This is encoded to the right-top region excluded by
the cosmological requirement that charged field (right-handed scalar
tau) should not be the LSP\@.

One of the notable features of the present scenario is that there
exists the parameter region (the top region in the figures) excluded
by the mass bounds of charginos and neutralinos. In other words, a
relatively small value of $\mu$ parameter is consistent with the
radiative EWSB\@. That is achieved with the $\xi$-dependent negative
contribution to $|\mu|^2$ in the solution~(\ref{mufit2}). In this
parameter region, the gaugino masses and holomorphic 
couplings $A$, $B$ and $\mu$ can be much smaller than non-holomorphic
scalar masses, and thus an approximate PQ and R symmetric spectrum is
realized. Such a hierarchical mass pattern was incompatible with the
minimal $SO(10)$-type unification in the previous section. These
symmetries tend to suppress the low-energy threshold correction to the
bottom quark mass. We find that, for both signs of 
the $\mu$ parameter, successful predictions of bottom quark mass are
obtained on the top margin of the allowed parameter region in 
Fig.~\ref{Nxi} where $\mu$ takes a relatively small value, and
the lighter chargino is gaugino-like and becomes lighter than 
about $120$~GeV\@.

The $b\to s\gamma$ branching ratio is decreased as the right-handed
scalar neutrino mass $N$, since the charged Higgs mediated 
amplitude $A_{H^+}$ is suppressed by the CP-odd neutral Higgs mass. The
branching ratio also has the $\xi$ dependence since a negative large
value of $\xi$ increases scalar quark masses. In the PQ and R
symmetric region, where a strong suppression of the threshold
correction $\Delta_b$ is obtained, the $b\to s\gamma$ decay constraint
apparently seems severe even for the positive $\mu$ case. This is
because the chargino contribution $A_{\tilde\chi^+}$ is enhanced to be
larger than $A_{H^+}$ by small mass eigenvalues of charginos. As a
result, large scalar masses are favored for both signs of $\mu$ in
order to avoid the experimental constraint from $b\to s\gamma$ rare
decay.

\begin{figure}[t]
\begin{center}
\includegraphics[width=5cm,clip]{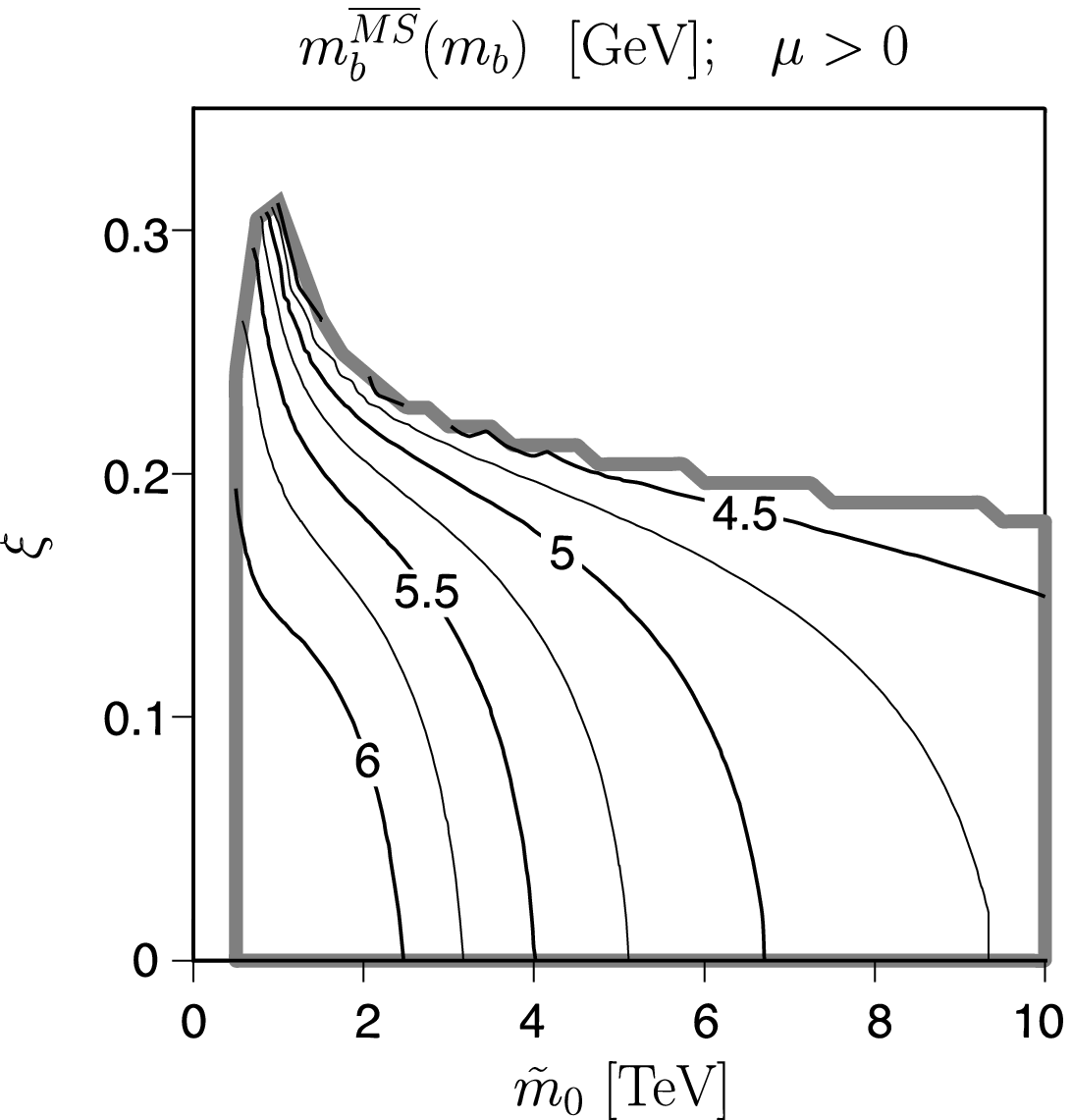}\hspace*{1cm}
\includegraphics[width=5cm,clip]{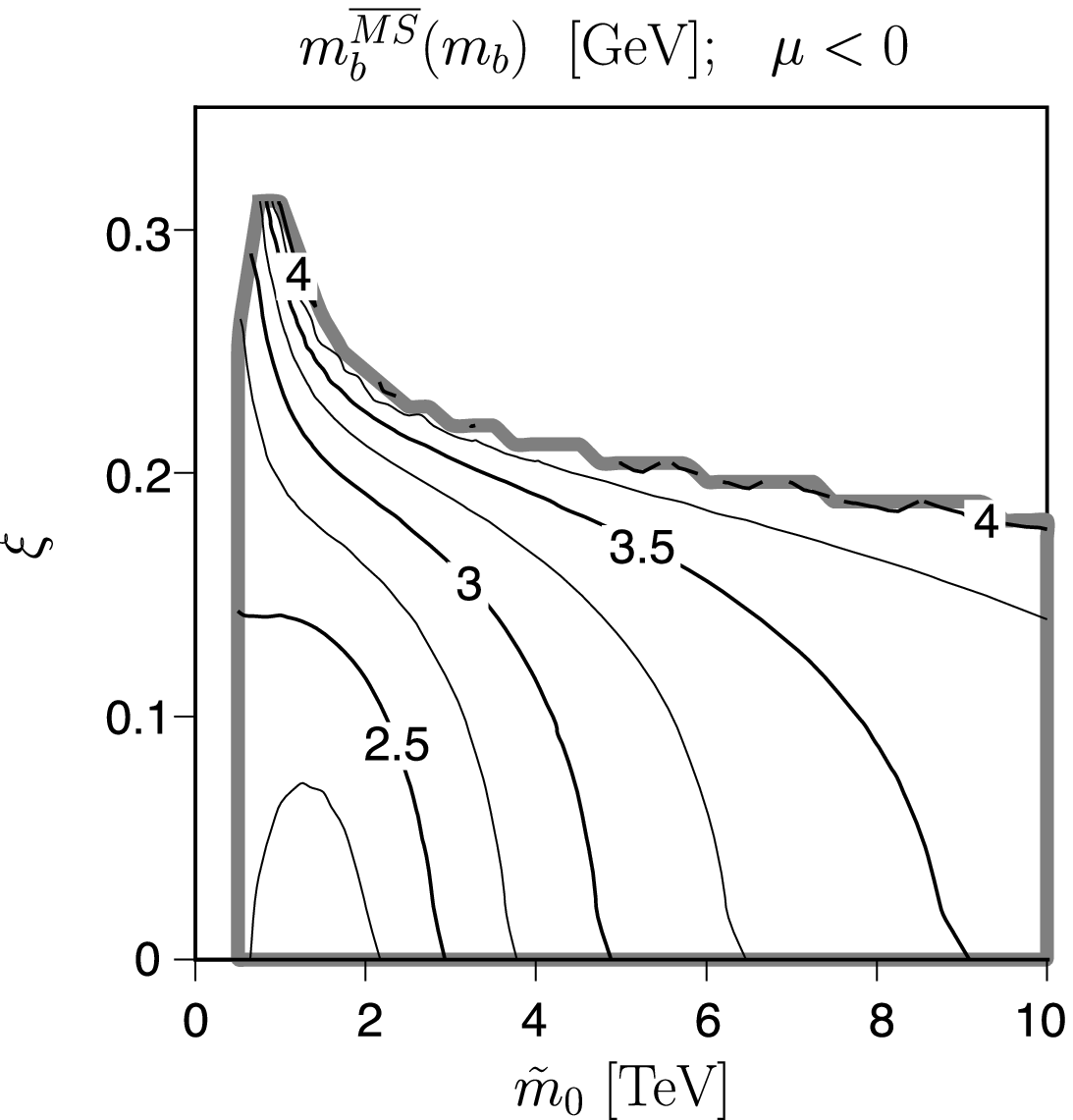}\\[5mm]
\includegraphics[width=5cm,clip]{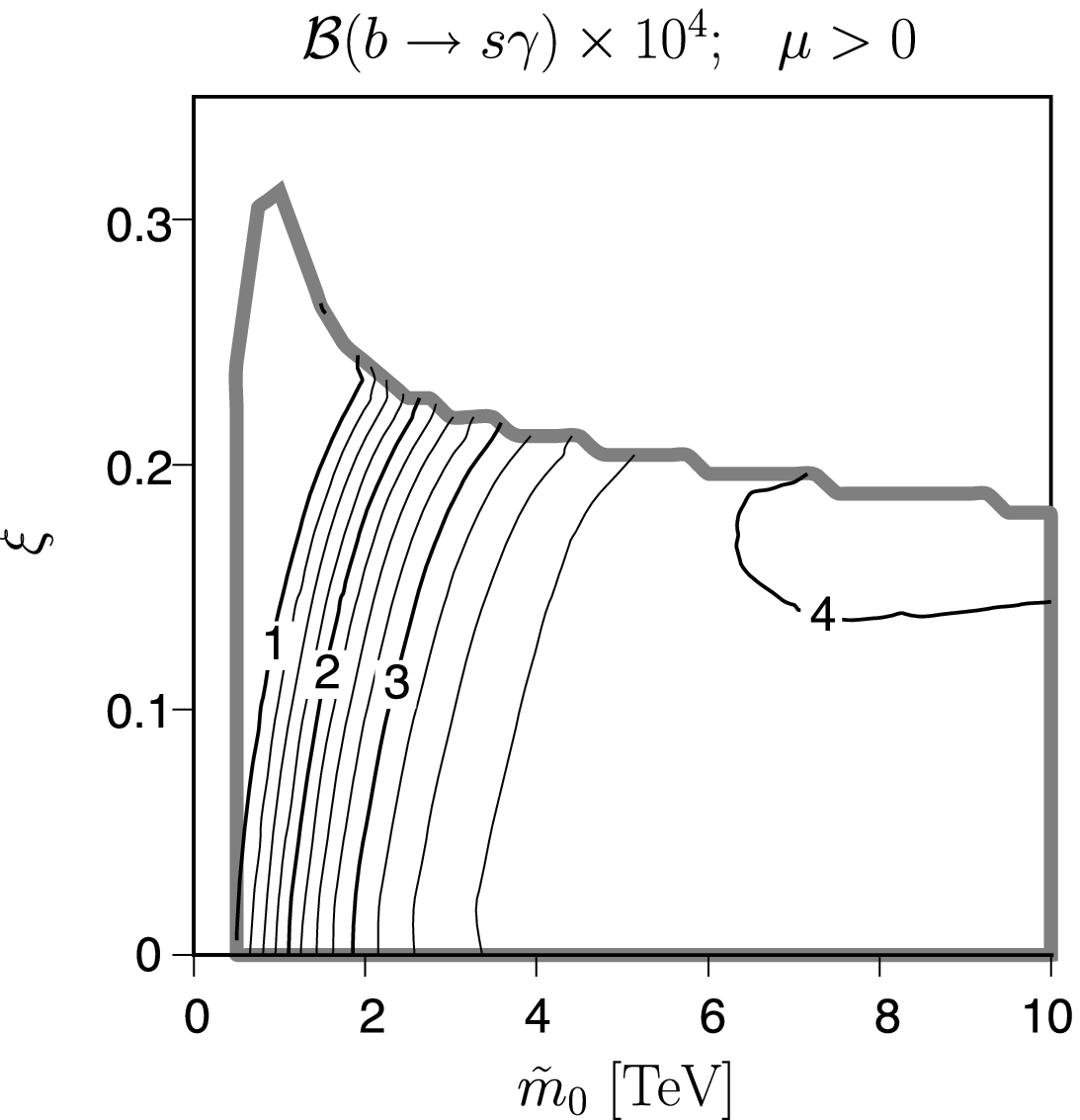}\hspace*{1cm}
\includegraphics[width=5cm,clip]{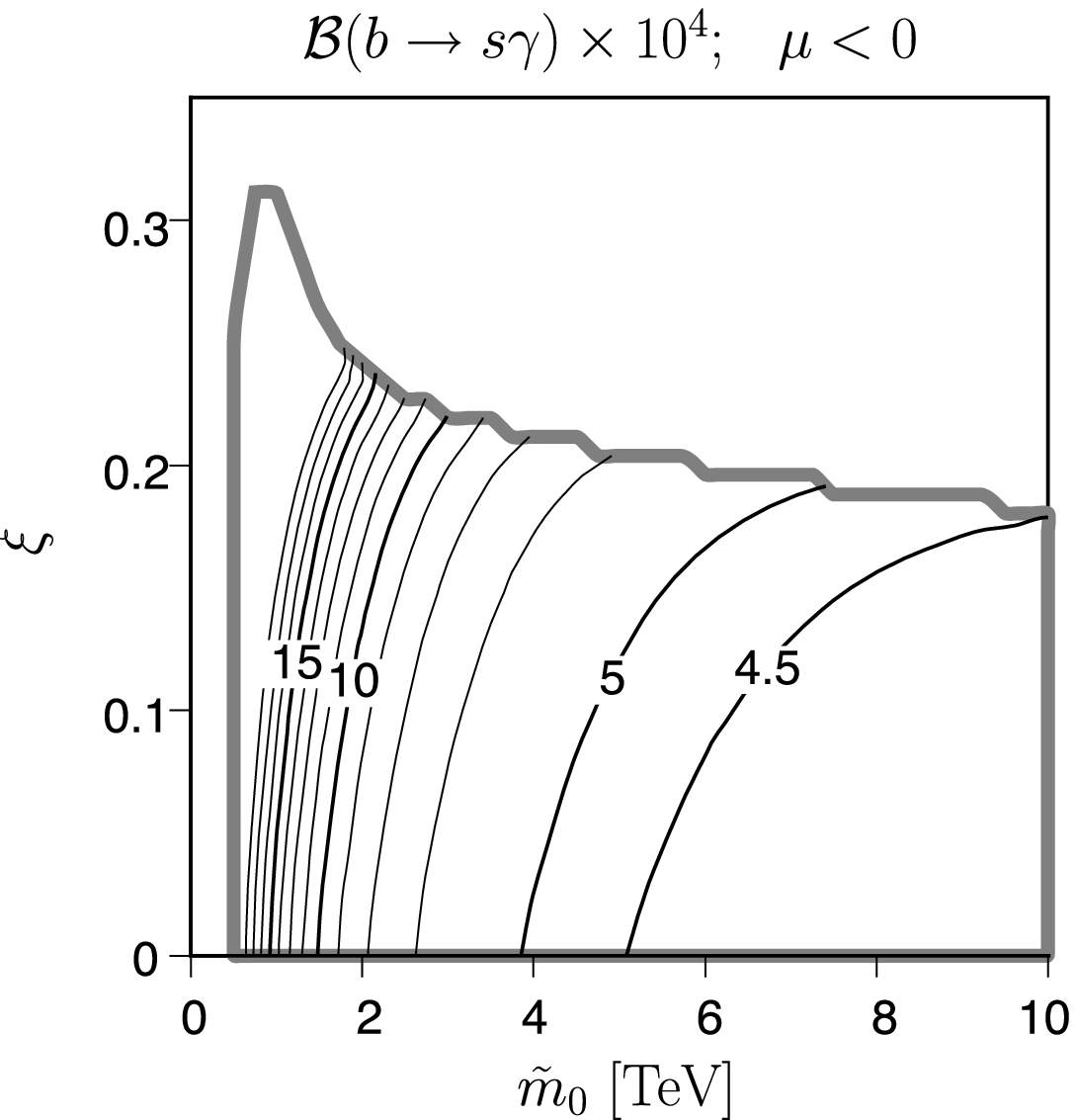}
\end{center}
\caption{The same as Fig.~\ref{Nxi} but 
for $\tilde m_0$--$\xi$ parameter space. The SUSY-breaking parameters
are set as $M_{1/2}=300$~GeV, $A_0=0$, and $N=2.5$ at the GUT
scale. In each figure, the narrow left region is excluded by the LSP
scalar tau lepton and the upper side is ruled out the experimental
mass bounds on CP-odd neutral Higgs boson, charginos, and
neutralinos.\bigskip}
\label{m0xi}
\end{figure}
The above discussion implies that the PQ and R symmetric superparticle
spectrum leads to phenomenologically preferred values of the bottom
quark mass and the $b\to s\gamma$ branching ratio if the matter
scalars become heavy. That can be seen in Fig.~\ref{m0xi} which
is the same as the previous figure~\ref{Nxi} but for 
the $\tilde m_0$--$\xi$ parameter space. In general, the left side
region is excluded by the LSP scalar tau lepton and the top one by the
experimental mass bounds of CP-odd neutral Higgs boson and chargino.

For the $\mu>0$ case, there is the parameter region consistent with
the experimental ranges of bottom quark mass and $b\to s\gamma$ decay
rate. In the allowed parameter region, the universal scalar mass is
found to become $\tilde m_0\gtrsim2.5$~TeV\@. The universal gaugino
mass is much smaller than scalar masses 
as $M_{1/2}\lesssim0.12\tilde m_0$, and $\mu$ is also suppressed 
as $\mu\lesssim0.15\tilde m_0$ at the electroweak scale. Thus the
approximate PQ and R symmetries are easily realized in the spectrum,
which suppresses SUSY threshold corrections and reproduces the
observed value of bottom quark mass. For $\tilde m_0\lesssim 4$~TeV,
relatively a small $\mu$ parameter is required to obtain an enough
suppression of $\Delta_b$, and therefore lighter chargino and
neutralinos contain non-negligible higgsino components. For heavier 
scalars ($\tilde m_0\gg$ a few TeV), the lighter chargino and
neutralinos can be either gaugino-like or higgsino-like. The
prediction of CP-odd neutral Higgs mass is correlated with the initial
scalar mass $\tilde m_0$ as $M_A\sim0.1\tilde m_0$. If one considers
the case that $\tilde m_0$ is less than a few~TeV, the charged Higgs 
loop $A_{H^+}$ gives a sizable contribution to the 
total $b\to s\gamma$ decay width. Also the chargino 
loop $A_{\tilde\chi^+}$ gives a non-negligible contribution. We find
that, in the parameter region consistent with the experimental 
bounds, $A_{H^+}$ and $A_{\tilde\chi^+}$ are of the same order and
have opposite signs. For $\tilde m_0\gtrsim2$~TeV, the cancellation
of partial amplitudes is enough to satisfy the experimental constraint
from $b\to s\gamma$ decay.

As for the negative $\mu$ case, the above cancellation of diagrams
cannot be obtained, though PQ and R symmetric spectrum is still
available and the $b\to s\gamma$ branching ratio is suppressed by
assuming a relatively large scalar mass $\tilde m_0$. In the region
that the bottom quark mass prediction is within the experimental
range, the PQ and R symmetries imply light charginos which lead to an
enhanced contribution to the $b\to s\gamma$ branching ratio. For 
example, for $m_t^{\rm pole}=178$~GeV as in Fig.~\ref{m0xi},
the threshold correction must be more suppressed than 
the $\mu>0$ case, which requires a tiny value of $\mu$ parameter. That
makes it difficult to predict the observed value of bottom quark mass
without introducing very large scalar masses. When the top quark mass
is taken to be a smaller value, the bottom quark mass is increased and
then the required value of $|\mu|$ becomes larger. Furthermore a
smaller value of the top quark mass increases scalar quark masses for
a fixed value of $\tilde m_0$ and hence the $b\to s\gamma$ constraint
is relaxed. We find, for example, 
that $m_t^{\rm pole}=172.7$~GeV~\cite{Tevatron} is consistent 
with $N=3$ and $\tilde m_0\gtrsim6$~TeV.

Finally we comment on the lepton flavor violating decay of charged
leptons. It is known in the Yukawa unification 
scenario~\cite{LFV} that a large value of $\tan\beta$ enhances the
decay amplitudes of charged leptons such 
as $\mu\to e\gamma$ and $\tau\to\mu\gamma$. However in the present
analysis we do not specify small elements of lepton Yukawa couplings
which control the generation mixing of charged leptons. Furthermore
the generation mixing from Yukawa couplings is affected by the
structure of right-handed neutrino Majorana mass matrix, the detail of
which is also irrelevant to the present analysis of EWSB\@. For these
reasons the prediction of flavor-violating rare decay of charged
leptons is not under control and could easily be consistent with the
current experimental bounds.

To summarize, we have found that the right-handed neutrino couplings
induce new types of radiative EWSB scenarios. The low-energy
superparticle mass spectrum is significantly modified by sizable
contributions of neutrino couplings in the RG evolution down to the
decoupling scale of right-handed neutrinos. In particular, the PQ and
R symmetric spectrum is available to achieve the observed values of
bottom quark mass and $b\to s\gamma$ branching ratio with heavy
scalars of a few~TeV\@. The $\mu$ parameter can also take a small
value as $|\mu|\lesssim0.15\tilde m_0$ and lighter chargino and
neutralinos contain a sizable amount of higgsino component, which
may be cosmologically favorable in that the LSP provides dark matter
component of the present universe~\cite{LSP_DM}. If the top quark
mass is taken to be smaller, the phenomenological requirements are
more easily satisfied.

%%%%%%%%%%%%%%%%%%%%%%%%%%%%%%%%%%%%%%%%%%%%%%%%%%%%%%%%%%%%%%%%%%%%%%
\bigskip
\section{$\boldsymbol{SO(10)}$ Unification with Large Lepton Mixing}
\label{sec:lop}
%%%%%%%%%%%%%%%%%%%%%%%%%%%%%%%%%%%%%%%%%%%%%%%%%%%%%%%%%%%%%%%%%%%%%%
The recent experimental results of solar and atmospheric neutrinos
have revealed that there exist large flavor mixings in the lepton
sector, while the corresponding mixing angles in the quark sector are
observed to be small. In typical GUT scenarios, quarks and leptons are
unified into a large multiplet and consequently their Yukawa couplings
satisfy some simplifying relations. Thus the observed difference
between the flavor structures of quarks and leptons is confusing but
exciting issue in particle physics. For example, in the 
minimal $SU(5)$ GUT scenario where down-type quarks and lepton
doublets belong to the same multiplets $5^*$, their Yukawa couplings
are related as $Y_d=Y_e^T$ at the GUT-breaking scale. If $Y_e$ has
large off-diagonal elements which are suitable for suggested large
generation mixing, it is naturally expected that the quark mixing
matrix also contains large angles which are not compatible with the
observation.

One of the attractive approaches to this problem is to consider the
following generation asymmetric form of Yukawa couplings:
\begin{equation}
  Y_d \;\simeq\; Y_e^{\rm T} \;\propto\;
  \left(\begin{array}{ccc}
   ~~~ & & \\
   & & \\
   & a' & a
  \end{array}\right),
\end{equation}
where $a$, $a'$ are of a similar order and the other blank entries are
small compared to $a$ and $a'$. The similarity 
between $Y_d$ and $Y_e^{\rm T}$ is a consequence of GUT gauge symmetry
such as $SU(5)$ or larger unified group. This asymmetric form of
Yukawa couplings is referred to as the lopsided form in the
literature~\cite{ABB,SY,Ramond,NY,BK,others}.\footnote{A systematic
analysis has recently been performed in~\cite{asym} for asymmetric
forms of quark and lepton mass matrices taking account of generation
mixing and neutrino physics.} \
A key ingredient of lopsided mass matrices is that the observed large
leptonic 2-3 generation mixing is explained by dominant two elements
in the charged-lepton Yukawa matrix with $a\simeq a'$, while
preserving small quark generation mixing because only right-handed
down-type quarks are largely mixed and it does not contribute to the
physical quark mixing. Various types of GUT scenarios with lopsided
mass matrices have been studied. In dynamical models based on 
the $SO(10)$ group, some non-minimal field contents are involved to
realize the lopsided form of Yukawa couplings. For example, the MSSM
matter and Higgs fields do not have ordinary high-energy origins such
as $16_i$ and $10_H$ adopted in the previous sections. This
possibility has also been used to construct realistic models with
larger unified symmetry than $SO(10)$.

In this section, we study phenomenological issues such as the
radiative EWSB in the $SO(10)$ unification which accommodates lopsided
mass matrices for neutrino physics. We typically consider the
following form of Yukawa couplings at the GUT scale:
\begin{equation}
  Y_u \,=\, Y_\nu \,=\,
  \left(\begin{array}{ccc}
    ~~~ & & \\
    & ~~ & \\
    & & y_{{}_G}
  \end{array}\right), \qquad
  Y_d \,=\, Y_e^{\rm T} \,=\,
  \left(\begin{array}{ccc}
    ~~~~ & & \\
    & & \\
    & y'\, & y_{{}_G}\cos\theta\!
  \end{array}\right),
  \label{lopY}
\end{equation}
where blank entries in each matrix are small compared to the filled
entries. The near-maximal atmospheric neutrino mixing is explained by
assuming that $y'$ is of similar order to $y_{{}_G}\cos\theta$. In
the following analysis, we simply 
take $y'=y_{{}_G}\cos\theta$, leading to the maximal 2-3 mixing angle
from the charged-lepton sector that is the central value of the
current experimental data~\cite{neuexp}.

It may be instructive here to illustrate a dynamical explanation of
the existence of angle $\theta$ in the Yukawa matrix (\ref{lopY}). A
crucial observation for lopsided matrix form is the multiplicity 
of $5^*$ components in the theory. That is, while right-handed down
quarks and lepton doublets are combined into $5^*$ representations 
of $SU(5)$, there is no way to identify to which multiplet 
this $5^*$ should be embedded in larger symmetry than $SU(5)$. In fact
there are several sources 
of $5^*$ in $SO(10)$ theory; $10$, $16$, $120$ representations,
etc. The simplest case for realizing lopsided generations is to
suppose the second and third generation $5^*$'s have different origins
in more fundamental theory like $SO(10)$. For example, in an 
explicit $SO(10)$ model~\cite{NY}, the second generation fields
in $5^*$ come unusually from a decuplet $10$ of $SO(10)$. In this
case, low-energy down-type Higgs field should be a mixed state 
of $10$ and $16$ Higgs multiplets, otherwise several fermions become
massless. The angle $\theta$ parametrizes the degree of such Higgs
mixing; the down-type Higgs $5^*_{H_d}$ is composed 
as $5^*_{H_d}=5^*(10_H)\cos\theta+5^*(16_H)\sin\theta$ where
$5^*(10_H)$ and $5^*(16_H)$ are the anti quintuplets contained 
in $10_H$ and $16_H$ Higgs fields, respectively. The angle $\theta$ is
dynamically controlled in terms of mass parameters in GUT Higgs
potential. An $SO(10)$ invariant superpotential 
term $f16_316_H10$ gives a lopsided matrix with $y'=f\sin\theta$. It
is interesting to notice that such $5^*$ flipping just corresponds to
the $SU(2)_R$ rotation in $SO(10)$ or higher theory. In the 
minimal $SO(10)$-type unification, the $SU(2)_R$ invariance in the
Higgs sector is only weakly violated by small couplings, which makes
the radiative EWSB difficult to be achieved. However in the present
case, $SU(2)_R$ is strongly broken by the mixing of two types 
of $5^*$'s, and the radiative EWSB is expected to be easily made
successful. A similar analysis may also be performed for GUT models
based on higher gauge groups 
than $SO(10)$, e.g.\ the $E_6$ unification. An important difference
appears in that, in the $E_6$ unification, the third-generation fields
in $5^*$ have high-energy embedding into a $10$-plet 
of $SO(10)$ group. That implies (i) the matching conditions of the
second and third-generation anti-quintuplets to low-energy multiplets
are altered (i.e.\ exchanged from those in the $SO(10)$ case), 
(ii) $\tan\beta\sim\frac{m_t}{m_b}\sin\theta\sim{\cal O}(1)$, and 
(iii) an additional $D$-term contribution arises from $E_6$ breaking
down to $SU(5)\times U(1)\times U(1)'$. Therefore the induced
low-energy phenomenology might be rather different. In particular, the
above property (ii) could make the radiative EWSB much easier to occur.

%%%%%%%%%%%%%%%%%%%%%%%%%%%%%%%%%%%%%%%%%%%%%%%%%%%%%%%%%%%%%%%%%%%%%%
\subsection{$\boldsymbol{SO(10)}$ Unification with Asymmetrical Yukawa
Matrices} 
%%%%%%%%%%%%%%%%%%%%%%%%%%%%%%%%%%%%%%%%%%%%%%%%%%%%%%%%%%%%%%%%%%%%%%
\subsubsection{Bottom Quark Mass}
%%%%%%%%%%%%%%%%%%%%%%%%%%%%%%%%%%%%%%%%%%%%%%%%%%%%%%%%%%%%%%%%%%%%%%
Let us first examine the prediction of bottom quark mass in 
the $SO(10)$ unification with the lopsided Yukawa 
matrices (\ref{lopY}), in which we take for 
simplicity $y'=y_{{}_G}\cos\theta$. The lopsided form of Yukawa
matrices generally leads to the prediction of third-generation fermion
masses quite different from that of the minimal $SO(10)$-type 
unification. As we have seen in the previous sections, the low-energy
threshold correction $\Delta_b$ at SUSY-breaking scale plays a central
role for predicting the bottom quark mass in Yukawa unified theory. In
the present lopsided case, there is an additional important 
factor $\theta$, which determines the mixture of high-energy Higgs
fields and induces neutrino large generation mixing. That is, since
the low-energy down-type Higgs in $5^*$ representation is now
controlled by the angle $\theta$, the predicted value 
of $m_b^{\overline{\text{MS}}}(m_b)$ also has possible large
dependence on $\theta$.
\begin{figure}[t]
\begin{center}
\includegraphics[width=5.5cm,clip]{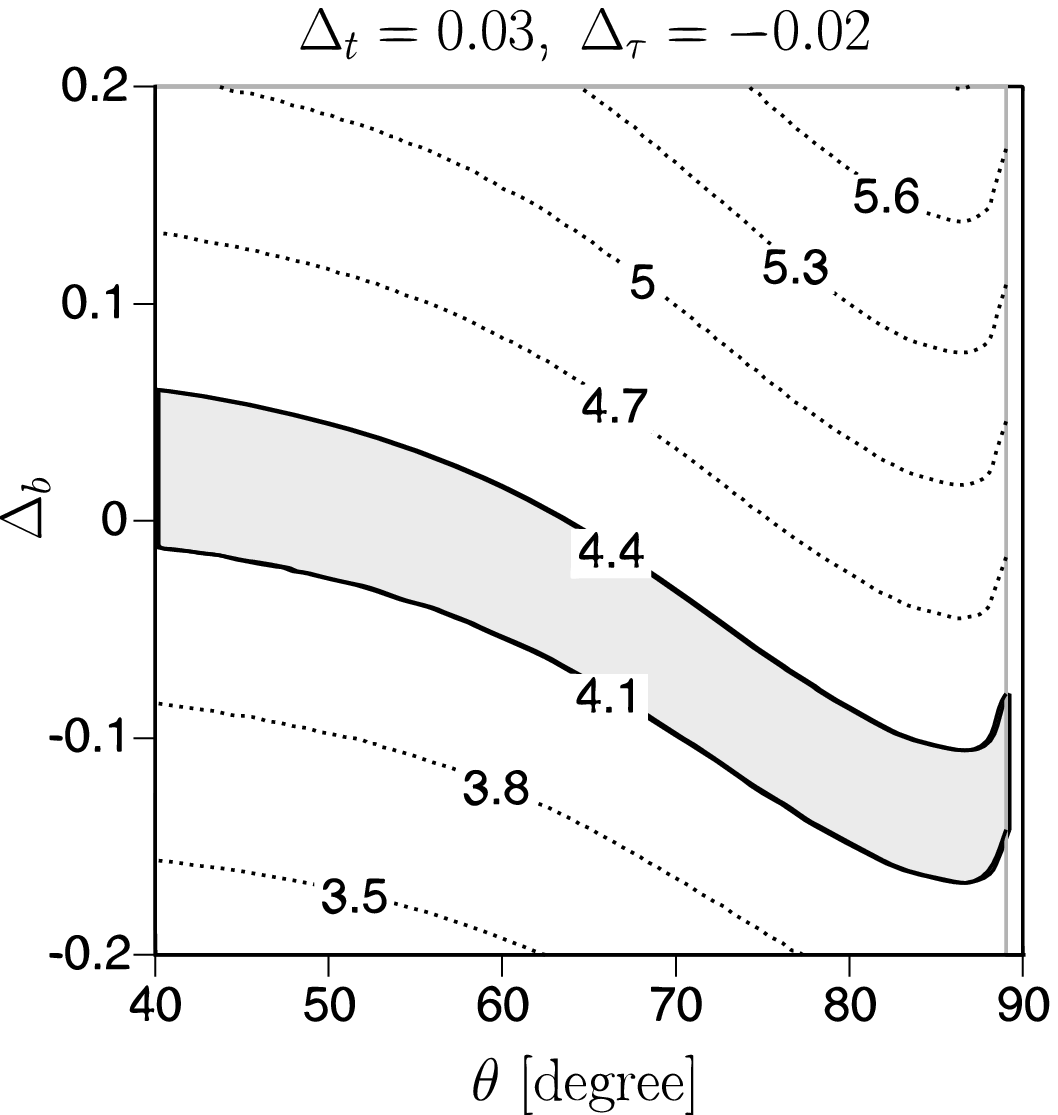}\hspace*{1cm}
\includegraphics[width=5.5cm,clip]{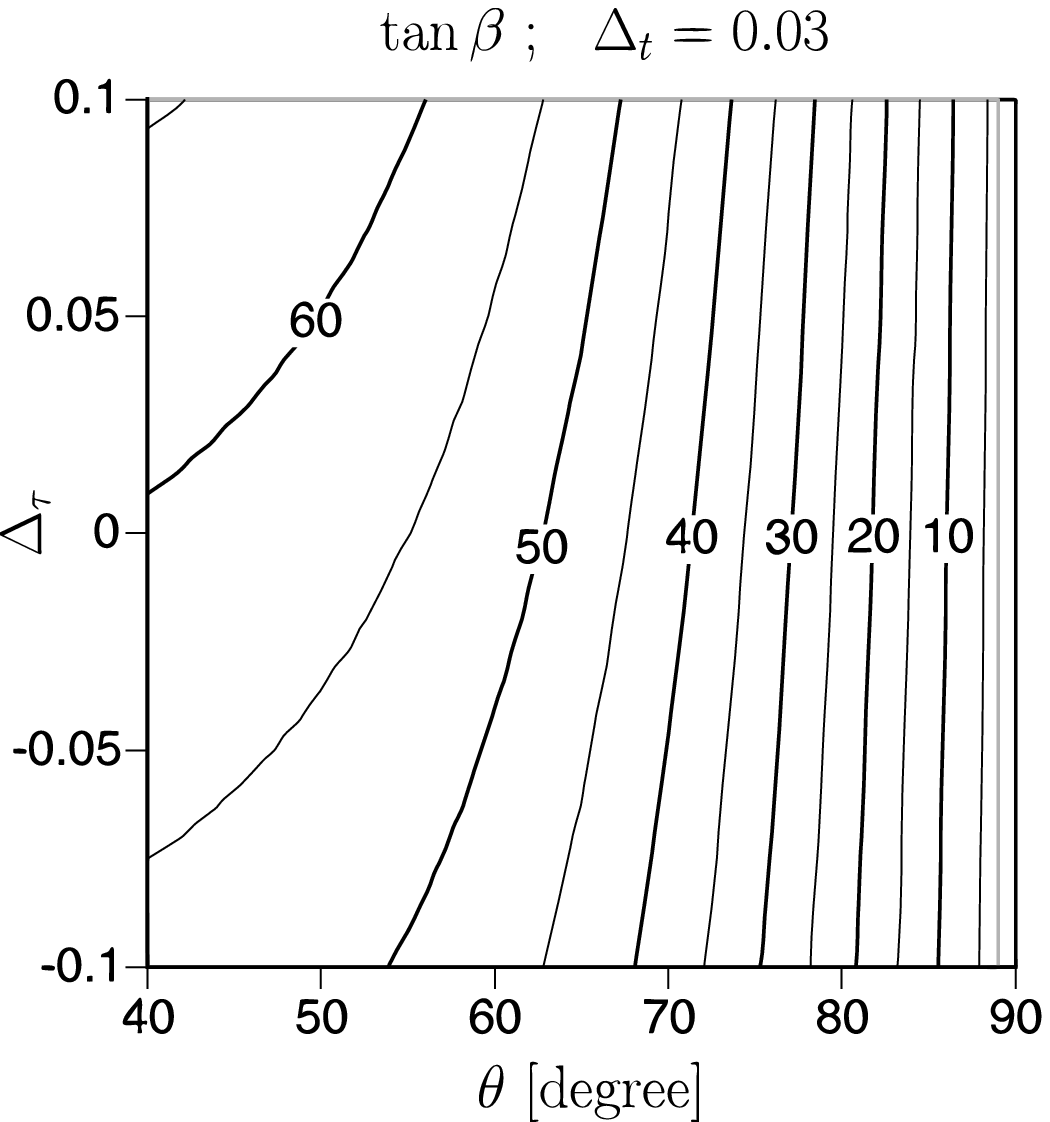}
\end{center}
\caption{The predictions of bottom quark mass 
and $\tan\beta$ in $SO(10)$ unification with lopsided mass
matrices (\ref{lopY}). The left figure 
shows $m_b^{\overline{\text{MS}}}(m_b)$ as the function of mixing
angle $\theta$ and SUSY threshold correction $\Delta_b$. The right
figure shows typical value of $\tan\beta$ in this model. Here we take
input parameters the same as in the previous figure~\ref{mbfig} as
well as $m_t^{\rm pole}=178$~GeV and $M_\nu=10^{14}$~GeV\@. The
low-energy threshold corrections are set 
as $\Delta_t=0.03$ and $\Delta_\tau=-0.02$ in the left figure 
and $\Delta_t=0.03$ in the right one.\bigskip}
\label{thetaDb}
\end{figure}
We show in Fig.~\ref{thetaDb} the prediction of bottom quark mass
as the function of these two important 
factors $\Delta_b$ and $\theta$. In the figures, we 
have $m_t^{\rm pole}=178$~GeV, $\Delta_t=0.03$, $\Delta_\tau=-0.02$,
and $M_\nu=10^{14}$~GeV\@. It is also found that $\tan\beta$ is another
important quantity affected by the mixing 
parameter $\theta$, since $\tan\beta$ is fixed by the prediction of
tau lepton mass and the RG solution of tau Yukawa coupling is
sensitive to $\theta$.

It is seen from the figure 
that $m_b^{\overline{\text{MS}}}(m_b)$ is increased for a larger value
of $\theta$. This is because the magnitude of bottom-quark and
tau-lepton Yukawa couplings are suppressed by the 
factor $\cos\theta$. For a moderate value of $\tan\beta$, the
bottom/tau mass ratio (without threshold corrections) is reached in
the RG evolution down to low-energy regime to a larger value than that
in large $\tan\beta$ case. This is because the Yukawa dependent
contributions are reduced in the RG evolution of Yukawa couplings for
a fixed value of $m_t^{\rm pole}$. With this behavior, a negative
threshold correction $\Delta_b$ is required 
for $\theta\gtrsim60^\circ$ (see Fig.~\ref{thetaDb}). Therefore a
negative value of $\mu$ parameter is preferred in this model. We
explicitly checked that this qualitative behavior is unchanged by
varying input parameters in appropriate 
range. For $\tan\beta\lesssim3$, the top quark 
mass $m_t^{\rm pole}$ is decreased as $\sin\beta$ and $y_{{}_G}$ is
consequently increased. Thus the above discussion is not applied to
the case of near-maximal value $\theta\simeq90^\circ$. In the
following analysis we do not consider such a small value 
of $\tan\beta$.

Notice also from the figure that the absolute value of threshold
correction to bottom quark mass must still be smaller than its naively
expected size. Thus the argument of PQ and R symmetries may be
helpful for suppressing $\Delta_b$ and attaining the
experimentally-allowed value of bottom quark mass similarly to the
minimal $SO(10)$-type unification. However there is an important
difference between them; in the present model, the desired suppression
factor of threshold correction $\Delta_b$ and also $\tan\beta$ have 
significant dependences on $\theta$, which induces neutrino large
mixing. In particular, even a bit large value of $|\Delta_b|$ is
possible for small $\cos\theta$. Thus a relatively weaker PQ and/or
R symmetric low-energy spectrum can be consistent with the
experimental bound on bottom quark mass rather than the 
minimal $SO(10)$-type unification.

%%%%%%%%%%%%%%%%%%%%%%%%%%%%%%%%%%%%%%%%%%%%%%%%%%%%%%%%%%%%%%%%%%%%%%
\subsubsection{Radiative EWSB and PQ, R Symmetric Limits}
%%%%%%%%%%%%%%%%%%%%%%%%%%%%%%%%%%%%%%%%%%%%%%%%%%%%%%%%%%%%%%%%%%%%%%
To determine the complete structure of SUSY-breaking parameters at the
GUT scale, we take the following simple assumptions: (i) the theory is
just the MSSM with right-handed neutrinos below the GUT-breaking
scale. (ii) the MSSM matter fields, except for $5^*_2$, come 
from $16_i$ ($i=1,2,3$). (iii) the matter fields in $5^*_2$ originate
from an additional matter $10$. (iv) the up-type Higgs is included 
in $10_H$, but the down-type one is a linear combination 
of $5^*(10_H)$ and $5^*(16_H)$ with a mixing angle $\theta$. With this
situation at hand, the matching conditions of MSSM SUSY-breaking
parameters at the GUT scale are given by the following form:
{\allowdisplaybreaks%
\begin{alignat}{2}
 & m_{\tilde Q}^2(M_G)_{ij} && \,=\; m_{\tilde u}^2(M_G)_{ij}
 \,=\, m_{\tilde e}^2(M_G)_{ij} \,=\, 
 (m_0^2+\Delta-D)\,\delta_{ij}, \\[1mm]
 & m_{\tilde d}^2(M_G)_{11} && \,=\; m_{\tilde L}^2(M_G)_{11} 
 \,=\, m_0^2+\Delta+3D, \\
 & m_{\tilde d}^2(M_G)_{22} && \,=\; m_{\tilde L}^2(M_G)_{22} 
 \,=\, m_0^2+\frac{4}{5}\Delta-2D, \\[1mm]
 & m_{\tilde d}^2(M_G)_{33} && \,=\; m_{\tilde L}^2(M_G)_{33} 
 \,=\, m_0^2+\Delta+3D, \\[2.5mm]
 & m_{\tilde \nu}^2(M_G)_{ij} && \,=\, 
 (m_0^2+\Delta-5D)\,\delta_{ij}, \\[2mm]
 & m_{H_u}^2(M_G) && \,=\; m_0^2+\frac{4}{5}\Delta+2D, \\
 & m_{H_d}^2(M_G) && \,=\; m_0^2 
 +\Big(\frac{4}{5}\cos^2\theta +\sin^2\theta\Big)\Delta 
 +\left(-2\cos^2\theta +3\sin^2\theta\right)D.
\end{alignat}}%
Here we have written down rather generic expressions for the boundary
conditions, including (i)~the usual (flavor-blind) universal scalar
mass $m_0^2$ given at the gravitational scale $M_P$, (ii)~the $D$-term
contribution denoted by $D$, which potentially arises from the GUT
symmetry breaking $SO(10)\to SU(5)\times U(1)$, and (iii) a radiative
effect $\Delta$ generated via RG running from $M_P$ down to $M_G$. If
one neglects Yukawa-dependent contribution, $\Delta$ is given by
\begin{equation}
  \Delta \;=\; \frac{45}{4b_{10}}\bigg[\Big
  (1-\frac{b_{10}g_{{}_G}^2}{8\pi^2}\ln\frac{M_P}{M_G}\Big)^{-2}\!-1
  \bigg]M_{1/2}^2,
\end{equation}
where $b_{10}$ is the beta function for $SO(10)$ gauge coupling 
and $M_{1/2}$ is the $SO(10)$ gaugino mass parameter evaluated at the
GUT-breaking scale. It is found from this expression 
that $\Delta$ always takes a positive value and independent 
of $M_{1/2}$ due to the unknown beta-function factor. For comparison
to the analyses in the previous sections, we parametrize the scalar
masses by $\tilde m_0$ and $\xi$ defined as
{\allowdisplaybreaks%
\begin{eqnarray}
  \tilde m_0^2 &\;\equiv\;& \frac{m_{10}^2+m_{16}^2}{2}\bigg|_{D=0}
  \!=\; m_0^2+\frac{9}{10}\Delta, \\
  \xi &\;\equiv\;& 
  \frac{m_{10}^2-m_{16}^2}{m_{10}^2+m_{16}^2}\bigg|_{D=0}
  \!=\; \frac{-\Delta}{10m_0^2+9\Delta}.
\end{eqnarray}}%
In the following discussion, we use $\tilde m_0^2$ and $\xi$ instead
of the original parameters $m_0^2$ and $\Delta$. It is noted that,
contrary to the previous analyses, the parameter $\xi$ is now limited
in the range $-1/9<\xi<0$ because of the positiveness 
of $m_0^2$ and $\Delta$. Such a bound is however not so strict but
might be relaxed by including RG effects of GUT Yukawa couplings
and/or by assuming negative $m_0^2$~\cite{m0nega}. We also assume for
simplicity that the GUT-scale scalar trilinear couplings are flavor
universal, referred to as $A_0$. Thus the independent variables
describing SUSY-breaking parameters at the GUT scale are
\begin{equation}
  \tilde m_0,\; \xi,\; M_{1/2},\; A_0,\; B_0,\; D.
\end{equation}
In addition to these, the mixing parameter $\theta$ is an important
factor for determining superparticle mass spectrum. As for the
radiative EWSB, a smaller value of $\cos\theta$ suppresses 
the $Y_{d,\tau}$ effects in the RG evolution and 
makes $m_{H_u}^2$ lower than $m_{H_d}^2$ in the infrared. Consequently
the experimental mass bound of CP-odd neutral Higgs boson is expected
to be satisfied in a wider parameter region and the radiative EWSB is
operative more easily than the ordinary Yukawa unification. That is
contrasted with the previous result that, in the Yukawa unification, a
difficulty in realizing the difference $m_{H_u}^2<m_{H_d}^2$ excludes
a large portion of parameter space.

\begin{figure}[t]
\begin{center}
\includegraphics[width=5cm,clip]{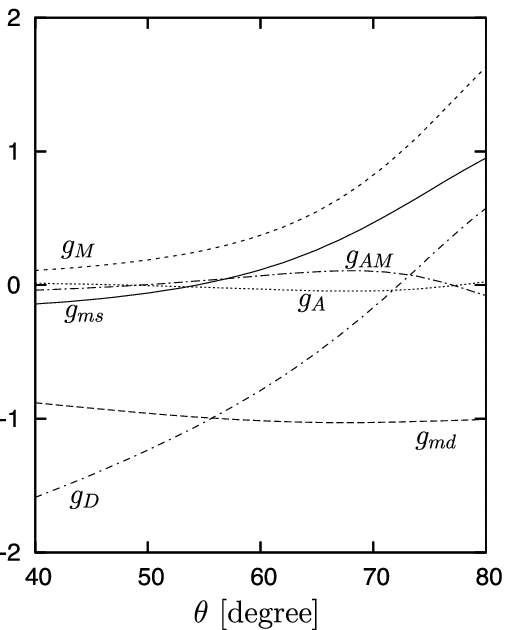}\hspace*{2cm}
\includegraphics[width=5cm,clip]{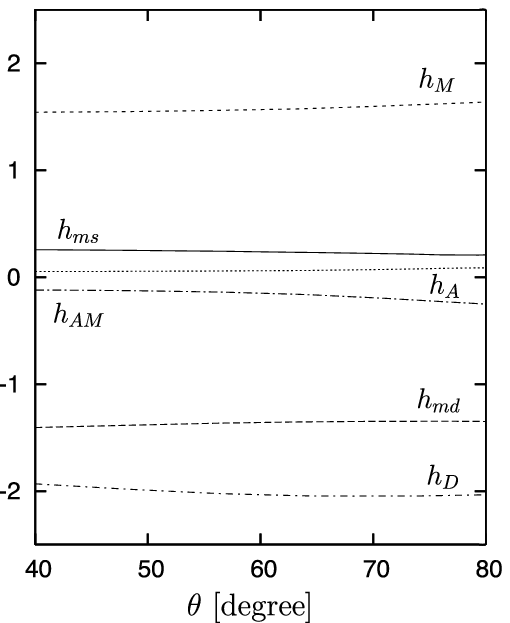}
\end{center}
\caption{The $\theta$ dependences of the RG 
solutions (\ref{mafit3}) and (\ref{mufit3}) evaluated at the
renormalization scale $Q=1$~TeV\@. Here we take input parameters as
the same as in the previous figure~\ref{fit2} in addition 
to $m_t^{\rm pole}=178$~GeV, $M_\nu=10^{14}$~GeV, and the SUSY
threshold corrections $\Delta_t=0.03$ and $\Delta_\tau=-0.02$.\bigskip}
\label{fit3}
\end{figure}

To explicitly confirm such EWSB property and to examine the
possibility of having PQ and R symmetries, we numerically solve the
MSSM RG equations with right-handed neutrino couplings and evaluate
the masses of physical particles in the electroweak symmetry broken
vacuum;
{\allowdisplaybreaks%
\begin{eqnarray}
  M_A^2 &\;=\;& (g_{ms}+g_{md}\xi)\tilde m_0^2 +g_MM^2_{1/2} 
  +g_{AM}A_0M_{1/2} +g_AA_0^2 +g_DD -M_Z^2,  \label{mafit3} \\
  |\mu|^2 &\;=\;& (h_{ms}+h_{md}\xi)\tilde m_0^2 +h_MM_{1/2}^2 
  +h_{AM}A_0M_{1/2} +h_AA_0^2 +h_DD -\frac{M_Z^2}{2}.  \label{mufit3}
\end{eqnarray}}%
Typical behaviors of the coefficients $g$'s and $h$'s are shown in 
Fig.~\ref{fit3} as the functions of mixing parameter $\theta$. In
the calculation, the input parameters are taken as the same as in the
previous solutions (Fig.~\ref{fit2}) for comparison 
and $m_t^{\rm pole}=178$~GeV\@. The EWSB conditions are solved at the
renormalization scale $Q=1$~TeV\@. It is found from the figure that
the coefficients $g$'s in the RG solution (\ref{mafit3}) obviously
depend on $\theta$. In particular, $g_{ms}$, $g_M$, and $g_D$ are
rather sensitive and increase 
as $\theta$. For $\theta\gtrsim55^\circ$, both the scalar ($g_{ms}$)
and gaugino ($g_M$) effects become positive and therefore an R
symmetric radiative EWSB is viable. This confirms the RG evolution
behavior of Higgs mass parameters mentioned above. The CP-odd neutral
Higgs mass squared is positive for a smaller value 
of $\cos\theta$. The $D$-term effect ($g_D$) is mainly controlled by
the initial $\theta$ dependence of the $D$-term contribution to
down-type Higgs mass parameter at the GUT scale.

In contrast to the uneven profile of $g$'s, the coefficients $h$'s in
the RG solution (\ref{mufit3}) are rather insensitive to the mixing
angle $\theta$. This is understood from the EWSB conditions
that $|\mu|$ is almost determined only by $m_{H_u}^2$ for a
not-so-small value of $\tan\beta$; $|\mu|^2\sim-m_{H_u}^2$, and also
from the fact that $\theta$ controls only $5^*$, i.e.\ the down-type
Higgs mass $m_{H_d}^2$. As a result, in a wide range of $\theta$, the
RG solution $|\mu|^2$ is found to receive positive contributions from
scalar ($h_{ms},h_{md}$) and gaugino ($h_M$) terms (notice here 
that $\xi$ only takes a negative value). The $A_0$-dependent
contributions cannot be largely negative because $h_M,h_A>0$. We thus
find that only the $D$-term contribution is relevant for 
decreasing $|\mu|^2$ and realizing a PQ symmetric EWSB\@. The $D$-term
effect ($h_D$) can be either positive or negative depending on the
sign of $D$, but has little dependence on $\theta$ since the up-type
Higgs mass is no connection with $\theta$. To make $|\mu|$ small
requires a positive value of $D$. However that makes the CP-odd
neutral Higgs boson lighter if $\theta\lesssim72^\circ$. Thus it is
not obvious whether approximate PQ and R symmetries are consistently
realized in the superparticle spectrum in order for the
third-generation fermion masses and the $b\to s\gamma$ constraint
being acceptable.

For an illustrative purpose, let us focus on the exact limits of PQ
and R symmetries. We have found in the above discussion that the PQ
and R symmetries require a small $\cos\theta$ and a 
positive $D$ term. To see the required values of $D$ and $\theta$, it
is useful to examine the following functions:
{\allowdisplaybreaks%
\begin{eqnarray}
  \hat D_{\rm PQ}(\theta) &\;=\;& \frac{-1}{h_D}(h_{ms}+h_{md}\xi), \\
  \hat M^2_A(\theta) &\;=\;& g_{ms}-\frac{h_{ms}}{h_D}g_D+
  \Big(g_{md}-\frac{h_{md}}{h_D}g_D\Big)\xi,
\end{eqnarray}}%
where $\hat D_{\rm PQ}$ satisfies the exact PQ symmetric 
equation $|\mu|=0$ in the R symmetric limit and $\hat M_A^2$ denotes
the CP-odd neutral Higgs mass (normalized by the scalar 
mass $\tilde m_0^2$) evaluated in these symmetric limits.
\begin{figure}[t]
\begin{center}
\includegraphics[width=6.5cm,clip]{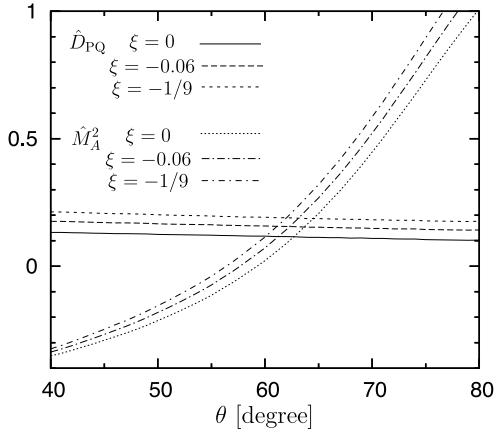}
\end{center}
\caption{The solutions $\hat D_{\rm PQ}$ and $\hat M^2_A$ for the
exact PQ and R symmetric limits (normalized by the scalar 
mass $\tilde m_0^2$). In this figure we take input parameters the same
as in the previous figure~\ref{fit3}.\bigskip}
\label{PQR}
\end{figure}
Fig.~\ref{PQR} shows the 
solutions $\hat D_{\rm PQ}$ and $\hat M^2_A$ as the functions 
of $\theta$ and $\xi$. While $\hat M^2_A$ has 
large $\theta$ dependence, $\hat D_{\rm PQ}$ is not so changed 
with $\theta$. We find in the figure that the PQ and R symmetric EWSB
with a positive mass squared of CP-odd neutral Higgs boson is achieved
for $0<D\lesssim0.2\tilde m_0^2$ and $\theta\gtrsim55^\circ$.

To summarize the discussion about the radiative EWSB and the mixing
angle dependence of bottom quark mass, there are two different options
available; One is the case $\theta\gtrsim60^\circ$ with a 
negative $\mu$ and the other is $\theta\lesssim60^\circ$ with both
signs of $\mu$. The former is consistent with PQ and R symmetries and
the suppression of large threshold correction $\Delta_b$ to bottom
quark mass is obtained to have experimentally allowed bottom quark
mass. In the latter option, only the R symmetric mass spectrum is
possible but the suppression of the threshold correction would be
still viable. To see these issues more explicitly, we will turn to the
numerical parameter analysis in the next section.

%%%%%%%%%%%%%%%%%%%%%%%%%%%%%%%%%%%%%%%%%%%%%%%%%%%%%%%%%%%%%%%%%%%%%%
\subsubsection{Parameter Space Analysis}
%%%%%%%%%%%%%%%%%%%%%%%%%%%%%%%%%%%%%%%%%%%%%%%%%%%%%%%%%%%%%%%%%%%%%%
As we have found in the previous subsection, there are two types of
solutions for the radiative EWSB conditions and an enough suppression
of excessive threshold correction to bottom quark mass. They are
parametrized by the angle $\theta$ which induces neutrino large
generation mixing. The solutions are divided into two 
regions, $\theta\gtrsim60^\circ$ and $\theta\lesssim60^\circ$,
depending on whether PQ symmetric spectrum is viable or not.

Before proceeding to the numerical analysis, it is worth discussing
the $b\to s\gamma$ process 
qualitatively. The $b\to s\gamma$ constraint is also affected by the
angle $\theta$ through the initial values of couplings and RG
evolution. As clarified in the analysis of Yukawa unification, for a
negative $\mu$, the $b\to s\gamma$ rare process generally gives
stronger constraints on SUSY-breaking parameters than for a 
positive $\mu$. Such qualitatively behavior of the branching ratio can
also be applied to the present model as long as $\tan\beta$ is not so
small. Thus the region $\theta\lesssim60^\circ$ with a 
positive $\mu$ is expected to easily avoid 
the $b\to s\gamma$ constraint than the other regions in which a
negative $\mu$ is required to obtain the successful prediction of
bottom quark mass. It is also noted that the gluino loop contribution
to $b\to s\gamma$ decay is important~\cite{bsggluino}. This is due to
the lopsided form of mass matrices and the non-universality of
SUSY-breaking scalar masses. We now have a large 2-3 mixing of
right-handed down quarks and flavor-dependent mass parameters of
right-handed scalar down quarks. These flavor dependences cannot be
rotated away by field redefinition and some imprint may appear, for
example, as a large generation mixing of scalar down
quarks. Furthermore the mass difference of up and down-type scalar
quarks is induced by a negative value of $D$ which mass difference is
known to enhance the $b\to s\gamma$ amplitude via the gluino 
diagram $A_{\tilde g}$. 

First we analyze the region of a small value 
of $\cos\theta\,$ ($\theta\gtrsim60^\circ$). Such a 
large $\theta$ raises the tree-level bottom quark mass as seen in
Fig.~\ref{thetaDb} and accordingly the negative sign 
of $\mu$ parameter is required. The threshold correction must be
roughly $|\Delta_b|\sim{\cal O}(0.1)$ which is smaller than its
naively expected size. Thus the approximate PQ and R symmetries are
useful to obtain the experimentally allowed bottom quark mass.
\begin{figure}[t]
\begin{center}
\includegraphics[width=5cm,clip]{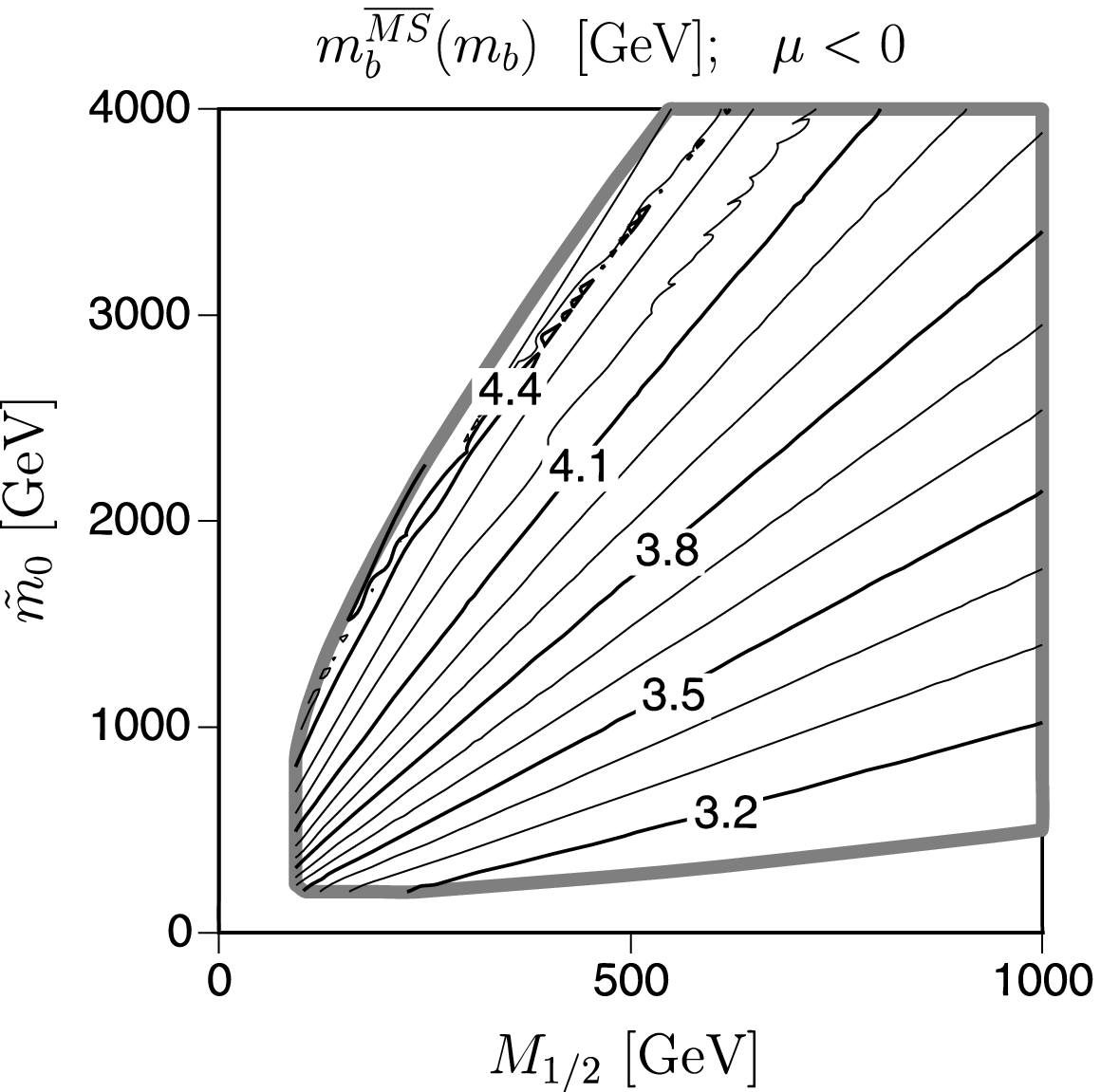}\hspace*{1cm}
\includegraphics[width=5cm,clip]{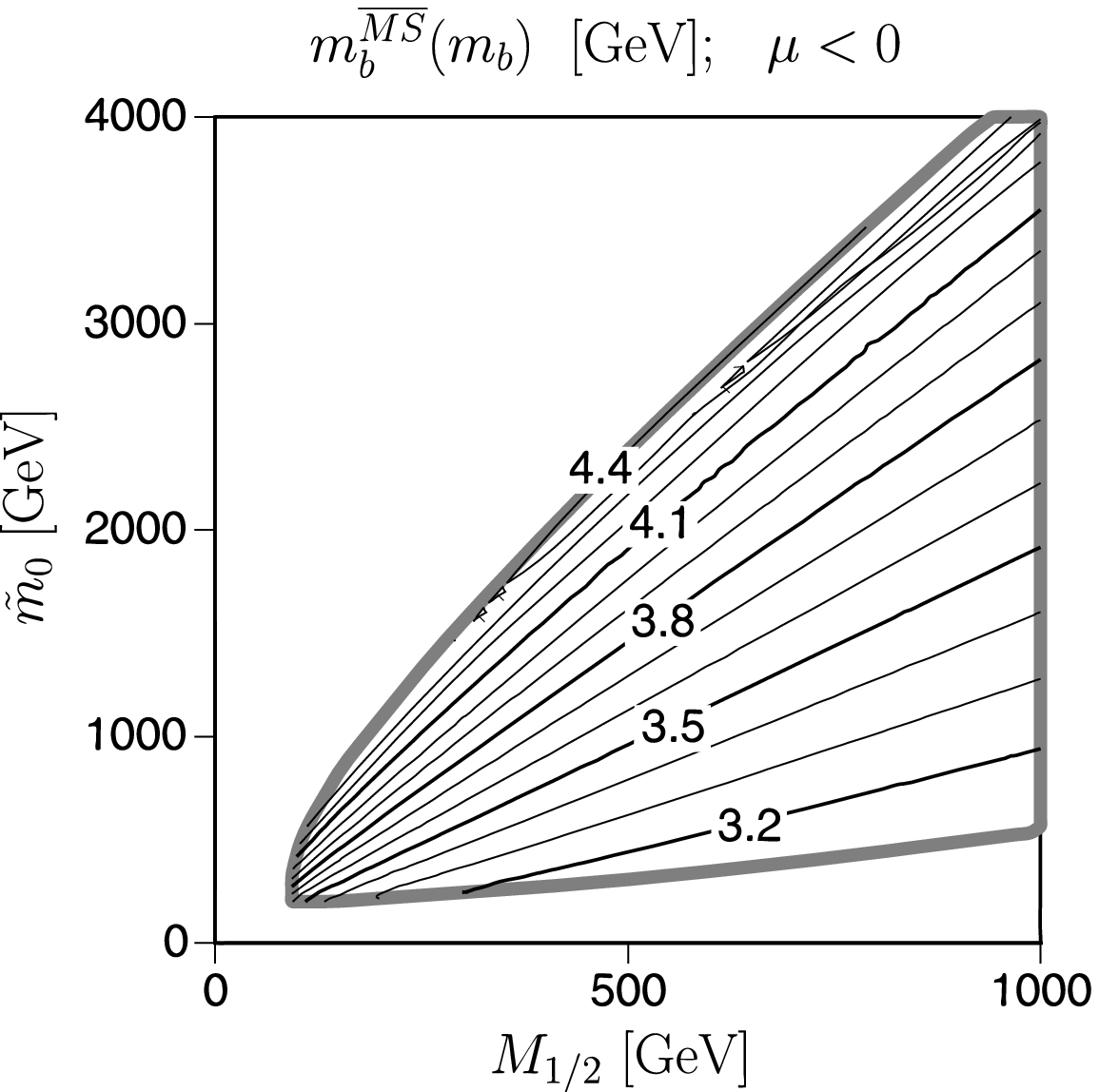}\\[5mm]
\includegraphics[width=5cm,clip]{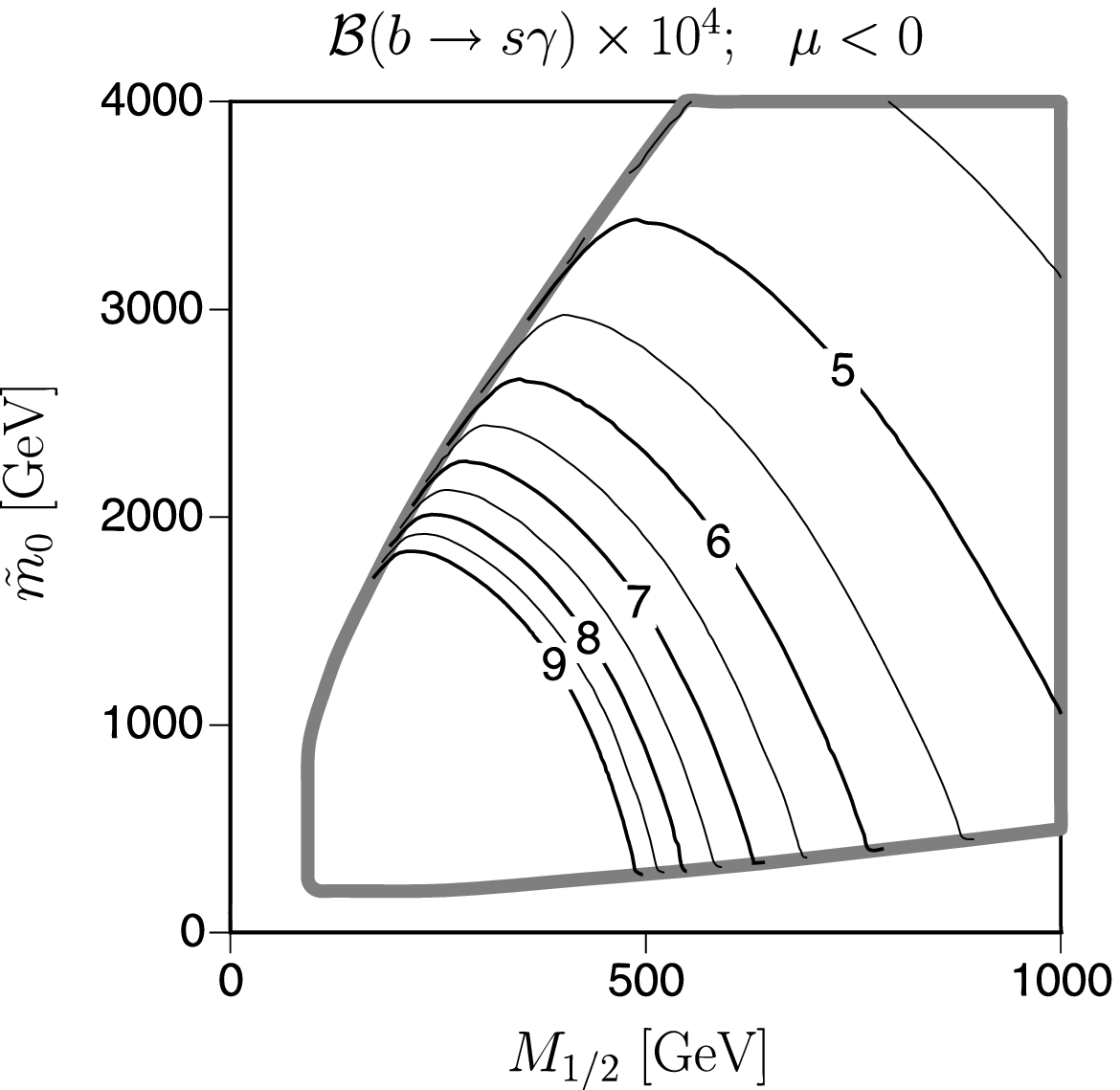}\hspace*{1cm}
\includegraphics[width=5cm,clip]{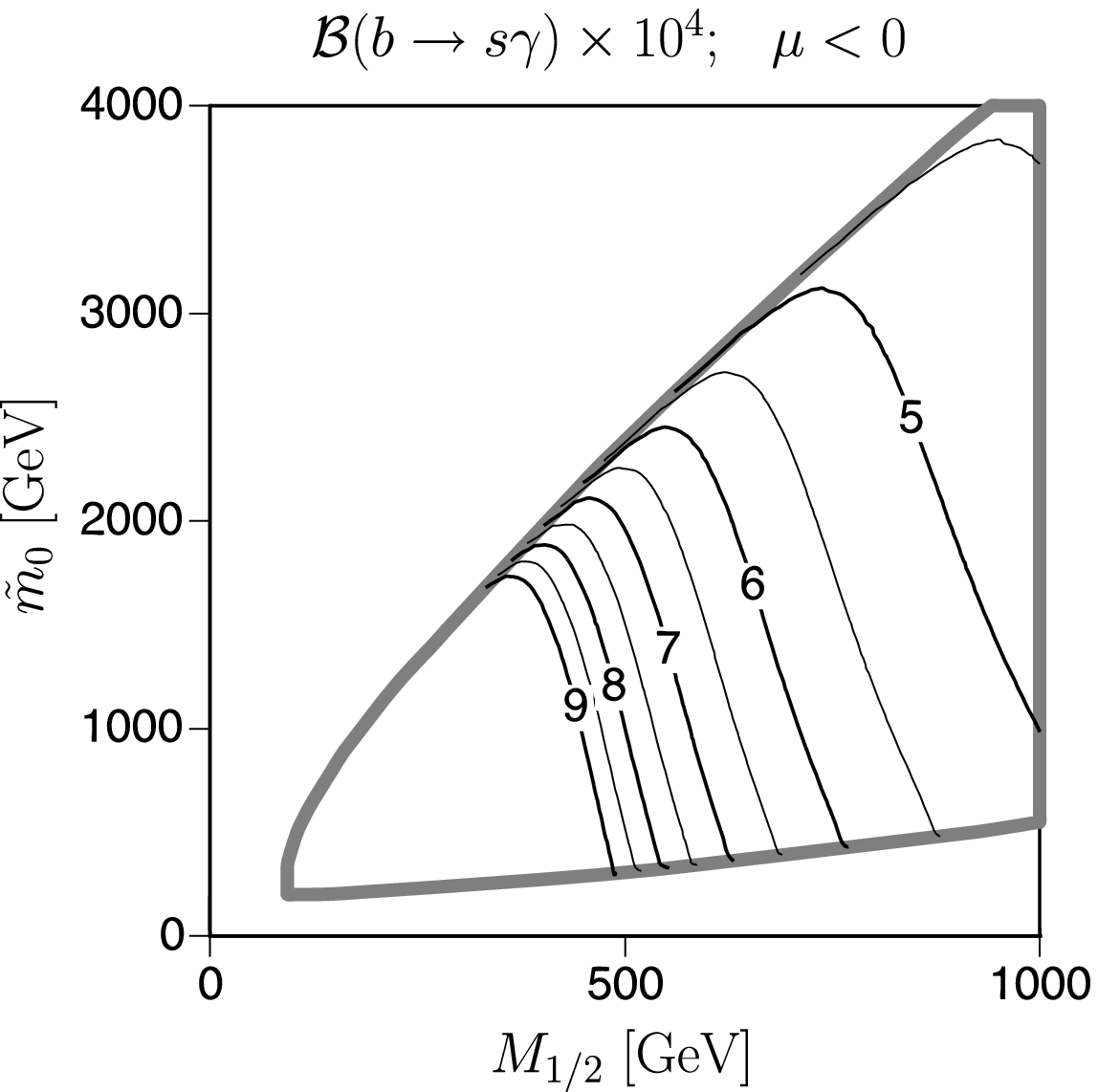}
\end{center}
\caption{The $M_{1/2}$--$\tilde m_0$ parameter space consistent with
the radiative EWSB conditions, the experimental mass bounds of
superparticles, and the requirement of neutral LSP 
in $SO(10)$ unification with the lopsided mass matrices. The bottom
quark mass $m_b^{\overline{\text{MS}}}(m_b)$ and 
the $b\to s\gamma$ branching ratio are also shown in the figures. Here
we take input parameters as the same as in the previous 
figure~\ref{Mgm0} in addition to $M_\nu=10^{14}$~GeV 
and $\theta=65^\circ$. The scalar mass parameters 
are $\xi=0$, $D=0.1\tilde m_0^2$ in the left-sided figures 
and $\xi=-1/9$, $D=0.2\tilde m_0^2$ in the right-sided ones. In each
figure, the narrow right-bottom region is excluded by the scalar tau
LSP and the left-upper side is ruled out by the current mass bounds of
charginos and neutralinos.\bigskip}
\label{Mgm0_lop}
\end{figure}
In Fig.~\ref{Mgm0_lop}, we show 
the $M_{1/2}$--$\tilde m_0$ parameter space consistent with the EWSB
conditions, the experimental mass bounds on superparticles, and the
requirement that the LSP is charge neutral. In the figures, the
predictions of bottom quark mass $m_b^{\overline{\text{MS}}}(m_b)$ and
the $b\to s\gamma$ branching ratio are shown in the allowed parameter
regions. As an example, we set $\theta=65^\circ$ and $\xi$ has the
maximal value ($\xi=0$) in the left-sided figures and the minimum
value ($\xi=-1/9$) in the right-sided ones. In both cases, we 
obtain $\tan\beta\simeq45$. In each figure, the narrow right-bottom
region is excluded by scalar tau lepton being the LSP and the
left-upper side is ruled out by the current mass bounds of charginos
and neutralinos. In both extreme cases $\xi=0$ and $-1/9$, we find
that there exist the parameter spaces which reproduce the observed
bottom quark mass, since a positive $D$ term 
suppresses $|\mu|$. Also the $b\to s\gamma$ branching ratio is found
to be within the experimentally observed range. This is achieved with
a few hundred GeV gaugino masses and a few TeV scalar masses in both
parameter spaces. Such a heavy scalar spectrum is due to the negative
sign of $\mu$. In the parameter region in Fig.~\ref{Mgm0_lop}, a
positive $D$ increases the bottom scalar quark mass compared to
those of up-type ones. Thus the gluino-loop contribution 
to $b\to s\gamma$ process becomes smaller than the chargino-loop
contribution.

We have found in this first case that, in the parameter space where
the bottom quark mass and the $b\to s\gamma$ branching ratio are in
agreement with the experimental range, superparticle mass spectrum
exhibits approximate PQ and R symmetries. In particular, lighter
chargino and neutralinos contain significant components of
higgsinos. Moreover the mass bound of CP-odd neutral Higgs boson does
not lead to strong constraints on the GUT-scale mass parameters. In
fact, the CP-odd neutral Higgs mass takes as large as 2~TeV in the
allowed parameter regions. These features are quite different from
those obtained in the minimal $SO(10)$-type unification.

\begin{figure}[t]
  \begin{center}
\includegraphics[width=5.5cm,clip]{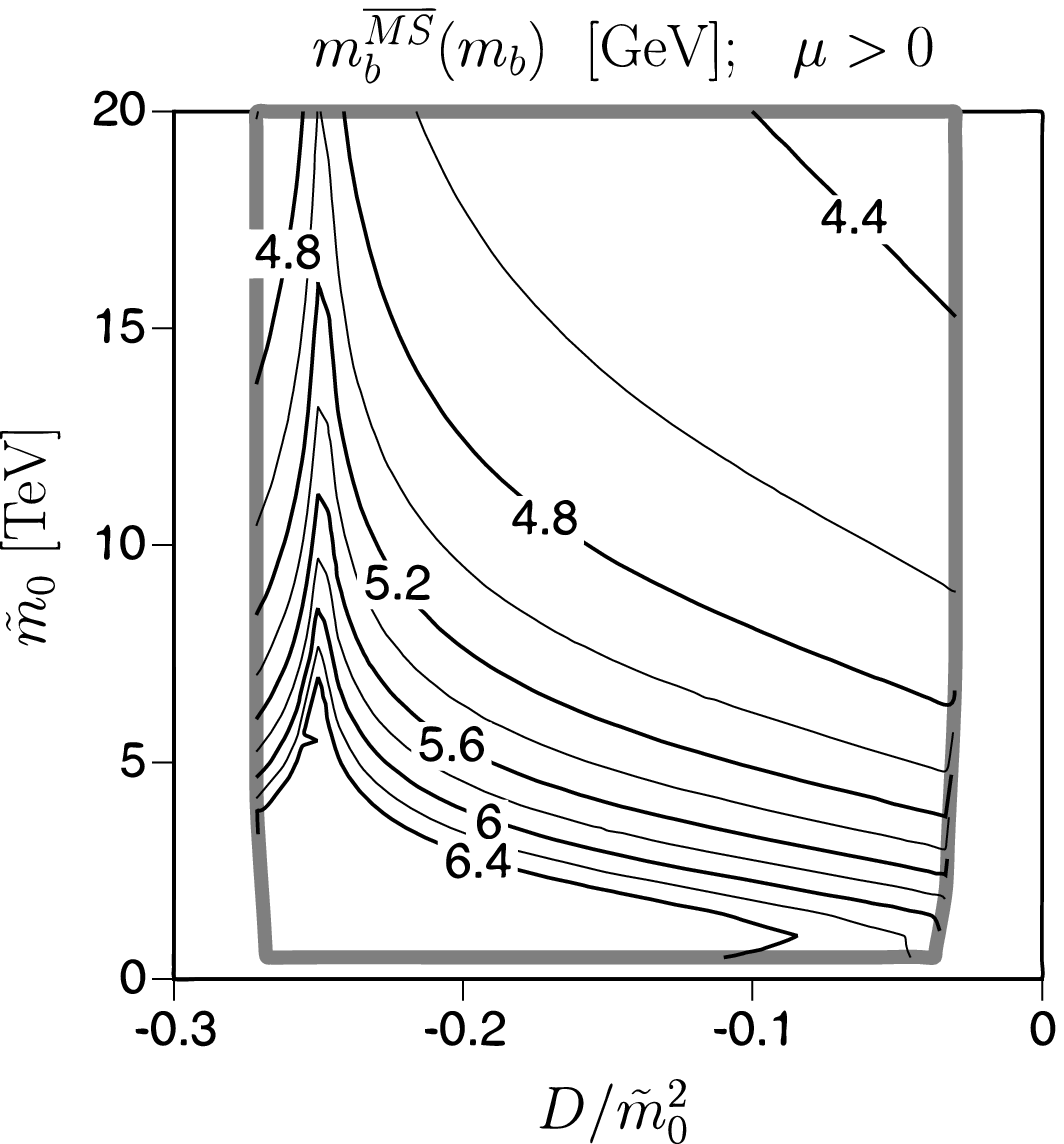}\hspace*{1cm}
\includegraphics[width=5.5cm,clip]{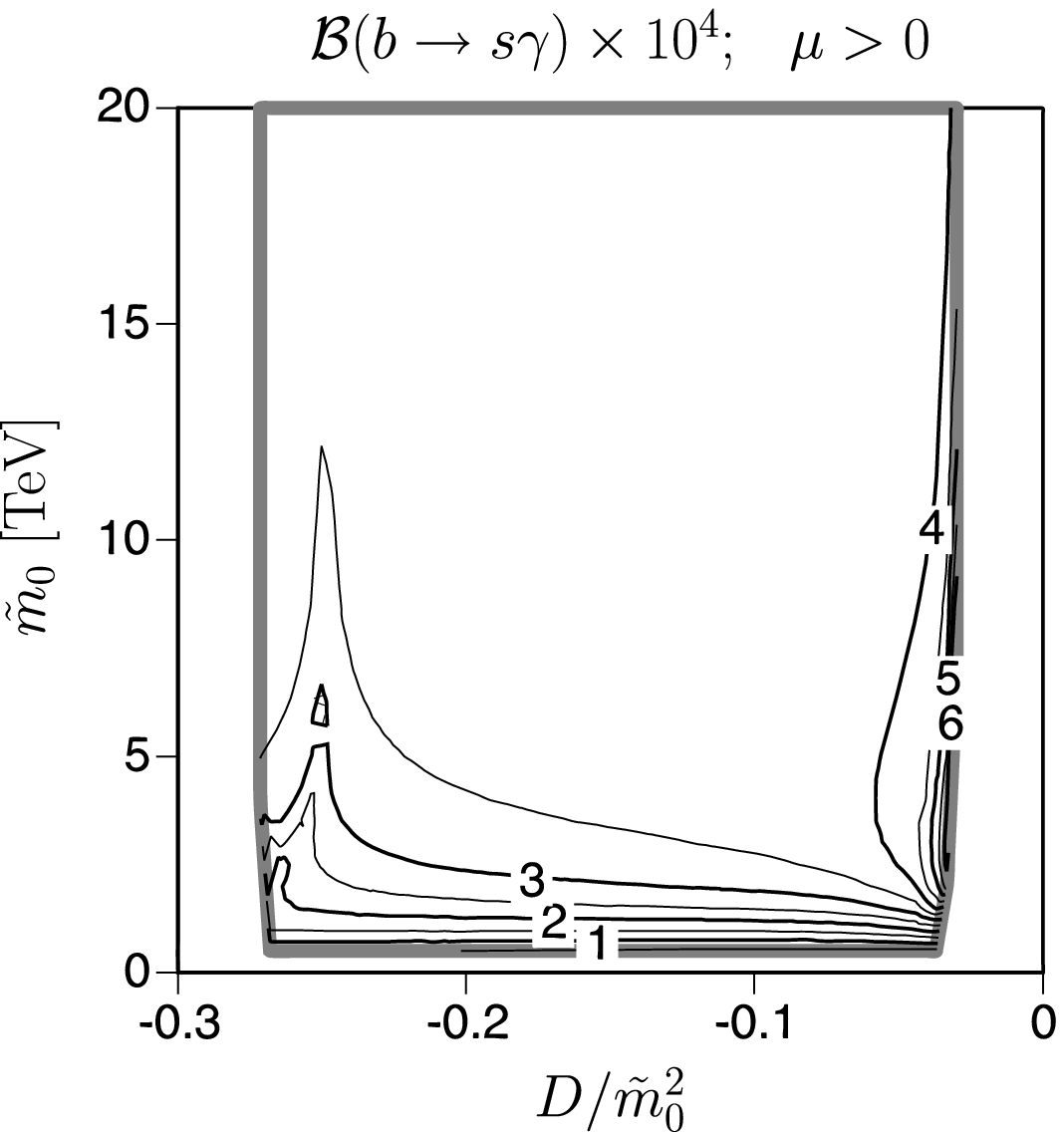}
\end{center}
\caption{The same as Fig.~\ref{Mgm0_lop} but for the parameter
space of $D$-term contribution and scalar masses. Here we 
set $\theta=50^\circ$, $\xi=0$, $M_{1/2}=250$~GeV, and $A_0=0$. In
each figure, the left and bottom regions are excluded by the LSP being
charged and the right one is excluded by the mass bound of CP-odd
neutral Higgs boson.\bigskip}
\label{Dm0}
\end{figure}

Next let us turn to studying another case
with $\theta\lesssim 60^\circ$. Both signs of $\mu$ parameter are
allowed in this case. In the following, we take a positive $\mu$ which
is advantageous to avoid the $b\to s\gamma$ constraint. The threshold
correction to the bottom quark mass must be again small. Since the PQ
symmetry is largely violated in this region, to reduce the size 
of $|\Delta_b|$ is obtained, for example, by large scalar 
masses $\tilde m_0\gg M_{1/2}$. We show in Fig.~\ref{Dm0} the
bottom quark mass $m_b^{\overline{\text{MS}}}(m_b)$ and 
${\cal B}(b\to s\gamma)$ as the functions of scalar mass 
parameters $\tilde m_0$ and $D$. In each figure, the left and bottom
regions are excluded by the charged LSP and the right one is excluded
by the mass bound of CP-odd neutral Higgs boson.
From Fig.~\ref{Dm0}, one can see that the prediction of
bottom quark mass is decreased as $\tilde m_0$ and also has a 
large $D$ dependence. This is because the RG evolution and EWSB
conditions leads to a suppressed $|\mu|$ with raising $D$, and then
the threshold correction $|\Delta_b|$ tends to be small. Therefore a
large size of negative $D$-term contribution is disfavored. For
example, the mass eigenvalue of the lighter scalar bottom quark has a
significant dependence on $D$ and is minimized 
around $D\simeq-0.25\tilde m_0^2$. From such property, the bottom
quark mass and the $b\to s\gamma$ branching ratio are highly enhanced
through the gluino--scalar bottom diagrams (the peaks in 
Fig.~\ref{Dm0}).

We have found in this second case that large scalar masses and 
small $D$-term contribution are suitable for low-energy
phenomenology. The $b\to s\gamma$ constraint is easily evaded with
enough heavy scalar quarks to suppress the superparticle
contributions. While the scalars become heavy, the gauginos are
relatively light and R symmetric spectrum is obtained with which
the threshold correction to bottom quark mass is suppressed. Due to a
large value of $|\mu|$, the lightest chargino and neutralino are
gaugino-like. The CP-odd neutral Higgs is much lighter than the scalar
quarks due to a small size of $D$. The charged Higgs boson tends to be
lighter but can easily be made up to a few~TeV and the charged Higgs
contribution $A_{H^+}$ is suppressed.

Finally we comment on the lepton flavor violating decay of charged
leptons. As in the $SO(10)$ unification scenario in
Section~\ref{sec:wRHnu}, the $\mu\to e\gamma$ decay rate is not under
control as long as Yukawa couplings for the first and second
generations and right-handed neutrino Majorana masses are
unspecified. On the other hand, the $\tau\to\mu\gamma$ amplitude is
calculable and expected to be large due to the Yukawa-induced large
mixing for explaining the atmospheric neutrino anomaly. In the minimal
supergravity boundary conditions, the $\tau\to\mu\gamma$ decay rate is
sometimes marginal to the current experimental upper
bound~\cite{ST}. We have however found in the present scenario 
that $\tan\beta$ can be lowered and scalar leptons are relatively
heavy. That makes the constraints from lepton flavor violation rather
weakened.

To summarize the results of $SO(10)$ unification with the lopsided
mass matrices~(\ref{lopY}), there are two types of allowed parameter
regions classified by $\theta$ which is the mixing parameter of Higgs
fields and control neutrino large generation mixing. The first region
is defined by $\theta\gtrsim 60^\circ$ which leads to the PQ and R
symmetric radiative EWSB and the prediction of bottom quark mass is
well within the experimental range. Such a large $\theta$ requires a
negative value of $\mu$. The $b\to s\gamma$ constraint is also avoided
if scalar masses are a few~TeV\@. The approximate PQ and R symmetries
lead to the lightest chargino and neutralino being 
higgsino-like (possibly the LSP dark matter), that does not appear in
the minimal $SO(10)$-type unification. In the other 
region, $\theta\lesssim 60^\circ$, the R symmetric mass spectrum
allows the prediction of bottom quark mass well within the
experimental range, while PQ symmetry is not 
realized. The $\mu$ parameter can be either positive or 
negative. The $b\to s\gamma$ constraint is satisfied with heavy scalar
quarks. In this mass spectrum, only the gauginos are expected to be
relatively light. A crucial difference between the 
minimal $SO(10)$-type unification and the present model is the
twisting of $5^*$ fields which affects Yukawa couplings and
SUSY-breaking parameters at the GUT scale. That strongly violates 
the $SU(2)_R$ symmetry and thus the radiative EWSB is operative easier
than the minimal $SO(10)$-type unification. The degree of $5^*$ mixing
also affects the prediction of bottom quark mass
which restricts the sign of $\mu$ parameter. In the present scenario,
a moderate value of $\tan\beta$ implies a negative value of $\mu$.

%%%%%%%%%%%%%%%%%%%%%%%%%%%%%%%%%%%%%%%%%%%%%%%%%%%%%%%%%%%%%%%%%%%%%%
\subsection{Asymmetrical Yukawa Matrices Modified}
%%%%%%%%%%%%%%%%%%%%%%%%%%%%%%%%%%%%%%%%%%%%%%%%%%%%%%%%%%%%%%%%%%%%%%
\subsubsection{Bottom Quark Mass}
%%%%%%%%%%%%%%%%%%%%%%%%%%%%%%%%%%%%%%%%%%%%%%%%%%%%%%%%%%%%%%%%%%%%%%
The $SU(5)$ gauge symmetry implies the down and charged-lepton Yukawa
matrices are exactly same; $Y_d=Y_e^{\rm T}$. However to reproduce the
observed mass pattern including the first and second generations
requires some violation of $SU(5)$ symmetry in the Yukawa sector. A
well-known example of symmetry-violating sources is the
group-theoretical factor arising from Higgs fields in
higher-dimensional representations such as $45$ of $SU(5)$ and 
$126$ of $SO(10)$. That introduces a relative factor $-3$ between
down-quark and charged-lepton Yukawa couplings~\cite{GJ} (the 
factor $3$ means the number of colors and the negative sign comes from
the traceless property of irreducible representations).

Keeping this issue in mind, here we consider an example of lopsided
form of mass matrices which does not respect the $SU(5)$ symmetry,
that is, $Y_d\neq Y_e^{\rm T}$. In this section, the following 
simplified form is assumed for the down-quark and charged-lepton
Yukawa couplings at the GUT scale:
\begin{equation}
  Y_d \;=\; y_{{}_G}\cos\theta 
  \left(\begin{array}{ccc}
    ~~ & & \\
    & & \\
    & \frac{-1}{3} & \;1
  \end{array}\right),\qquad\quad
  Y_e \;=\; y_{{}_G}\cos\theta
  \left(\begin{array}{ccc}
    ~~ & ~~ & \\
    & & 1 \\
    & & 1
  \end{array}\right), 
  \label{lopY2}  
\end{equation}
where the blank entries are negligibly small compared to the filled
entries. The large two elements in the charged-lepton Yukawa matrix
are responsible for the atmospheric neutrino mixing and they are now
assumed to be equal, which leads to the maximal mixing angle from
the charged-lepton sector that is the central value of the current
experimental data. Compared with the $SU(5)$ symmetric lopsided 
form (\ref{lopY}), the 3-2 element in $Y_d$ now involves a relative
factor $-1/3$. This factor may originate from, for example,
higher-dimensional representations of Higgs fields or
higher-dimensional operators effectively inducing Yukawa terms.

\begin{figure}[t]
\begin{center}
\includegraphics[width=5.5cm,clip]{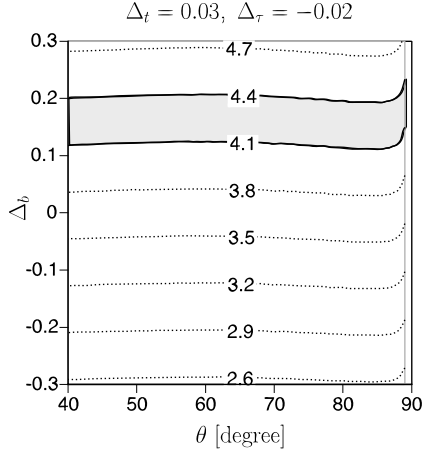}
\end{center}
\caption{The prediction of bottom quark 
mass $m_b^{\overline{\text{MS}}}(m_b)$ in the $SO(10)$ unification
with the modified lopsided mass matrices (\ref{lopY2}). Here we take
input parameters the same as in the $SU(5)$ symmetric 
case (Fig.~\ref{thetaDb}).\bigskip}
\label{thetaDb2}
\end{figure}

An important effect of group-theoretical factor appears in the
bottom/tau mass ratio. With the modified asymmetrical Yukawa 
matrices (\ref{lopY2}) at hand, the initial mass ratio at the GUT
scale is estimated as
\begin{equation}
  \frac{m_b(M_G)}{m_\tau(M_G)} \;=\; \frac{\sqrt{5}}{3} \;\simeq\; 0.75.
  \label{b-tau}
\end{equation}
Therefore the RG evolution predicts a low-energy value of the
bottom/tau mass ratio without including SUSY threshold corrections
smaller than the previous $SU(5)$ symmetric lopsided model. The
modified bottom/tau mass ratio (\ref{b-tau}) indicates that the
tree-level bottom quark mass at low energy is decreased from the 
case $Y_d=Y_e^{\rm T}$ and now requires non-vanishing SUSY threshold
correction $\Delta_b$. In Fig.~\ref{thetaDb2}, we show the
prediction of bottom quark mass as the function 
of $\Delta_b$ and $\theta$, where the input parameters are taken as
the same as in the previous figure~\ref{thetaDb}. We find that the
bottom quark mass is rather insensitive to $\theta$ and a sizable and
positive $\Delta_b$ is needed to attain the observed bottom quark mass
in a wide range of parameter 
space; $0.1\lesssim\Delta_b\lesssim0.2$. We numerically checked that
this result is not changed qualitatively by varying other input
parameters. The $\theta$ dependence of $\tan\beta$ is also examined
and found to be almost the same as the previous $SU(5)$ symmetric case
(Fig.~\ref{thetaDb}).

The above result, i.e.\ a positive $\Delta_b$, implies that 
the $\mu$ parameter must be positive in the wide range of $\theta$ to
reproduce the correct bottom quark mass. That is quite contrast to 
the $SU(5)$ symmetric lopsided case $Y_d=Y_e^{\rm T}$ with which a
small $\cos\theta$ prefers a negative value of $\mu$. An important
point is that a positive $\mu$ parameter makes the cancellation
possible among different $b\to s\gamma$ decay amplitudes via the
charged Higgs boson and superparticles. The total branching ratio of
the $b\to s\gamma$ rare process can be suppressed. The compatibility
of a positive $\mu$ parameter and a relatively small value 
of $\cos\theta$ implies that, if one considers PQ and R symmetric mass
spectrum, the $b\to s\gamma$ constraint is easily avoided than 
the $SU(5)$ symmetric case. That is explicitly shown by a detailed
analysis in the following subsection.

%%%%%%%%%%%%%%%%%%%%%%%%%%%%%%%%%%%%%%%%%%%%%%%%%%%%%%%%%%%%%%%%%%%%%%
\subsubsection{Radiative EWSB}
%%%%%%%%%%%%%%%%%%%%%%%%%%%%%%%%%%%%%%%%%%%%%%%%%%%%%%%%%%%%%%%%%%%%%%
As in the previous analyses, it is useful to solve the EWSB conditions
about the CP-odd neutral Higgs 
mass $M_A$ and the $\mu$ parameter. They are determined by the
GUT-scale SUSY-breaking parameters through the MSSM RG equations with
right-handed neutrino couplings. The solutions therefore implicitly
depend on the mixing angle $\theta$ which relates to neutrino large
generation mixing. We here focus on the compatibility of successful
radiative EWSB with approximate PQ and R symmetric mass spectrum. As
has been discussed, these symmetries are favorable for obtaining
acceptable bottom quark mass, $b\to s\gamma$ decay rate, and suitable
amount of dark matter component of the universe.

\begin{figure}[t]
\begin{center}
\includegraphics[width=5.5cm,clip]{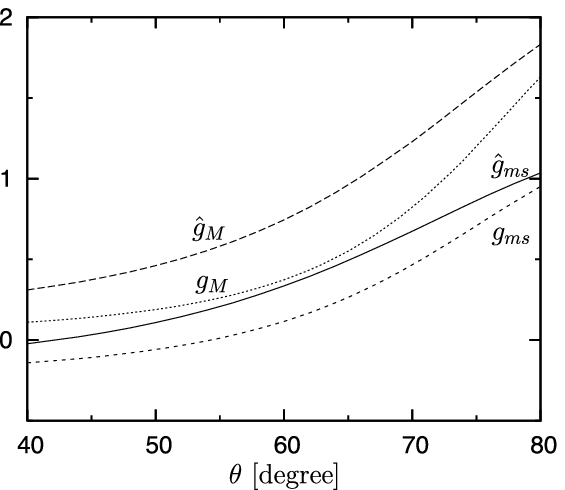}\hspace*{1cm}
\includegraphics[width=5.5cm,clip]{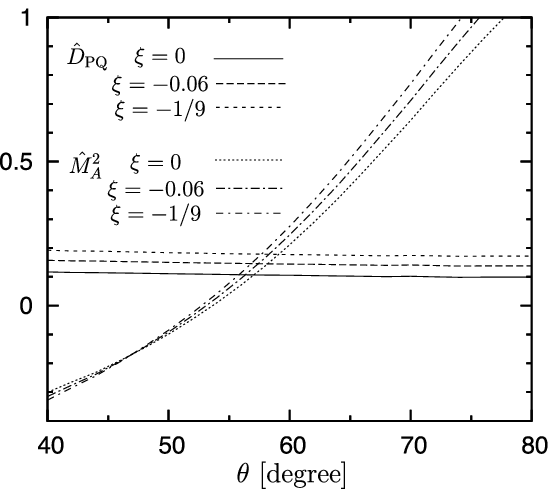}
\end{center}
\caption{The left figure shows the RG solution for the CP-odd neutral
Higgs mass, in particular, the scalar and gaugino 
pieces $\hat g_{ms}$ and $\hat g_M$ [for the parametrization, see
Eq.~(\ref{mafit3})]. For comparison, the corresponding 
coefficients $g_{ms}$ and $g_M$ are also presented. The right figure
is the same as Fig.~\ref{PQR} but for the modified Yukawa 
matrices (\ref{lopY2}). In these figures, we take input parameters as
the same as in Fig.~\ref{PQR}.\bigskip}
\label{fit4}
\end{figure}

The matching conditions for the MSSM SUSY-breaking parameters at the
GUT-breaking scale are supposed to be the same as those in the
previous $SU(5)$ symmetric case.\footnote{The existence of the 
factor $-1/3$ in the modified matrix $Y_d$ requires some
higher-representation fields or higher-dimensional operators in the
Higgs sector. In the latter case, the MSSM matter and Higgs fields
apparently have the same forms of couplings as in 
the $SU(5)$ symmetric lopsided model, and all the matching conditions
are unchanged. In the former case, however, higher-representation
multiplets modify radiative effects. For example, if a $144$-plet
Higgs of $SO(10)$ is adopted for the factor $-1/3$, only the matching
condition for down-type Higgs mass is modified 
to $m_{H_d}^2(M_G)=m_0^2+\big(\frac{4}{5}\cos^2\theta
+\frac{17}{9}\sin^2\theta\big)\Delta 
+(-2\cos^2\theta +3\sin^2\theta)D$. We found that the modification is
so small that the results are unchanged almost quantitatively.} 
The solutions to the RG equations and EWSB conditions are therefore
almost similar to the previous ones (\ref{mafit3}) and
(\ref{mufit3}). We here denote the coefficients in the present RG
solutions as $\hat g$'s and $\hat h$'s corresponding 
to $g$'s and $h$'s in (\ref{mafit3}) and (\ref{mufit3}). We show in
Fig.~\ref{fit4} the 
coefficients $\hat g_{ms}$ and $\hat g_M$ in the RG solutions 
for $M_A^2$. For comparison we also re-present the corresponding
factors $g_{ms}$ and $g_M$. The other $\hat g_x$'s and $\hat h_x$'s
are found to have no sizable differences from corresponding quantities; 
$\hat g_x\simeq g_x$ and $\hat h_x\simeq h_x$ ($x\neq ms, M$).
As in the previous $SU(5)$ symmetric 
case, $\hat g_{ms}$ and $\hat g_M$ have large $\theta$ dependence and
increase as $\theta$, since it directly controls the relative strength
of Yukawa couplings. Thus a small value of $\cos\theta$ generally
leads to a positive mass squared of CP-odd neutral Higgs
boson. Moreover in the present model, the 3-2 element in $Y_d$ is
smaller than the $SU(5)$ symmetric case (\ref{lopY}). That 
makes $m_{H_d}^2$ larger in the infrared regime through smaller
effects of Yukawa terms in the RG evolution 
of $m_{H_d}^2$. Consequently, the CP-odd neutral Higgs mass is easily
raised. This is numerically understood from the behavior 
of $\hat g_{ms}$ and $\hat g_{M}$ in Fig.~\ref{fit4}; they are always
larger than the corresponding $g$'s for fixed values 
of $\theta$ and $m_t^{\rm pole}$.

Unlike $\hat g$'s, the coefficients $\hat h$'s in the RG solution 
for $|\mu|^2$ are insensitive to $\theta$ and the results in 
the $SU(5)$ symmetric case are equally applied to the present
case. This implies that the PQ symmetric spectrum, that is, a small
size of $|\mu|$, is achieved with a positive $D$ term. The coexistence
of approximate PQ symmetry and a well-lifted CP-odd neutral Higgs mass
is allowed depending on the mixing angle $\theta$. With the modified
lopsided Yukawa couplings, we are able to have a 
smaller $\theta$ consistent with the experimental bound on $M_A$, as
shown in Fig.~\ref{fit4} (the right figure). The figure shows that a
weak bound $\theta\gtrsim 55^\circ$ and a positive $D$-term
contribution make PQ and R symmetric radiative EWSB available.

%%%%%%%%%%%%%%%%%%%%%%%%%%%%%%%%%%%%%%%%%%%%%%%%%%%%%%%%%%%%%%%%%%%%%%
\subsubsection{Parameter Space Analysis}
\begin{figure}[t]
\begin{center}
\includegraphics[width=5.5cm,clip]{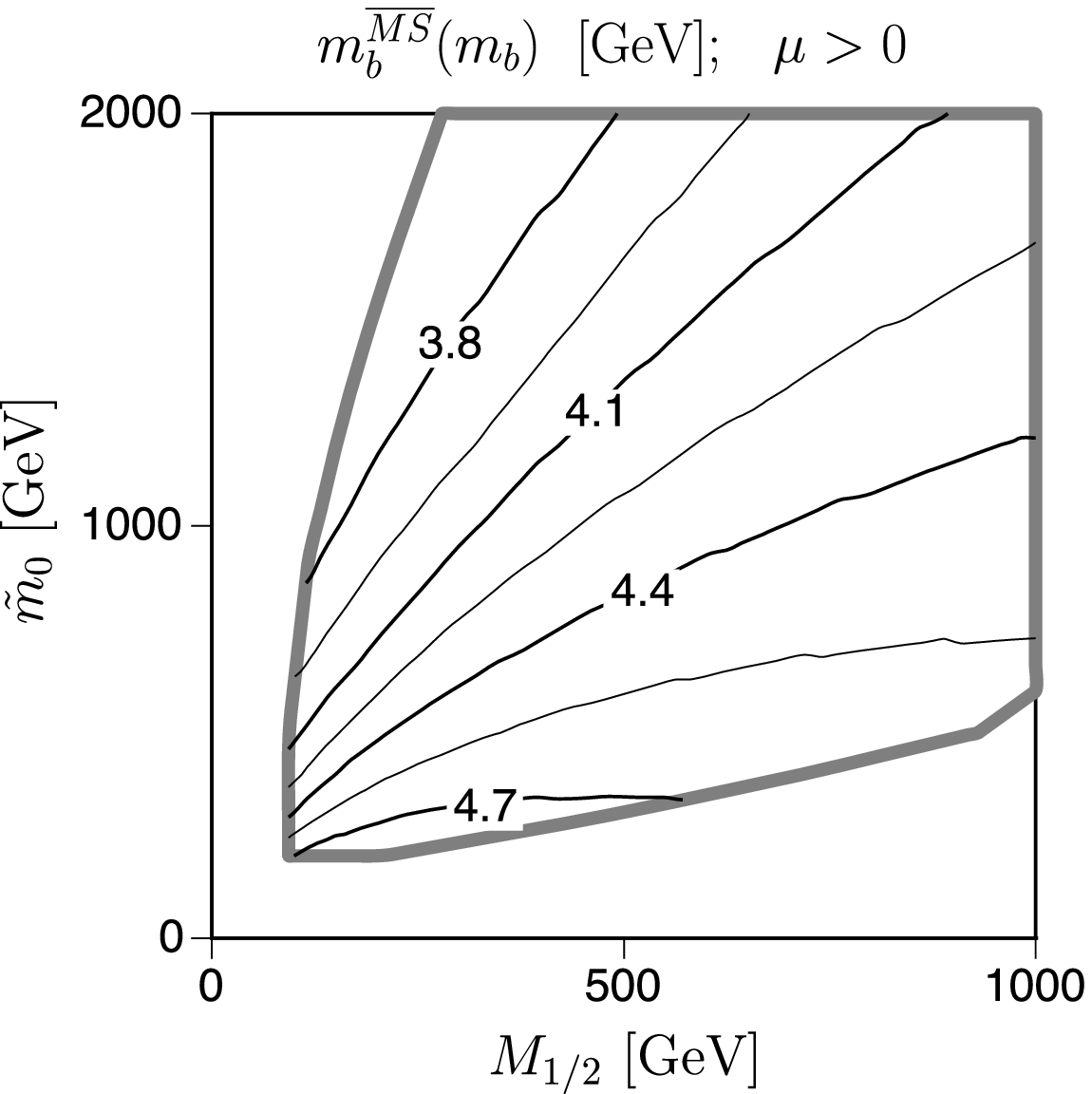}\hspace*{1cm}
\includegraphics[width=5.5cm,clip]{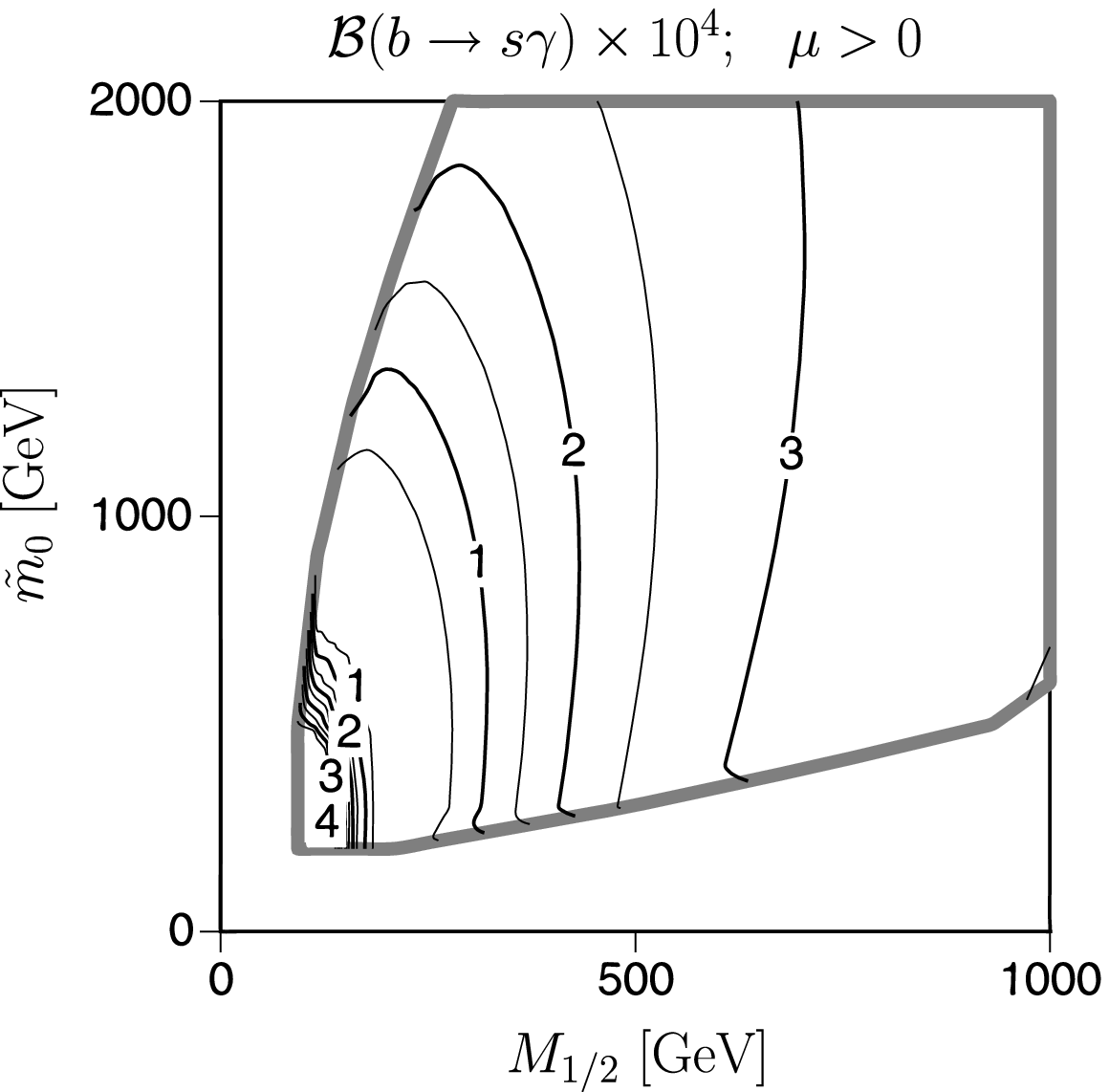}
\end{center}
\caption{The $M_{1/2}$--$\tilde m_0$ parameter space consistent with
the EWSB conditions, the experimental mass bounds on superparticles,
and the requirement that the LSP is charge neutral, i.e.\ the same as
Fig.~\ref{Mgm0_lop} but for different Yukawa forms and a
resultant positive $\mu$ parameter. The bottom quark 
mass $m_b^{\overline{\text{MS}}}(m_b)$ and 
the $b\to s\gamma$ branching ratio are also shown in the figures. Here
we set $\theta=65^\circ$ and the GUT-scale SUSY-breaking parameters 
as $A_0=0$, $\xi=0$, and $D=0.1\tilde m_0^2$. In each figure, the
right-bottom region is excluded by the scalar tau LSP and the
left-upper side is ruled out by the experimental mass bounds of
charginos and neutralinos. In this parameter space, we 
obtain $\tan\beta\simeq45$.\bigskip}
\label{Mgm0_lop2}
\end{figure}

%%%%%%%%%%%%%%%%%%%%%%%%%%%%%%%%%%%%%%%%%%%%%%%%%%%%%%%%%%%%%%%%%%%%%%
It is now clear that a crucial discrepancy between 
the $SU(5)$ symmetric and modified lopsided Yukawa matrices is 
the (signature of) threshold correction to bottom quark mass. This is
a direct consequence of $SU(5)$ breaking in the Yukawa 
sector, $Y_d\neq Y_e$. The modification of Yukawa couplings lowers the
tree-level bottom quark mass and hence requires a 
positive $\mu$, which in turn is consistent with PQ and R symmetric
radiative EWSB\@. A more important implication of 
positive $\mu$ parameter is that an excessive $b\to s\gamma$ branching
ratio can be reduced to be consistent to the observation with a
cancellation among various partial amplitudes. In
Fig.~\ref{Mgm0_lop2}, we show 
the $M_{1/2}$--$\tilde m_0$ parameter space consistent with the EWSB
conditions, the experimental mass bounds on superparticles, and the
requirement that the LSP is charge neutral, i.e.\ the same as
Fig.~\ref{Mgm0_lop} but for different Yukawa forms and a
positive $\mu$ parameter. In the figures, the predictions of bottom
quark mass $m_b^{\overline{\text{MS}}}(m_b)$ and 
the $b\to s\gamma$ branching ratio are shown in the allowed parameter
regions. As an example, we set $\theta=65^\circ$ and $\xi=0$. The
right-bottom region is excluded by the LSP scalar tau lepton and the
left-upper side is ruled out by the current mass bounds of charginos
and neutralinos.

From the left figure for the bottom quark mass prediction, we find
that there exists a large parameter region to reproduce the observed
bottom quark mass. Roughly speaking, the region 
around $\tilde m_0\sim2M_{1/2}$ is preferred and fat scalar
particles are not needed. If one wanted to consider R symmetric
spectrum $\tilde m_0>2M_{1/2}$, a larger $D$-term contribution should
be included.

On the other hand, the right figure indicates that the constraint 
from $b\to s\gamma$ rare decay is much weakened and is readily made
within the experimentally allowed range. We also find two separate
regions consistent with the 
observation; $M_{1/2}\sim200$~GeV and $M_{1/2}\gtrsim400$~GeV\@. In
both cases, superparticles are relatively light, a few hundred
GeV\@. This is a sharp contrast to the other scenarios discussed in
this paper. Such a light spectrum is due to the positive sign 
of $\mu$ and the resultant cancellation of $b\to s\gamma$ decay
amplitudes from the charged Higgs boson and superparticles. In the
narrow parameter region between these two separate ones, the branching
ratio tends to be too small since a relatively small gaugino mass
enhances the chargino-loop contribution and the cancellation is too
effective. We also note that, in
Fig.~\ref{Mgm0_lop2} for ${\cal B}(b\to s\gamma)$, the left allowed
region with a tiny gaugino mass is excluded by the experimental lower
bound of the lightest Higgs boson mass, since the radiative corrections
from the top sector~\cite{HiggsRC} are not sufficient to meet the bound.

To summarize, in the $SO(10)$ unification with modified lopsided form
of Yukawa couplings, superparticles exhibit light and non-hierarchical
mass spectrum to satisfy the observed values of bottom quark mass 
and $b\to s\gamma$ decay rate in the EWSB vacuum. This behavior is due
to the fact that the threshold correction to bottom quark mass is
needed to be positive and a bit large. That requires a 
positive $\mu$ parameter with which the $b\to s\gamma$ decay rate is
made suppressed via diagram cancellations. The PQ and R symmetries are
weakly attained in this model. The lightest neutralino and chargino
consist of gaugino components, but possibly, a sizable amount of
higgsino components is involved, which may be suitable for
cosmological issues such as LSP dark matter. These features are quite
different from the other scenarios in this paper and a detailed
analysis is left to future work.

%%%%%%%%%%%%%%%%%%%%%%%%%%%%%%%%%%%%%%%%%%%%%%%%%%%%%%%%%%%%%%%%%%%%%%
\bigskip
\section{Summary}
\label{sec:sum}
%%%%%%%%%%%%%%%%%%%%%%%%%%%%%%%%%%%%%%%%%%%%%%%%%%%%%%%%%%%%%%%%%%%%%%
In this work we have investigated the low-energy phenomenology of
supersymmetric $SO(10)$ unification with neutrino effects suggested by
its tiny mass scale and large generation mixing in the lepton
sector. The analysis includes the radiative electroweak symmetry
breaking, the third-generation fermion masses, and the flavor-changing
rare processes.

In the general Yukawa unification with large $\tan\beta$, the observed
fermion masses, especially the bottom quark mass, require suppressed
threshold corrections at low-energy decoupling scale of
superparticles. The smallness of the corrections is ensured with
superparticle mass spectrum which is approximately PQ and R symmetric,
that is, the gaugino masses and supersymmetric Higgs mass parameter
should be smaller than scalar masses. However in the 
minimal $SO(10)$-type unification without including neutrino
couplings, the successful radiative EWSB leads to a large gaugino 
mass $M_{1/2}\gtrsim\tilde m_0$ to make the CP-odd neutral Higgs mass
experimentally allowed. Consequently, low-energy SUSY-breaking
parameters are strongly correlated to the gaugino masses, following
which the threshold correction to bottom quark mass tends to be large
and unacceptable.

Then we have included the effects of neutrino couplings in RG
evolution down to the intermediate scale where right-handed neutrinos 
are decoupled. The newly introduced parameters are the neutrino Yukawa
coupling of similar order of the other third-generation Yukawa
couplings, the mass and trilinear couplings of right-handed scalar
neutrinos. We have found that any of these three types of neutrino
couplings is of great use for a successful EWSB\@. The parameter 
space $M_{1/2}\ll \tilde m_0$ gets allowed and the bottom mass
threshold correction and the $b\to s\gamma$ decay rate are
suppressed. The CP-odd neutral Higgs mass squared receives several
positive contributions from the neutrino couplings in addition to the
usual gaugino mass effect. Consequently the PQ and R symmetric
superparticle spectrum can be consistent with the successful
EWSB\@. For the positive $\mu$ case, excessive threshold corrections
to bottom quark mass are suppressed for such type of  
superparticle spectrum. The $b\to s\gamma$ branching ratio is made
within the experimental range, e.g.\ for $M_{1/2}=300$~GeV 
and $\tilde m_0\gtrsim 2.5$~TeV\@. The $\mu$ parameter can also be
made small and there appears the parameter region which accommodates
the higgsino-like lightest neutralino. For the negative $\mu$ case,
rather heavy scalars are inevitable because $\Delta_b$ should be
highly suppressed. The constraint from $b\to s\gamma$ also requires
heavy scalars $\tilde m_0\gtrsim10$~TeV\@. If the top quark is taken
to be lighter, the phenomenological constraints become satisfied by
lighter scalars, $\tilde m_0\gtrsim 6$~TeV\@. In general, the
low-energy superparticle spectrum is preferred to have hierarchical
structure: light gauginos/higgsinos and heavy scalars are expected.

Finally we have taken into account the observed large generation
mixing of neutrinos. In particular we have focused on the case that
the maximal mixing between the second and third generations arises
from the charged-lepton sector. An important factor is the 
parameter $\theta$ in high-energy Higgs sector which determines the
neutrino large generation mixing. In the exact unification of down and
charged-lepton Yukawa couplings, the tree-level bottom quark mass
increases as $\theta$, and a negative $\mu$ parameter is required. In
addition, the CP-odd neutral Higgs mass is raised with a smaller value
of $\cos\theta$. Thus the PQ and R symmetric radiative EWSB is
possible with $\theta\gtrsim60^\circ$ and a positive $D$-term
contribution. The observed bottom quark mass and 
the $b\to s\gamma$ constraint are easily satisfied with a few TeV
scalar quarks. The lightest neutralino and chargino contain a sizable
amount of higgsino components, which may be suitable for cosmological
issues such as LSP dark matter. For $\theta\lesssim60^\circ$, while
the PQ symmetric mass spectrum is not consistent with the positiveness
of CP-odd neutral Higgs mass squared, the bottom quark mass is made
within the experimental range only with help of R symmetry and 
relatively heavy scalars $\tilde m_0\gtrsim15$~TeV\@. We have also
examined the modification of $SU(5)$ symmetric Yukawa couplings by
introducing a group-theoretical factor for the masses of the
second-generation fermions being properly reproduced. A crucial
consequence of this modification is that the low-energy bottom quark
mass without threshold correction is turned out to be reduced and a 
positive $\mu$ parameter is predicted. In this case, 
the $b\to s\gamma$ decay rate is made suppressed via diagram
cancellations. Superparticles also exhibit light and non-hierarchical
mass spectrum. These features are quite different from the other
scenarios discussed in this paper.

In any case, our study has shown that the neutrino coupling effects
induce new types of EWSB in $SO(10)$ unification consistent with
various experimental constraints. Physical implications of these
scenarios such as predicted superparticle spectrum would be tested in
the future experimental searches of supersymmetry.

\bigskip
%%%%%%%%%%%%%%%%%%%%%%%%%%%%%%%%%%%%%%%%%%%%%%%%%%%%%%%%%%%%%%%%%%%%%%
\subsection*{Acknowledgments}
The authors would like to thank Atsushi Watanabe for helpful
discussions. This work is supported by scientific grant from the
Ministry of Education, Science, Sports, and Culture of Japan
(No.~17740150) and by grant-in-aid for the scientific research on
priority area (\#441) "Progress in elementary particle physics of the
21st century through discoveries of Higgs boson and supersymmetry"
(No.~16081209).

%%%%%%%%%%%%%%%%%%%%%%%%%%%%%%%%%%%%%%%%%%%%%%%%%%%%%%%%%%%%%%%%%%%%%%

\end{document}